\documentclass[12pt,a4paper,final]{iopart}

\usepackage{iopams}  
\usepackage{graphicx}
\usepackage[breaklinks=true,colorlinks=true,linkcolor=blue,urlcolor=blue,citecolor=blue]{hyperref}
\usepackage{amsfonts}

\expandafter\let\csname equation*\endcsname\relax
\expandafter\let\csname endequation*\endcsname\relax
\usepackage{amsmath}

\usepackage{amssymb}
\usepackage{bm}
\usepackage{dcolumn}
\usepackage{graphicx}
\usepackage{graphics}
\usepackage[latin1]{inputenc}
\usepackage{latexsym}
\usepackage{rotating}
\usepackage{hyperref}
\usepackage[all]{hypcap} 
\usepackage{xspace} 
\usepackage[usenames,dvipsnames]{color}
\usepackage{mathrsfs}
\usepackage{subfigure}

\usepackage{ulem}
\usepackage{outlines}
\usepackage{enumitem}
\setenumerate[1]{label=\Roman*.}
\setenumerate[2]{label=\Alph*.}
\setenumerate[3]{label=\roman*.}
\setenumerate[4]{label=\alph*.}

\definecolor {darkgreen}{rgb}{0.2,0.7,0.2}

\usepackage{etoolbox}

\makeatletter
\def\@mkboth#1#2{}
\newlength\appendixwidth
\preto\appendix{\addtocontents{toc}{\protect\patchl@section}}
\newcommand{\patchl@section}{%
  \settowidth{\appendixwidth}{\textbf{Appendix }}%
  \addtolength{\appendixwidth}{1.5em}%
  \patchcmd{\l@section}{1.5em}{\appendixwidth}{}{\ddt}%
}
\makeatother

\newcommand\be{\begin{equation}}
\newcommand\ba{\begin{eqnarray}}
\newcommand\ee{\end{equation}}
\newcommand\ea{\end{eqnarray}}

\newcommand\bw{\begin{widetext}}
\newcommand\ew{\end{widetext}}

\newcommand{\nn}{\nonumber}

\newcommand{\ISCO}{{\mbox{\tiny ISCO}}}
\newcommand{\LSO}{{\mbox{\tiny LSO}}}

\newcommand{\K}{{\mbox{\tiny K}}}
\newcommand{\R}{{\mbox{\tiny R}}}
\newcommand{\A}{{\mbox{\tiny A}}}
\newcommand{\orb}{{\mbox{\tiny orb}}}
\newcommand{\cj}{{\mbox{\tiny c}}}
\newcommand{\sj}{{\mbox{\tiny s}}}
\newcommand{\N}{{\mbox{\tiny N}}}
\newcommand{\IMF}{{\mbox{\tiny IMF}}}
\newcommand{\OA}{{\mbox{\tiny OA}}}
\newcommand{\Osc}{{\mbox{\tiny Osc}}}
\newcommand{\DE}{{\mbox{\tiny DE}}}
\newcommand{\osc}{{\mbox{\tiny osc}}}
\newcommand{\Sec}{{\mbox{\tiny sec}}}
\newcommand{\lum}{{\mbox{\tiny L}}}
\newcommand{\PA}{{\mbox{\tiny PA}}}
\newcommand{\I}{{\mbox{\tiny I}}}
\newcommand{\BG}{{\mbox{\tiny BG}}}

\newcommand{\spa}{{\mbox{\tiny spa}}}

\begin{document}
\title{The Eccentric Behavior of Inspiraling Compact Binaries}

\author{Nicholas Loutrel}
\address{Department of Physics, Princeton University, Princeton, NJ, 08544, USA}
\ead{nloutrel@princeton.edu}

\author{Samuel Liebersbach}
\address{eXtreme Gravity Institute, Department of Physics, Montana State University, Bozeman, MT 59717, USA.}
\ead{samuel.liebersbach@gmail.com}

\author{Nicol\'as Yunes}
\address{eXtreme Gravity Institute, Department of Physics, Montana State University, Bozeman, MT 59717, USA.}
\ead{nicolas.yunes@montana.edu}

\author{Neil Cornish}
\address{eXtreme Gravity Institute, Department of Physics, Montana State University, Bozeman, MT 59717, USA.}
\ead{ncornish@montana.edu}

\date{\today}

\begin{abstract} 

Even if there is no gauge invariant definition of eccentricity, it has an important impact on the observed gravitational wave signal of such systems, generating power in all possible harmonics of the orbital period. We here clarify the possible discrepancies between different eccentricity parameters used to describe the orbital dynamics of binary systems across different approximations, specifically the post-Newtonian approximation, the self-force approximation, and numerical relativity. To this end, we highlight disparities between the typically used orbit averaged method of evolving binary systems under radiation reaction, and more direct techniques of solving the two-body problem in post-Newtonian theory. We show, both numerically and analytically, that the orbit averaged method breaks down in the late inspiral, failing to capture a strong secular growth in the Keplerian eccentricity parameter and producing a orbital de-phasing relative to direct integration of the two-body equations of motion. We show that the secular growth and de-phasing affect the observed gravitational wave signal, which could bias how accurately we may recover parameters for systems with signal-to-noise ratios $\gtrsim 100$. We further develop a frequency domain post-adiabatic waveform model to capture these effects, and study the precision to which we may estimate parameters with this model through a Fisher information matrix analysis.

\end{abstract}



\maketitle


\newpage
\section{Introduction}
\label{intro}

Gravitational waves (GWs) from compact binary systems have provided us with a keyhole through which to observe the Universe, and with new detectors both planned and in development, that keyhole is only expected to expand~\cite{GW150914, Abbott:2016nmj, Abbott:2017vtc, Abbott:2017gyy, Abbott:2017oio, TheLIGOScientific:2017qsa}. From a theoretical perspective, GWs, particularly from binary systems, have been studied extensively over the past fifty years. The existence of GWs was inferred indirectly for the first time through observations of the Hulse-Taylor binary~\cite{Hulse:1974eb}, a system composed of a neutron star (NS) and a pulsar. By accurately, tracking the arrival time of the pulses over several decades, Taylor \& Weisberg were able to show that the orbital period of the binary was decaying in a manner fully consistent with the prediction of GW emission from General Relativity~\cite{Weisberg:1981bh, Weisberg:1981mt, Taylor:1982zz, Weisberg:2004hi}.

As of today, the Hulse-Taylor binary exists in an elliptical orbit with eccentricity of $e \sim 0.7$. The effect of radiation reaction on such binary systems was first calculated by Peter \& Mathews~\cite{PetersMathews, Peters:1964zz}, who showed that the rates of energy and angular momentum loss cause the eccentricity to decay as the binary inspirals. A straightforward calculation using these results reveals that by the time the Hulse-Taylor binary enters the detection band of ground-based GW detectors, the binary's eccentricity will be $e \sim 10^{-5}$~\cite{Maggiore:1900zz}. Currently, ground-based detectors are expected to be able to measure eccentricities of $e > 10^{-2}$~\cite{PhysRevD.92.044034, Huerta:2017kez, Brown:2009ng, Gondan:2017hbp}. For all intensive purposes then, the Hulse-Taylor binary would be effectively circular and could be recovered using circular GW templates by the time it begins to emit  GWs in the detection band of ground-based detectors.

Systems such as the Hulse-Taylor binary provided the following picture: widely separated binaries, even with large (close to unity) eccentricity, will ``circularize'' by the time they emit GWs in the frequency band of ground-based detectors. This was the canonical view of binary sources for many years, until recent population synthesis studies began to question the canon. These studies have shown that although most sources will indeed be approximately circular, globular clusters and galactic nuclei could host a non-negligible population of moderate ($e > 0.1$) and high ($e \sim 1$) eccentric binaries that emit in the band of ground-based detectors~\cite{Belczynski:2016obo, Park:2017zgj, Samsing:2017xmd, Samsing:2013kua, 2009MNRAS.395.2127O, O'Leary:2007qa, 2013ApJ...773..187N}. Currently, LIGO~\cite{Abramovici:1992ah, Harry:2010zz, ligo} and Virgo~\cite{Caron:1997hu, TheVirgo:2014hva} are not sensitive enough to allow for the extraction of small eccentricities, and thus, the roughly dozen current detections are fully consistent with circular binaries. But as detectors are improved, future observations could allow for the extraction of eccentricity, and thus, provide us with one more piece of the puzzle to elucidate the origin of the currently observed systems.

The very notion of eccentricity in tight compact binaries, however, is difficult to define because in General Relativity elliptical orbits do not close. In Newtonian gravity, orbits close between consecutive pericenter passages and \textit{orbital eccentricity} is a well defined parameter associated with the ellipticity of the orbit. In General Relativity, elliptical orbits undergo precession of periastron and, thus, do not necessarily close unless the ratio of the orbital period to the precession timescale is a rational number. Further, GWs induce dissipation of the system, and so, even in the absence of precession, the binary's orbit does not, in general, close. The typical Newtonian notion of orbital eccentricity is, thus, not well defined in General Relativity.

There are, however, multiple ways of solving the binary problem in General Relativity, and multiple ways of defining eccentricity in specific orbital parameterizations that resemble the Newtonian definition. The post-Newtonian (PN) formalism~\cite{Chandrasekhar:1969, Chandrasekhar:1965, Chandrasekhar:1970, PW, Blanchet:2013haa} seeks to solve the Einstein field equations in a slow motion, weak field expansion, specifically $v^{2}/c^{2} \sim G M/c^{2} R \ll 1$, where $M, R,$ and $v$ are the characteristic mass, length, and velocity of the system. The binary problem in the PN formalism has been well studied~\cite{Bernard:2015njp, Bernard:2016wrg, Bernard:2017bvn, Bernard:2017mac, Marchand:2017pir}, with the equations of motion generically written as $\vec{a} = \vec{f}_{\rm cons} + \vec{f}_{\rm diss}$, where $\vec{a}$ is the relative acceleration of the binary, $\vec{f}_{\rm cons} = \vec{f}_{\rm N} + {\cal{O}}(c^{-2})$ is the conservative (time-reversal even) part of the relative force, and $\vec{f}_{\rm diss} = c^{-5} \vec{f}_{\rm 2.5PN} + {\cal{O}}(c^{-7})$ is the dissipative (time-reversal odd) part of the relative force due to radiation reaction. The problem is then to find a solution for the binary's motion under these equations of motion.

Generally, the method of solving the PN equations of motion for binary systems is to first solve for the binary's motion in the absence of GW emission, and then to use perturbation theory to promote the parameters of said orbit, such as the orbital energy and angular momentum, to functions of time. The orbit can be described in two different ways. We will refer to the first as the Lincoln \& Will formalism~\cite{Lincoln:1990ji, Mora:2003wt, Will:2016pgm}, where the orbit is treated through a Newtonian parameterization. The Newtonian orbital parameters, such as the semi-latus rectum $p$ and eccentricity $e$, are then promoted to functions of time that change due to a perturbing force $\delta \vec{f}_{\rm cons} = c^{-2} \vec{f}_{\rm 1PN} + {\cal{O}}(c^{-4})$, in exactly the same way radiation reaction is treated. This method benefits from having only one parameter that is interpreted as eccentricity, but suffers from the fact that circular and parabolic orbits do not necessarily correspond to $e = 0$ and $e = 1$, respectively. The second is the formalism of Damour \& Deruelle~\cite{zbMATH03938612, zbMATH04001537}, also known as the quasi-Keplerian (QK) formalism~\cite{Schaefer1993196, 0264-9381-12-4-009, Memmesheimer:2004cv, Damour:2004bz, Konigsdorffer:2006zt}, which obtains a PN accurate parameterization of the orbit by solving $\vec{a} = \vec{f}_{\rm cons}$ analytically order by order in a PN expansion. The QK formalism has three eccentricities: the time eccentricity $e_{t}$ (associated with the orbital period), the radial eccentricity $e_{r}$ (associated with the radial motion), and the azimuthal eccentricity $e_{\phi}$ (associated with the azimuthal motion), which take the appropriate limit for circular orbits, i.e. $e_{t} = e_{r} = e_{\phi} = 0$.

Complementary to the PN formalism, the self-force formalism~\cite{Barack:2018yvs} seeks to solve the Einstein field equations to all orders in the orbital velocity, but in a small mass ratio expansion. This formalism is suitable for extreme-mass-ratio inspirals (EMRIs), in which a small compact object zooms and whirls into a supermassive one, as it emits GWs. To leading order in the mass ratio, the small object behaves like a test mass moving in the background generated by a Schwarzschild or Kerr black hole, and its trajectory is described by a geodesic of the background geometry~\cite{Thorne:1984mz}. At next order in the mass ratio, the spacetime is perturbed by the smaller mass, with the perturbation generating the emission of GWs, which causes back reaction on the orbit and for the small mass to inspiral into the supermassive BH~\cite{Mino:1996nk, Quinn:1996am}. This (leading-order dissipative) self-force on the small mass, like in the PN formalism, leads to the equations of motion $\vec{a} = \vec{f}_{\rm BG} + \vec{f}_{\rm SF}$, where $\vec{f}_{\rm BG}$ is due to the curved background spacetime of the massive object, and $\vec{f}_{\rm SF}$ is the (dissipative) self-force acting on the smaller mass.
   
To solve for the evolution of EMRIs, one must simultaneously solve the equations of motion of the small mass and the perturbations to the spacetime metric, a complicated problem due to the presence of many disparate scales that need to be resolved in the problem. These systems of equations are commonly solved using computational methods~\cite{Barack:2009ux, Wardell:2015kea, Barack:1999wf, Barack:2001bw, Barack:2007jh, Barack:2007we, Vega:2007mc}, as well as perturbative approaches, such as the self-consistent approximation~\cite{Gralla:2008fg,Pound:2009sm} and multiscale expansions~\cite{Bender, Hinderer:2008dm, Pound:2010pj, Pound:2015wva}. Eccentricity parameters can be constructed from geodesic motion, regardless of the background. A similar method to the Lincoln \& Will formalism was employed by Pound \& Poisson~\cite{Pound:2005fs, Pound:2007th, Pound:2007ti} to calculate a time varying eccentricity of the smaller body's orbit. Contrary to the usual behavior of the eccentricity parameter in the PN radiation reaction problem, the eccentricity parameter in the self-force formalism exhibits growth near the separatrix in the adiabatic limit~\cite{Cutler:1994pb}.

Separately, numerical relativity (NR)~\cite{Bishop:2016lgv} seeks to solve the full Einstein field equations numerically. This method relies on a 3+1 decomposition where the spacetime manifold is foliated with spacelike hypersurfaces~\cite{toolkit}, and the field equations and locations of punctures are solved for on each hypersurface. No orbital parameterization is assumed in this method. Historically, eccentricity was considered an undesirable feature in NR simulations for two reasons. First, the importance of eccentricity to GW sources for ground based detectors is only a recent development. Second, eccentricity introduces additional scales to the problem, specifically the periastron timescale, which can be orders of magnitude smaller than the orbital period if the eccentricity is sufficiently close to unity. The presence of such small timescales can complicate numerical simulations, since it requires very fine resolution to probe the smallest scales in the problem. However, the development of initial data for quasi-circular binaries proved to be complicated, since the radial velocity of the binary typically has a non-zero oscillatory piece for general initial data, indicating the presence of eccentricity. Methods have been developed to approximate the eccentricity, which are applied to eccentricity reduction methods, whereby initial data is varied to reduce this eccentricity parameter to below the level of numerical error~\cite{Boyle:2007ft, Pfeiffer:2007yz, Tichy:2010qa, Buonanno:2010yk}. Along with the active research pursuing eccentricity in source populations, recent simulations have sought mergers of eccentric binaries, and advances in algorithms and hardware have made such simulation possible with even moderate eccentricity~\cite{SXS, Chaurasia:2018zhg, Hinder:2017sxy, Dietrich:2015pxa, Moldenhauer:2014yaa, Gold:2011df, Hinder:2008kv, Hinder:2007qu, Maione:2016zqz, Kyutoku:2014yba, Sperhake:2007gu,PhysRevD.98.104028}. 

While these methods are distinct, observables should agree in a suitable overlapping region of the approximations. Eccentricity is, however, not a gauge invariant quantity, and these methods need not agree on the behavior of the eccentricity as the binary inspirals. Further, even in the PN formalism of the two-body problem, different approximations do not agree on the behavior of the eccentricity under radiation reaction. For example, in the PN formalism, one usually orbit averages the GW fluxes of energy and angular momentum, which cause the eccentricity to monotonically decrease. However, if one does not work in this orbit-averaged approximation, and instead calculates the effect of radiation reaction in a multiple scale analysis, one finds that the eccentricity can grow secularly in the late inspiral~\cite{Loutrel:2018ssg}, which is also seen within the self-force formalism. Yet, this seemingly disagrees with well known results in numerical relativity simulations, specifically that the eccentricity decrease throughout the coalescence. While eccentricity may itself not be a gauge invariant quantity, it does have an observable effect of the GWs emitted by the binary system. How does one reconcile these differences between formalisms and approximations in an unambiguous way? This work seeks to clarify the notion of eccentricity in inspiraling binaries and its impact on the observable GW signal. 

\subsection{Executive Summary}

We here extend the discussion that we started in~\cite{Loutrel:2018ssg}, related to the presence of secular growth in eccentricity in the relative-Newtonian-order radiation reaction (rN-RR) problem. Our goal is to provide a more in depth description of this effect, provide a mathematical framework for the computation of secular growth in inspiraling binaries, and determine its potential impact on GW observations. Primarily, we work in the PN formalism, but provide comparisons to both self-force and NR calculations where suitable and possible. 

We consider the evolution of binary systems in different approximations of the rN-RR problem, where $\vec{f}_{\rm cons} = -(M/r^{2}) \vec{n}$ and $\vec{f}_{\rm diss} = \vec{f}_{\rm 2.5PN}$. First, we numerically evolve the trajectory through the acceleration equation $\vec{a} = \vec{f}_{\rm cons} + \vec{f}_{\rm diss}$ to obtain the relative coordinates as a function of time. Complementary to this, we compute the evolution of the binary using the method of osculating orbits, where the orbits are treated as Keplerian ellipses, with their parameters functions of time rather than constants. Finally, we consider the orbit-averaged approximation, which hinges on the averaged balance laws, specifically, that the average GW fluxes must be balanced by the rate at which orbital energy and angular momentum are lost by the binary. By directly comparing the trajectories computed via these three different methods, we show that there is a dephasing of the orbit-averaged solution relative to the direct numerical integration of the acceleration equation for systems with small, and even vanishingly small, initial Keplerian eccentricity. Further, this comparison reveals that the osculating trajectory agrees with the direct numerical integration to double precision, even though the Keplerian eccentricity experiences a strong secular growth in the late inspiral. This indicates a breakdown in the orbit-averaged approximation in small eccentricity binaries that has not been previously considered.

Once we establish the breakdown of the orbit-averaged approximation, we develop the analytic framework necessary to describe the evolution of the binary beyond this simple approximation. We work in a multiple scale analysis (MSA), where we use the fact that the orbital period $T_{\orb}$ of the binary is much shorter than the radiation reaction timescale $T_{\R\R}$. This allows us to work perturbatively in $\zeta \sim T_{\rm orb}/T_{\R\R} \ll$ and solve for the evolution of the Keplerian orbital elements, such as the Keplerian eccentricity, as functions of time analytically. This also allows us to develop a \textit{post-adiabatic} approximation for the rN-RR problem, with the leading order solution, called the \textit{adiabatic} approximation, being consistent with the orbit-averaged approximation. First-order, post-adiabatic corrections then scale as ${\cal{O}}(\zeta)$, while second-order terms scale as ${\cal{O}}(\zeta^{2})$, and so on.  Since we are most interested in the small eccentricity limit where we have previously observed secular growth, we obtain solutions in a small eccentricity expansion to ${\cal{O}}(e_{\I}^{2})$, where $e_{\I}$ is the initial Keplerian eccentricity of the binary. From these solutions, we show that one can recover the secular growth of eccentricity observed in~\cite{Loutrel:2018ssg} by considering the definition of the Keplerian eccentricity in terms of the components of the Runge-Lenz vector.

With the evolution of the Keplerian eccentricity in hand, we consider the impact of post-adiabatic corrections and the secular growth on GW observations. We begin said discussion by considering the power contained in each harmonic of the waveform. It is well known that eccentricity creates extra harmonic content in the waveform of a binary system relative to the quasi-circular GWs from the same system. We show that the distribution of power across the harmonics does indeed contain evidence of the secular growth through comparison of our three different approximations for solving the rN-RR problem. We continue this discussion by direct comparisons of the waveforms computed via these different methods through a match calculation. We show that the mismatch, defined as one minus the match, between the direct numerical integration and the orbit-averaged approximation is typically around $10^{-4}$, indicating that orbit-averaged waveforms can bias recovered parameters for sufficiently high signal-to-noise ratio (SNR) observations. As a result, the recovered parameters may not be accurate relative to the true parameters of the binary.

We also seek to quantify how precise we may be able to recover eccentricity from GW observations using post-adiabatic effects. Toward this, we develop a post-adiabatic Fourier domain waveform through the application of the stationary phase approximation (SPA). We refer to this as the Post-Adiabatic eCcentric Multi-scale-analysis Next-to-leading-order (PACMAN) waveform. The PACMAN waveform takes the usual frequency domain waveform structure $\tilde{h}(f) = {\cal{A}}(f) e^{i \Psi(f)}$, where the phase $\Psi(f)$ and amplitude ${\cal{A}}(f)$ can be written as
\begin{align}
\Psi(f) &= \Psi_{0}(f) \left[1 + (\pi {\cal{M}} f)^{5/3} \delta \Psi_{\PA}(f)\right]
\\
{\cal{A}}(f) &= {\cal{A}}_{0}(f) \left[1 + (\pi {\cal{M}} f)^{5/3} \delta {\cal{A}}_{\PA}(f) + (\pi {\cal{M}} f)^{10/3} \delta {\cal{A}}_{{\mbox{\tiny 2}}\PA}(f)\right]
\end{align}
with ${\cal{M}}$ the chirp mass of the binary, $[\Psi_{0}(f), \tilde{{\cal{A}}}_{0}(f)]$ the orbit averaged phase and amplitude, and $[\delta \Psi_{\PA}(f), \delta {\cal{A}}_{\PA}(f), \delta {\cal{A}}_{{\mbox{\tiny 2}}\PA}(f)]$ post-adiabatic corrections to the phase and amplitude, respectively. We study the precision to which the eccentricity of the binary can be recovered using GW observations from LIGO through a Fisher information matrix analysis, which is applicable in the high SNR limit. We show, through comparison to orbit-averaged Fourier domain waveforms, that the PACMAN waveform does not provide significant improvements on the precision to which eccentricity can be measure with ground-based detectors.

The remainder of this paper presents the details of the conclusions summarized above. In Sec.~\ref{primer}, we review orbital parameterizations for eccentric binaries and methods of defining the eccentricity. We study the evolution of eccentricity under radiation reaction in Sec.~\ref{dynamics} and compare the different approximations of describing radiation-reaction effects in Sec.~\ref{comp}, with the results displayed through Figs.~\ref{traj-fig} and~\ref{traj-fig2}. We further contrast the growth detailed here with secular growth reported in other approximations in Sec.~\ref{comp2}. In Sec.~\ref{msa}, we seek a deeper understanding of the secular growth observed in the PN formalism, with analytic expressions for the time-domain evolution of Keplerian orbital elements available in Eqs.~\eqref{eq:b0-sol},~\eqref{eq:p0-sec},~\eqref{eq:phi-1-sec}-\eqref{eq:t-1-sec},~\eqref{eq:p1-osc}-\eqref{eq:t0-osc}, and~\eqref{eq:p1-sec}-\eqref{eq:t0-sec}. We show how to recover the Keplerian eccentricity in Eq.~\eqref{eq:rec-e}, with a comparison to the numerical evolution of the Keplerian eccentricity using the osculating approximation in Fig.~\ref{pa-fig}. Finally, in Sec.~\ref{waves}, we study the effect the growth has on GWs emitted by binary systems in the PN formalism. We compare the value of harmonic coefficients between the different methods of solving the rN-RR problem in Figs.~\ref{psi-fig-1}-\ref{psi-fig-2}, and compare the matches in Table~\ref{match}. We develop the PACMAN waveform in Sec.~\ref{pacman}, with the waveform given by Eqs.~\eqref{eq:h-pacman}-\eqref{eq:psi-pa}, and~\eqref{eq:calA-pa}-\eqref{eq:calA-2pa-rel}. We perform a Fisher analysis on the PACMAN waveform in Sec.~\ref{pe}, with the results given in Table~\ref{fisher-tab}. We set $G = c = 1$ for the remainder of this work.

\section{Conservative Mechanics of Eccentric Binaries}
\label{primer}

Before considering the evolution of eccentricity under radiation reaction, it is useful to study the conservative dynamics of the two-body problem, and the conservative definitions of eccentricity. This is what we will do in this section, starting with a definition of the Kepler problem to establish some notation, and then introducing the QK parameterization and other definitions of eccentricity. 

\subsection{Kepler Problem}
\label{newt}

We begin by studying the classical two-body problem in Newtonian gravity, which is equivalent to the leading PN order two-body problem. We consider a non-spinning binary system with component masses $m_{1}$ and $m_{2}$, and work in an effective one-body description, where a smaller mass $\mu = m_{1} m_{2}/ (m_{1} + m_{2})$ orbits around a larger mass $M = m_{1} + m_{2}$, fixed to the center of mass of the system. The smaller mass experiences an acceleration due to the gravitational force from the larger object, specifically $\vec{a} = - (M/r^{2}) \vec{n}$, where $\vec{n} = (\cos\phi, \sin\phi,0)$ is the radial unit vector, with $(r, \phi)$ the radial and azimuthal coordinates of the small mass, respectively. 

Symmetries can now be invoked to simplify the problem. Since the gravitational force acts on the smaller mass only along the radial direction (proportional to $\vec{n}$), the two-body problem admits a constant of motion associated with the azimuthal velocity $\dot{\phi}$. This constant is just the (reduced) orbital angular momentum $h = r^{2} \dot{\phi}$, and its constancy implies that the orbit of the small mass exists in a plane. This also implies that there are two additional constants of motion associated with the orientation of the orbital plane, relative to an arbitrary coordinate system. To define these, one can start with a planar coordinate system $(x,y,z)$ where the orbital angular momentum points in the z-direction. Using Euler angles, one can then rotate to a new spatial coordinate system $(X,Y,Z)$. The two constants associated with planar motion correspond to the inclination angle $\iota$, which is the angle between the Z-axis and the direction of the orbital angular momentum, and the longitude of the ascending node $\Omega$, the angle between the $X$-axis of the coordinate system and the ascending node of the orbital plane. A schematic of the orbital orientation can be found in Fig.~3.2 of~\cite{PW}.

With the conserved orbital angular momentum defined, the acceleration equation then becomes
\begin{equation}
\label{eq:accel}
\ddot{r} - \frac{h^{2}}{r^{3}} = - \frac{M}{r^{2}}\,.
\end{equation}
This equation can be integrated by multiplying both sides by the radial velocity $\dot{r}$ to obtain
\begin{equation}
\label{eq:en}
\frac{1}{2} \dot{r}^{2} = \epsilon - V_{\rm eff}(r) \,,
\end{equation}
where $\epsilon$ is the conserved (reduced) orbital energy, and $V_{\rm eff}(r)$ is the effective potential, specifically,
\begin{equation}
V_{\rm eff}(r) = \frac{h^{2}}{2 r^{2}} - \frac{M}{r}\,.
\end{equation}
The turning points of the orbit can be found by solving Eq.~\eqref{eq:en} when $\dot{r} = 0$. Alternatively, Eq.~\eqref{eq:en} can be factored such that $(1/2) \dot{r}^{2} = (1/r - 1/r_{+})(1/r - 1/r_{-})$, where 
\begin{equation}
r_{\pm} = \frac{M}{2 \epsilon} \left[1 \pm \left(1 - \frac{h^{2}}{M^{2}} \epsilon \right)^{1/2}\right]
\end{equation}
are the apocenter and pericenter distances, respectively. 

Constancy of the apocenter and pericenter are directly associated with the constancy of the orbital energy and angular momentum. From these quantities, we may define the semi-major axis 
\be
a = \frac{1}{2} (r_{-} + r_{+})\,,
\ee
and the Keplerian eccentricity 
\be
e_{\K} = \frac{r_{+} - r_{-}}{r_{+} + r_{-}}\,
\label{eq:eK-def}
\ee 
of the orbit. The orbits are then described as conic section with varying values of $e_{\K}$: $e_{\K} = 0$ corresponds to a circular orbit, $0<e_{\K}<1$ corresponds to an elliptical orbit, $e_{\K}=1$ corresponds to a parabolic orbit, and $e_{\K}>1$ corresponds to a hyperbolic orbit.

Returning to Eq.~\eqref{eq:accel}, we may perform a change of variables using $u = 1/r$ and change from time derivatives to azimuthal derivatives using $\dot{\phi}$, to obtain
\begin{equation}
\frac{d^{2}u}{d\phi^{2}} + u = \frac{M}{h^{2}}\,.
\end{equation}
This differential equation can be solved directly to reduce the two-body problem in Newtonian gravity to quadratures, specifically
\begin{align}
\label{eq:rkep}
r &= \frac{p}{1 + e_{\K} \cos\left(\phi - \omega\right)}
\\
\label{eq:phidotkep}
\dot{\phi} &= \left(\frac{M}{p}\right)^{3} \left[1 + e_{\K} \cos\left(\phi - \omega\right)\right]^{2}\,,
\end{align}
where $p = h^{2}/m$ is the semi-latus rectum of the orbit, and $\omega$ corresponds to the longitude of pericenter, the angle between the $x$-axis of the planar coordinate system and the direction of pericenter. 

The constancy of $\omega$, and thus, the constancy of the direction of pericenter is actually a special feature of the Kepler problem associated with a hidden SO(4) symmetry. This symmetry is only revealed through the conserved Runge-Lenz vector $\vec{A} = (1/M) \vec{v} \times \vec{h} - \vec{n}$, where $\vec{v}$ is the orbital velocity, $\vec{h}$ is the angular momentum vector, and $\times$ corresponds to the flat-space cross product between spatial vectors. The hidden symmetry can be understood by realizing that the Poisson brackets between the angular momentum and Runge-Lenz vector are given by,
\begin{align}
\label{eq:pb}
\{h_{i}, h_{j}\} &= \frac{1}{\mu} \epsilon_{ijk} h^{k}\,, \qquad \{h_{i}, A_{j}\} = \frac{1}{\mu} \epsilon_{ijk} A^{k}\,, 
\nn \\
\{A_{i}, A_{j}\} &= - \frac{2}{M^{2} \mu^{2}} \left(\frac{P^{2}}{2 \mu} - \frac{M \mu}{r}\right) \epsilon_{ijk} h^{k}
\end{align}
where $\vec{P} = \mu \vec{v}$ is the momentum of the mass $\mu$, and $\epsilon_{ijk}$ is the Levi-Civita symbol. The Poisson bracket between components of the orbital angular momentum defines the SO(3) symmetry associated with the planar motion of the binary, while the remaining Poisson brackets in Eq.~\eqref{eq:pb} reveal the Lie algebra of the hidden SO(4) symmetry. This symmetry is unique to central force problems with an inverse-square law, and makes the Kepler problem maximally superintegrable~\cite{JoseSaletan}.

The orbital motion of the binary system is fully described by Eqs.~\eqref{eq:rkep} and~\eqref{eq:phidotkep} and the five constants of motion $(p, e_{\K}, \omega, \iota, \Omega)$, but there is an alternative parameterization that is of relevance to the two-body problem in the PN formalism. This alternative parameterization arises due to difficulty in integrating Eq.~\eqref{eq:phidotkep} to obtain $t(\phi)$. Rather than working in terms of the azimuthal coordinate $\phi$, one can work in terms of the eccentric anomaly $u$, which is equivalent to the phase variable in an elliptical coordinate system. The mapping between the azimuthal coordinate and the eccentric anomaly, and the radial equation in terms of $u$, is given by
\begin{align}
\label{eq:ru}
r &= a \left(1 - e_{\K} \; \cos u\right)
\\
\label{eq:phiu}
\phi - \omega &= 2 \; {\rm tan}^{-1} \left[\left(\frac{1+e_{\K}}{1-e_{\K}}\right)^{1/2} \; {\rm tan}\left(\frac{u}{2}\right)\right]\,,
\end{align}
To complete the description of the orbit in terms of $u$, one needs to determine the mapping between time and $u$. This may be obtained by combining Eq.~\eqref{eq:phiu} with Eq.~\eqref{eq:phidotkep} to derive the expression for $\dot{u}$, which may then be directly integrated to obtain
\begin{equation}
\label{eq:Kepeq}
l = u - e_{\K} \; \sin u\,,
\end{equation}
where $l = (2\pi/T_{\orb}) (t - t_{0})$ is the mean anomaly, with $T_{\orb} = 2\pi a^{3/2}/M^{1/2}$ the orbital period and $t_{0}$ the time of pericenter passage. 

The above transcendental equation is known as Kepler's equation. First derived in 1609, there is still no closed-form expression of its inversion, $u(l)$. The most common inversion used within the field of gravitational wave modeling is the Fourier series solution,
\begin{equation}
u = l + 2 \sum_{q=1}^{\infty} J_{q}(q e) \; \frac{\sin(q l)}{q}\,,
\end{equation}
where $J_{q}$ are Bessel functions. This representation has the draw back of potentially needing a large number of terms in order to obtain an accurate representation of the function under consideration, especially when the argument of the Bessel functions is not much smaller than unity~\cite{PhysRevD.80.084001, Arun:2009mc, Arun:2007rg, Gondan:2017hbp}. There are multiple methods of avoiding this~\cite{Arun:2007rg, Tanay:2016zog, Forseth:2015oua, Loutrel:2016cdw}, and this draw back may not be as serious as previously considered when it comes to computing Fourier domain waveforms for eccentric binaries~\cite{Moore:2018kvz}. 

\subsection{Quasi-Keplerian Formalism}
\label{qk}

The Kepler problem is useful from a fundamental point of view to understand the motion of two objects under a mutual gravitational force. However, it is not an adequate model of the dynamics of binary systems within GR. As discussed in Sec.~\ref{intro}, the only way of currently studying the full two-body problem within GR is through NR, but there are ways of properly approximating the full solution. We here provide the details of an orbital parameterization for binary systems that has been worked out to high PN order. For simplicity in the discussion that follows, we restrict attention to the two-body problem at 1PN order and neglect radiation reaction.

The relative acceleration at 1PN order, or ${\cal{O}}(c^{-2})$ beyond Newtonian order, is given by~\cite{PW}
\begin{equation}
\label{eq:a1PN}
\vec{a} = - \frac{M}{r^{2}} \vec{n} - \frac{M}{r^{2}} \left\{\left[(1 + 3 \eta) v^{2} - \frac{3}{2} \eta \dot{r}^{2} - 2 (2 + \eta) \frac{M}{r}\right] \vec{n} - 2 (2 - \eta) \dot{r} \vec{v}\right\}\,,
\end{equation}
where $\eta = \mu/m$ is the symmetric mass ratio of the binary. These equations of motion, like the Kepler problem, are time symmetric and describe motion taking place in an orbital plane. As such, there exists a conserved orbital energy and angular momentum for the two-body problem at 1PN order, specifically
\begin{align}
\epsilon &= \frac{1}{2} v^{2} - \frac{M}{r} + \frac{3}{8} (1 - 3 \eta) v^{4} + \frac{M}{2 r} \left[(3 + \eta) v^{2} + \eta \dot{r}^{2} + \frac{M}{r}\right]\,,
\\
\vec{h} &= \left(\vec{r} \times \vec{v}\right) \left[1 + \frac{1}{2} (1 - 3 \eta) v^{2} + (3 + \eta) \frac{M}{r}\right]\,.
\end{align}
While the orbital motion of the binary still takes place in a plane, the relative force between the two binary components is no longer a simple inverse-square law. As a result, the SO(4) symmetry of the Kepler problem is broken, causing the Runge-Lenz vector to precess around the orbital angular momentum. Thus, in GR, binary systems undergo precession of periastron, an important feature that helped to provide observational support for GR from observations of Mercury's precession around the Sun~\cite{will-living}.

Generally, there are two ways of solving the PN two-body problem defined through Eq.~\eqref{eq:a1PN}. The first is the osculating method~\cite{PW, Lincoln:1990ji, Mora:2003wt, Will:2016pgm, Pound:2007th, Damour:2004bz, Konigsdorffer:2006zt}, a generic perturbative technique to account for perturbations of the Newtonian two-body problem. While one can apply this method at this PN order, it is also of relevance to the radiation-reaction problem, so we will detail it in the next section. Alternatively, one could seek a closed form analytic solution to Eq.~\eqref{eq:a1PN}, just as we did for the Newtonian two-body problem. This was achieved by Damour \& Deruelle~\cite{zbMATH03938612, zbMATH04001537}, who developed the QK formalism.

The QK formalism seeks to solve the conservative two-body problem at a given PN order through a Keplerian-style parameterization. At 1PN order, the solution takes the form
\begin{align}
\label{eq:rqk}
r &= a_{r} \left(1 - e_{r} \; \cos u\right)\,,
\\
\label{eq:lqk}
l &= u - e_{t} \; \sin u\,,
\\
\label{eq:phiqk}
\phi - \phi_{p} &= 2 K \; {\rm tan}^{-1}\left[\left(\frac{1+e_{\phi}}{1-e_{\phi}}\right)^{1/2} {\rm tan}\left(\frac{u}{2}\right)\right]
\end{align}
which resembles the solution to the Kepler problem in Eqs.~\eqref{eq:ru},~\eqref{eq:phiu}, and~\eqref{eq:Kepeq}, except for a few crucial differences. First, in the azimuthal equation, $\omega$ has been replaced with $\phi_{p}$, which is no longer the longitude of pericenter, but an overall integration constant associated with freedom in how we define the zero point of the azimuthal coordinate. This equation is also modified by the presence of the constant $K$, which describes the advance of pericenter; in one orbit, pericenter advances by an angle $\Delta \phi = 2 \pi (K-1)$. Second, the semi-major axis in the radial equation has been replace with the PN corrected semi-major axis $a_{r}$. Finally, the Keplerian eccentricity $e_{\K}$ has been replaced by three QK eccentricities $(e_{t}, e_{r}, e_{\phi})$. These quantities are all related to the energy and angular momentum of the binary, specifically~\cite{Blanchet:2013haa}
\begin{align}
a_{r} &= \frac{M}{\varepsilon} \left[1 + \frac{1}{4} \left(-7 + \eta\right) \varepsilon\right]\,,
\\
\label{eq:er}
e_{r} &= (1-j)^{1/2} + \frac{\varepsilon}{8 (1-j)^{1/2}} \left[24 - 4 \eta + 5 j(-3 + \eta)\right]\,,
\\
\label{eq:et}
e_{t} &= (1-j)^{1/2} + \frac{\varepsilon}{8 (1-j)^{1/2}} \left[-8 + 8 \eta + j (17 - 7 \eta)\right]\,,
\\
\label{eq:ephi}
e_{\phi} &= (1-j)^{1/2} + \frac{\varepsilon}{8 (1-j)^{1/2}} \left[24 + j (-15 + \eta)\right]\,,
\\
K &= 1 + \frac{3 \varepsilon}{j}\,,
\end{align}
where $\varepsilon = 2 \epsilon$ and $j = 2 \epsilon h^{2}$, completing the 1PN accurate description of the orbit.

The three QK eccentricities are not independent quantities, as can be seen from Eqs.~\eqref{eq:er}-\eqref{eq:ephi}, but are defined as given above such that the orbital parameterization of Eqs.~\eqref{eq:rqk}-\eqref{eq:phiqk} takes the same form as the Keplerian parameterization of Eqs.~\eqref{eq:ru}-\eqref{eq:Kepeq}. In any practical PN calculation, one will typically choose one of these three eccentricities and write down all expressions in terms of it. The most common choice within the PN literature has been $e_{t}$, but there is no strong requirement for making this choice. Parameterizing the orbit in terms of these eccentricities has the advantage of allowing one to take the appropriate circular limit, since specifically a circular orbit corresponds to $e_{t} = e_{r} = e_{\phi} = 0.$

While the QK eccentricities allow us to achieve a Keplerian-like parametrization of the orbit at 1PN order, this is not true at higher PN orders~\cite{Blanchet:2013haa}. When including the 2PN corrections of Eq.~\eqref{eq:a1PN}, the QK orbital parameterization of Eqs.~\eqref{eq:lqk} and~\eqref{eq:phiqk} becomes modified through the addition of extra harmonics of the eccentric and true anomalies. For example, the 2PN accurate Kepler's equations takes the form~\cite{0264-9381-12-4-009}
\begin{equation}
l = u - e_{t} \; \sin u + f_{t} \; \sin V + g_{t} \left(V - u\right)
\end{equation}
where $V = (\phi - \phi_{p})/K$ is the true anomaly, and $(f_{t}, g_{t})$ are known functions of the energy and angular momentum (see, for example, Eqs.~(7.7g) and~(7.7h) of \cite{Arun:2007sg}). Inversions of this are significantly more complicated then the Kepler problem, but have been achieved in recent years through 3PN order~\cite{Boetzel:2017zza}. However, there are still no accurate analytic Fourier domain waveforms for arbitrarily eccentric binaries at high PN order. This is still an open problem, although there have been studies that have considered this through a variety of approximations~\cite{Moore:2018kvz, Klein:2018ybm, Gondan:2017hbp}. 

\subsection{Other Measures of Eccentricity}

The preceding discussions have focused on eccentricity in the PN formalism, but it is not the only method of solving the two-body problem within GR, and the PN framework is not the only way to define eccentricity in the binaries under consideration. For completeness, we provide below a list of different eccentricity parameters and where they are primarily used, including some definitions we already introduced earlier in this section.

\begin{itemize}
	\item \textbf{Keplerian eccentricity} $e_{\K}$: Reviewed in Sec.~\ref{newt}, $e_{\K}$ is a parameter that enters a Keplerian orbital parametrization (see Eq.~\eqref{eq:eK-def}). This eccentricity is commonly used within osculating methods to solve for the motion of the binary system, and as a result, is commonly promoted to a function of time under some perturbing force (whether this be conservative or dissipative). While circular (parabolic) orbits correspond to $e_{\K} = 0$ ($e_{\K} = 1$) in the unperturbed problem, this is not necessarily so when there is a perturbing force. This eccentricity measure is used in both the PN and self-force formalisms.
	\item \textbf{Quasi-Keplerian eccentricities} $e_{t}, e_{r}, e_{\phi}$: Reviewed in Sec.~\ref{qk}, the QK eccentricities are parameters that enter the PN accurate orbital description first derived by Damour and Deruelle. These eccentricities are constants of the orbit when considering the conservative PN forces, as opposed to $e_{\K}$ which is then a function of time. However, under a dissipative perturbing force (such as that induced by radiation reaction), these eccentricities can also be considered functions of time calculated through the method of variation of constants. These measures are primarily only used within the PN formalism.
	\item \textbf{Angular velocity eccentricity} $e_{\Omega} = (\Omega_{p}^{1/2} - \Omega_{a}^{1/2})/(\Omega_{p}^{1/2} + \Omega_{a}^{1/2})$: An eccentricity parameter that is constructed from the angular frequencies of apocenter $\Omega_{a}$ and pericenter $\Omega_{p}$. When computing this, one does not require a specific orbital parameterization, but instead one only needs the orbital phase as a function of time. This measure is sufficiently general that it can be applied within any formalism, but it is most commonly used within the self-force formalism and in NR (see, for example,~\cite{Mora:2003wt, Buonanno:2006ui}).
	\item \textbf{Radial velocity eccentricity} $e_{\R} = {\rm max}\left[(r/M)^{1/2} \dot{r}\right]$: Similar to the previous measure, this parameter requires that one only know the radial separation of the binary as a function of time. The eccentricity is then recovered by finding the envelope of the $(r/M)^{1/2} \dot{r}$ time function. 
	\item {\textbf{Radial acceleration eccentricity}} $e_{\A} = r^{2} \ddot{r}$: An instantaneous NR eccentricity parameter, it can easily be recovered from the temporal evolution of the radial separation of the binary. It is sufficiently general to be applied to any formalism for the two-body problem~\cite{Heal:2017abq}.
	\item \textbf{Pfeiffer et. al. coordinate separation eccentricity} $e_{s}$: Developed in~\cite{Pfeiffer:2007yz, Boyle:2007ft}, this is a commonly used parameter in eccentricity reduction methods for NR. After setting up initial data, the binary is evolved for several orbits, which gives the coordinate separation $s(t)$. From this, one computes $ds/dt$, and fits this to
\begin{equation}
\frac{ds}{dt} = A_{0} + A_{1} t + B \sin(\omega t + \phi_{0})\,,
\end{equation}
where $(A_{0}, A_{1}, B, \omega, \phi_{0})$ are constants to be fitted. The eccentricity parameter is then calculated through $e_{s} = B/\omega s_{0}$ where $s_{0} = s(0)$. A more extensive discussion, and alternative methods of defining this eccentricity parameter, are discussed in~\cite{Buonanno:2010yk}.
	\item \textbf{Tichy \& Marronetti (TM) coordinate separation eccentricity} $e_{r}:$ Developed in~\cite{Tichy:2010qa} for setting up low eccentricity initial data in NR simulations, this eccentricity parameter is given by
\begin{equation}
\label{eq:e-TM}
e_{r}(t) = \frac{\Delta r_{\max}(t) - \Delta r_{\min}(t)}{2 r_{\rm avg}}\,,
\end{equation}
where $(\Delta r_{\max}, \Delta r_{\min}, r_{\rm avg})$ are defined in Eqs.~(2)-(4) therein. 
	\item \textbf{(2,2)-mode eccentricity} $e_{22}$: An eccentricity parameter applicable to setting up initial data in NR simulations that relies on the spin-weighted spherical harmonic decomposition of the gauge invariant Weyl scalar $\Psi_{4}$~\cite{Tichy:2010qa}. The amplitude of the $l=m=2$ mode is given by
\begin{equation}
\big|\Psi_{4}^{(2,2)}\big| = 32 \left(\frac{\pi}{5}\right)^{1/2} \eta \left(M \omega\right)^{5/3}\,,
\end{equation}
where $\omega$ is the orbital angular frequency. Kepler's third law is then invoked to derive a separation $r_{22}$ from the above equation, which is then used to define an eccentricity in a similar manner to the TM coordinate separation eccentricity given by Eq.~\eqref{eq:e-TM}, specifically
\begin{equation}
e_{22} = \frac{\Delta r_{22, \max}(t) - \Delta r_{22, \min}(t)}{2 r_{22, \rm avg}}\,.
\end{equation}
This is a common measure of eccentricity used in NR simulations.
	\item \textbf{Fitted angular frequency eccentricity} $e_{\omega}$: A common eccentricity parameter used for eccentricity reduction in NR simulations, it is found by fitting a fifth order polynomial to the angular velocity of the binary $\omega(t)$ to obtain $\omega_{\rm fit}(t)$~\cite{Husa:2007rh, Buonanno:2006ui, Baker:2006ha, Tichy:2010qa}. The eccentricity is then approximated using
\begin{equation}
e_{\omega} = \frac{\omega(t) - \omega_{\rm fit}(t)}{2 \omega_{\rm fit}(t)}\,.
\end{equation}
	\item \textbf{Fitted coordinate separation eccentricity} $e_{d}$: Identical to the previous eccentricity parameter, but using the coordinate separation of punctures instead of the angular velocity~\cite{Husa:2007rh, Buonanno:2006ui, Baker:2006ha}.
\end{itemize}

These eccentricity parameters can broadly be sorted into two categories: \textit{coordinate eccentricities} and \textit{variational eccentricities}. Coordinate eccentricities are parameters that are computed from the relative coordinates of the binary system and do not require a specific orbital parameterization. They may be reconstructed from envelopes of the velocities, such as $e_{\Omega}$ and $e_{\R}$, taken as instantaneous functions of the relative coordinates, such as $e_{\A}$, or even obtained as fits to numerical data, such as $(e_{s}, e_{r}, e_{22}, e_{\omega}, e_{d})$. Variational eccentricities are parameters that appear in a specific orbital parameterization, which typically solves a conservative part of the equations of motion to a given PN order, such as $e_{\K}$, $e_{t}$, $e_{r}$, and $e_{\phi}$. These are constants of the motion unless the binary is acted upon by a dissipative perturbing force or a conservative perturbing force of higher order than that considered when building the orbital parameterization. When this occurs, the eccentricity parameters are promoted to functions of time and are allowed to vary according to a set of evolution equations of the form $de/dt = {\cal{F}}(t)$; this is why we refer to them as variational eccentricities. We will discuss these evolution equations in more detail in the next section.

Since these definitions are distinct, the behavior of the eccentricity parameters need not necessarily agree with each other. In particular, it is clear that all of these definitions agree in the limit of infinite separation, but they will tend to disagree with each other as the binary separation decreases.  The concept of eccentricity, regardless of how one defines a parameter to quantify it, is related to the exterior curvature of the line defining the orbit in space, and is thus \emph{slicing dependent}. As a result, the concept of eccentricity is not a gauge invariant quantity, and the behavior of an eccentricity parameter under perturbations will depend on how one defines it. However, as we will detail later, the concept of eccentricity creates additional harmonic content in the GWs emitted by a binary system, and the latter is indeed observable. As long as one compares waveforms computed using the same approximations, one should measure the same waveform regardless of how the eccentricity parameter that enters the waveform model is defined. We touched on an example of this in~\cite{Loutrel:2018ssg}, but we will provide more details on this in later sections.

\begin{figure}[ht]
\includegraphics[clip=true, width=\columnwidth]{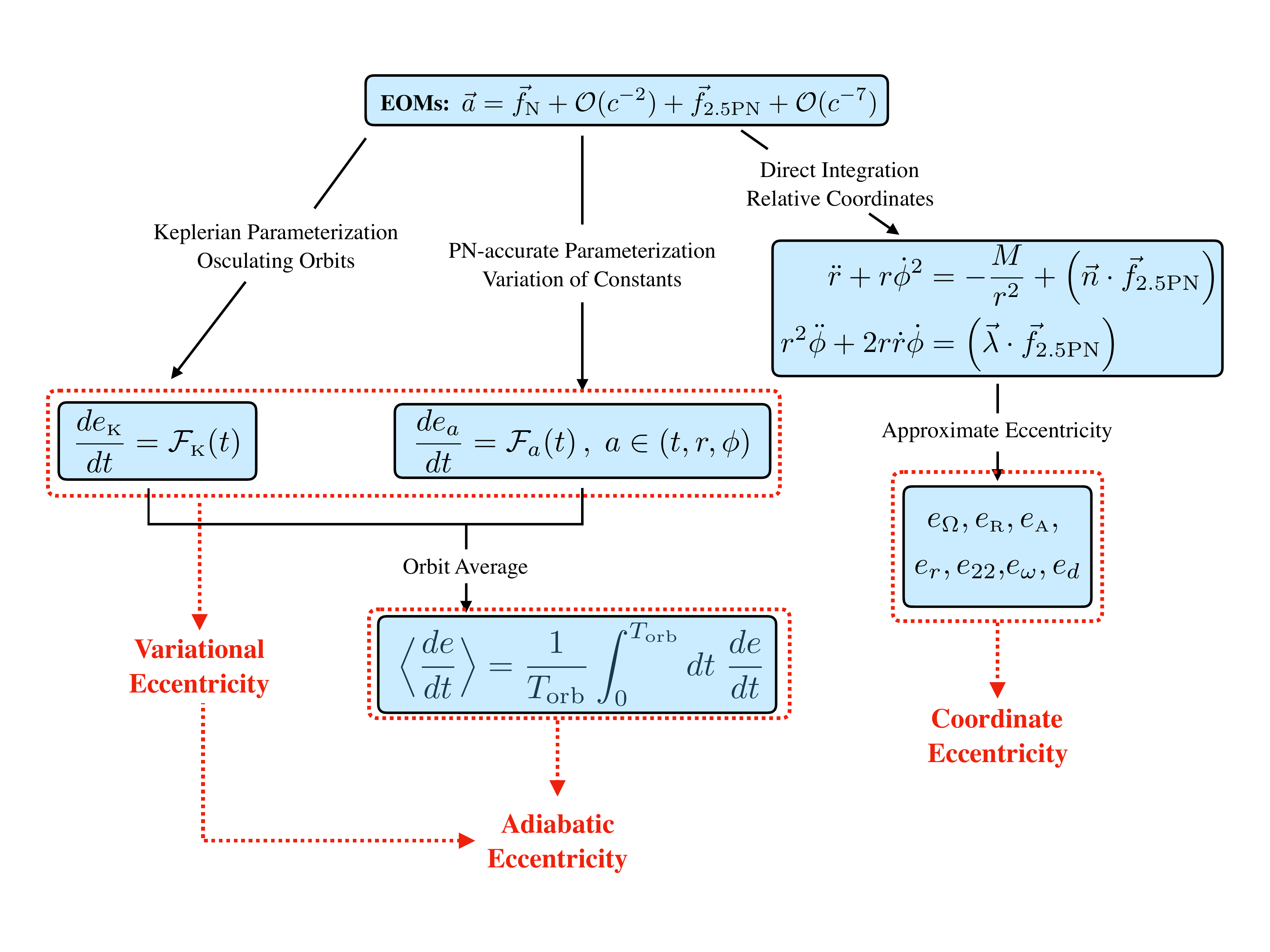}
\caption{\label{fig1} Methods of solving for the evolution of eccentricity parameters in the rN-RR problem in the PN formalism. Starting from the equations of motion, one can either choose an orbital parameterization (left and center branches) or directly integrate the relative coordinates (right branch). When one directly evolves the relative coordinates of the binary, one has to then define the eccentricity parameter in one of several ways that rely on the velocities or accelerations. These constitute the notion of \textit{coordinate eccentricity}. On the other hand, one can choose a specific orbital parameterization, and promote the orbital elements of said parameterization to functions of time. If the parameterization is Keplerian (Sec.~\ref{newt}), one computes the time varying Keplerian eccentricity parameter $e_{\K}(t)$. If one chooses a PN accurate parameterization (Sec.~\ref{qk}), one then computes one of the time-varying QK eccentricity parameters, $e_{t}(t)$ for example. These constitute the notion of \textit{variational eccentricity}. At Newtonian order, the evolutions for $e_{\K}(t)$ and $e_{t}(t)$ (or $e_{r}(t)$, or $e_{\phi}(t)$) agree, but not at higher PN order. A common tool used to solve  the radiation-reaction problem is the orbit average approximation. When one does this, one only recovers the leading order secular behavior of the full radiation-reaction equations in a multiple scale analysis (Sec.~\ref{msa}). These eccentricity parameters constitute the notion of \textit{adiabatic eccentricity}, a subset of variational eccentricity.}
\end{figure}

There is one other class of eccentricity parameters we will consider here. We refer to these as \textit{adiabatic eccentricities}, which are a subset of the variational eccentricities and are computed through the orbit-averaged version of the eccentricity evolution equations, $\langle de/dt \rangle = (1/2\pi) \int_{0}^{2\pi} d\phi \; (1/\dot{\phi}) \; (de/dt)$. This class of eccentricity parameters assume that perturbative effects only cause secular changes on timescales much longer than the orbital timescale. However, as we found in~\cite{Loutrel:2018ssg}, this is not necessarily true when considering the late inspiral of a binary system due to radiation reaction. We will here show that these definitions are not accurate representations of the dynamics of the binary under radiation reaction, and the waveforms computed using them become inaccurate at sufficiently high signal-to-noise ratio (SNR). We schematically present these notions of eccentricity in Fig.~\ref{fig1}.

%

\section{Eccentric Dynamics}
\label{dynamics}

Now that we introduced the dynamics of eccentric binaries, let us begin to consider the evolution of the binary, and  the different definitions of eccentricity, under radiation reaction. We will here consider the rN-RR problem, i.e we take both the conservative and dissipative forces to be only the leading PN order contributions. This amounts to solving the equation of motion $\vec{a} = - (M/r^{2}) \vec{n} + \vec{a}_{\R \R}$, where
\begin{equation}
\vec{a}_{\rm RR} = \frac{8}{5} \eta \frac{M^{2}}{r^{3}} \left[\left(3 v^{2} + \frac{17}{3} \frac{M}{r}\right) \dot{r} \vec{n} - \left(v^{2} + 3 \frac{M}{r}\right) \vec{v}\right]
\end{equation}
is the leading PN order radiation-reaction force in the harmonic gauge~\cite{PW, Damour:2004bz}. We will consider the dynamics of eccentric binaries under direct-evolution, osculating, and orbit-averaged methods of solving the equations of motion, and discuss similarities and differences between the three methods. 

\subsection{Direct-Evolution Method}
\label{de}

To begin, we consider the direct evolution of the relative coordinates of the binary's components through numerical methods. More specifically, we seek to solve for $r(t)$ and $\phi(t)$. Using the fact that $\vec{v} = \dot{r} \vec{n} + r \dot{\phi} \vec{\lambda}$, with $\vec{\lambda} = (-\sin\phi, \cos\phi, 0)$, the equations of motion can be separated into the following set
\begin{align}
\ddot{r} &= \frac{h^{2}}{r^{3}} - \frac{M}{r^{2}} + \frac{8}{5} \eta \frac{M^{2}}{r^{3}} \dot{r} \left(2 \dot{r}^{2} + 2 \frac{h^{2}}{r^{3}} + \frac{8}{3} \frac{M}{r}\right)\,,
\\
\dot{h} &= \frac{8}{5} \eta \frac{M^{2}}{r^{3}} \frac{h}{r} \left(\dot{r}^{2} + \frac{h^{2}}{r^{3}} + 3 \frac{M}{r}\right)\,,
\end{align}
where $h = r^{2} \dot{\phi}$. We provide the details of our method of numerically evolving these equations in Sec.~\ref{comp}.

Obtaining a notion of eccentricity is somewhat difficult using this approach, as it is not something that one can visualize immediately from the trajectories unless the binaries is sufficiently elliptical. To estimate the eccentricity as a function of time for this binary, we choose to compute the radial velocity eccentricity $e_{\R}$. To compute $e_{\R}(t)$, we must first properly account for a secular drift that appears in $\dot{r}(t)$ through an empirical mode decomposition (EMD), a technique often employed in Hilbert-Huang transforms~\cite{Huang903}. The method we apply is as follows:
\begin{enumerate}
	\item Define the function ${\cal{G}}(t) = (r/M)^{1/2} \dot{r}$ and find all local extrema of ${\cal{G}}(t)$.
	\item Interpolate the data points describing the extrema using cubic splines, generating functions describing the upper ${\cal{G}}_{+}(t)$ and lower ${\cal{G}}_{-}(t)$ envelopes of ${\cal{F}}(t)$.
	\item Compute the average of these interpolating functions to obtain $\langle {\cal{G}} \rangle(t) = (1/2) [{\cal{G}}_{+}(t) + {\cal{G}}_{-}(t)]$.
	\item Subtract off the average from the original function to obtain a new function ${\cal{G}}_{1}(t) = {\cal{G}}(t) - \langle {\cal{G}}(t) \rangle$.
\end{enumerate}
The cubic spline interpolation in the method above introduces a small amount of error, specially close to the end of the inspiral where all functions are changing more rapidly (as the dynamical time scale becomes comparable to the orbital time scale). This, in turn, can force the average of the function we wish to approximate to not vanish as we would like it to. The above method, however, can be repeated multiple times, and with each new iteration, the average will be closer and closer to zero. Therefore, the iteration of this method yields ${\cal{G}}_{k}(t) = {\cal{G}}_{k-1}(t) - \langle {\cal{G}}_{k-1} \rangle(t)$, with each iteration of the routine referred to as a sifting, and the index $k$ corresponding to how many siftings have been carried out. To determine when to stop the procedure, we compute the number of siftings $N$ wherein the extrema of the resulting function ${\cal{G}}_{k}(t)$ differ from the number of zero crossings by at most one. When $N>N_{\max}$, we stop the above procedure to obtain ${\cal{G}}_{\IMF}(t)$. For the system under consideration, we take $N_{\max} = 10$. The function ${\cal{G}}_{\IMF}(t)$ is referred to as the intrinsic mode function (IMF).

Once the sifting procedure completes and we obtain ${\cal{G}}_{\IMF}(t)$, we compute the function $e_{\R}(t)$ by finding all maxima of the IMF, and interpolating the resulting data using cubic splines. We will provide an example of this when we compare to the other evolution methods in Sec.~\ref{comp}.

\subsection{Osculating Method}
\label{osc}

As an alternative to the method of directly integrating the equations of motion of a binary system, one can use perturbation theory methods to evolve the orbits. In this method, the equations of motion take the form $\vec{a} = \vec{f}_{0} + \delta \vec{f}$, where $\vec{f}_{0}$ corresponds to the relative force in the unperturbed problem and $\delta \vec{f}$ is the perturbing force. One generally solves the unperturbed problem to obtain the solution $\vec{r} = \vec{r}_{0}(t, \mu^{a})$ and $\vec{v} = \vec{v}_{0}(t, \mu^{a})$, where $\mu^{a}$ is the set of orbital elements, which are constant in the absence of perturbations. To solve the perturbed problem, one promotes the orbital elements to functions of time, $\mu^{a} \rightarrow \mu^{a}(t)$. After inserting this back into the equations of motion, one obtains evolution equations for the orbital elements
\begin{align}
\frac{\partial{\vec{r}_{0}}}{\partial \mu^{a}} \frac{d\mu^{a}}{dt} = 0\,, \qquad \frac{\partial{\vec{v}_{0}}}{\partial \mu^{a}} \frac{d\mu^{a}}{dt} = \delta \vec{f}\,.
\end{align}
The above equations constitute six first order equations corresponding to the six initial conditions $\vec{r}_{0}(t=0)$ and $\vec{v}_{0}(t=0)$.

For non-spinning binaries in the PN formalism, these equations reduce to four independent equations for the orbital elements $(\epsilon, h, \phi_{0}, t_{0})$, provided the perturbing force has no non-zero components orthogonal to the orbital plane. The orbital parameterization for the unperturbed problem is dependent on what one chooses for $\vec{f}_{0}$. For $\vec{f}_{0} = \vec{f}_{N}$, the orbital parameterization is Keplerian (Sec.~\ref{newt}), while for $\vec{f}_{0} = \vec{f}_{\rm PN}^{\rm cons}$, the orbits are described by the QK parameterization (Sec.~\ref{qk}). Since we are working with the rN-RR problem in this section, these two parameterizations are equivalent. We take the orbital elements to be $\mu^{a} = (p, \vec{A})$, so as to avoid divergences at small eccentricities, of the form $e^{-1}$, in the osculating equations. With this parameterization, the osculating equations become
\begin{align}
\label{eq:dpdt-osc}
\frac{dp}{dt} &= \frac{1}{15} \eta \left(\frac{M}{p}\right)^{3} \sum_{j=0}^{5} \left[C_{p}^{j} \cos(j \phi) + S_{p}^{j} \sin(j \phi)\right]\,,
\\
\label{eq:dAxdt-osc}
\frac{d\alpha}{dt} &= \frac{1}{60} \frac{\eta}{M} \left(\frac{M}{p}\right)^{4} \sum_{j=0}^{7} \left[C_{\alpha}^{j} \cos(j \phi) + S_{\alpha}^{j} \sin(j \phi)\right]\,,
\\
\label{eq:dAydt-osc}
\frac{d\beta}{dt} &= \frac{1}{60} \frac{\eta}{M} \left(\frac{M}{p}\right)^{4} \sum_{j=0}^{7} \left[C_{\beta}^{j} \cos(j \phi) + S_{\beta}^{j} \sin(j \phi)\right]\,,
\\
\label{eq:dphidt-osc}
\frac{d\phi}{dt} &= \left(\frac{M}{p^{3}}\right)^{1/2} \sum_{j=0}^{2} \left[C_{\phi}^{j} \cos(j \phi) + S_{\phi}^{j} \sin(j \phi)\right]\,,
\end{align}
where $\alpha = e_{\K} \cos\omega$, $\beta = e_{\K} \sin\omega$, with $(e_{\K}, \omega)$ the Keplerian eccentricity and longitude of pericenter, respectively, and we provide the coefficients $C_{a}^{j}$ and $S_{a}^{j}$ in~\ref{coeffs}. As with the direct evolution, one can numerically integrate the above equations to obtain the evolution of the binary and the behavior of the eccentricity, in this case $e_{\K}$. One can also analytically solve these osculating equations through application of multiple scale analysis (MSA)~\cite{Bender}, which we pursue in Sec.~\ref{msa}

\subsection{Orbit-Averaged Approximation}
\label{oa}

There is one other method of evolving the binary under radiation reaction that we consider here, specifically the orbit-averaged approximation. In this approximation, one generally assumes that the effects of radiation reaction are small over any one orbit, so that we can orbit average the osculating equations to obtain the secular behavior of the orbital elements. The concept of averaging was first proposed by Isaacson~\cite{Isaacson:1968ra, Isaacson:1968gw} to remove oscillatory gauge effects from the stress energy tensor of GWs. If we apply orbit averaging to the osculating equations of Eqs.~\eqref{eq:dpdt-osc}-\eqref{eq:dAydt-osc}, we obtain
\begin{align}
\label{eq:dpdt-oa}
\Bigg\langle \frac{dp}{dt} \Bigg\rangle&= - \frac{64}{5} \eta \left(\frac{M}{p}\right)^{3} \left(1 - e_{\K}^{2}\right)^{3/2} \left(1 + \frac{7}{8} e_{\K}^{2}\right)\,,
\\
\label{eq:dedt-oa}
\Bigg\langle \frac{de_{\K}}{dt} \Bigg\rangle&= - \frac{304}{15} e_{\K} \frac{\eta}{M} \left(\frac{M}{p}\right)^{4} \left(1 - e_{\K}^{2}\right)^{3/2} \left(1 + \frac{121}{304} e_{\K}^{2}\right)\,,
\end{align}
where we have recovered the eccentricity through $e_{\K}^{2} = \alpha^{2} + \beta^{2}$. These equations match the well known results of Peters \& Mathews~\cite{PetersMathews, Peters:1964zz}. As one can see from the above differential equations, in this approximation the longitude of pericenter remains fixed, while the Keplerian eccentricity $e_{\K}$, and thus the magnitude of the Runge-Lenz vector, monotonically decreases throughout the coalescence. One can now obtain the temporal evolution of the eccentricity by solving Eqs.~\eqref{eq:dpdt-oa} and~\eqref{eq:dedt-oa} either analytically or numerically. 

The orbit-averaged approximation can also be derived through the balance laws formalism, which is how Eqs.~\eqref{eq:dpdt-oa}-\eqref{eq:dedt-oa} were originally derived by Peters \& Mathews. Here, the rate of loss of orbital energy and angular momentum is balanced by the averaged GW energy and angular momentum fluxes, respectively. The balance laws have been shown to hold through 3PN order~\cite{Blanchet:93, Blanchet:1996vx, Jaranowski:97, Pati:2002ux, Konigsdorffer:2003ue, Nissanke:2004er, Itoh:2003fy, Arun:2009mc, Arun:2007sg, Arun:2007rg}. Without the averaging procedure, the balance laws do not hold, and are corrected by 2.5PN order and higher order contributions to the orbital energy and the angular momentum, which are analogous to the Schott energy and angular momentum in electromagnetism~\cite{PW, Galley:2015kus, Gron2014}. We explore these notions in Sec.~\ref{comp}.

\subsection{Comparison of the Temporal Evolution of the Coordinate, Adiabatic and Variational Eccentricity Parameters}
\label{comp}

Now that we have described three different ways in which to evolve binary systems under radiation reaction and obtain the behavior of three different eccentricity parameters, let us compare them. For the direct integration, we compute the radial velocity eccentricity parameter using the method detailed in Sec.~\ref{de}. For the osculating method, we numerically evolve Eqs.~\eqref{eq:dpdt-osc}-\eqref{eq:dphidt-osc}, and reconstruct the Keplerian eccentricity parameter $e_{\K}$ as the magnitude of the Runge-Lenz vector. For the orbit-averaged evolution, we numerical evolve Eqs.~\eqref{eq:dpdt-oa}-\eqref{eq:dedt-oa}, which automatically gives us the evolution of the orbit-averaged Keplerian eccentricity parameter. All of the numerical integrations are performed with \texttt{Mathematica}'s \texttt{NDSolve} command, using the \texttt{ImplicitRungeKutta} method. We take the accuracy and precision tolerances to be $10^{-13}$ and evolve the binary up to the last stable orbit for test masses around a Schwarzschild BHs, specifically up to $p_{\LSO} = 2 M (3 + e_{\K})$.

\begin{figure*}[ht]
\includegraphics[clip=true, scale=0.4]{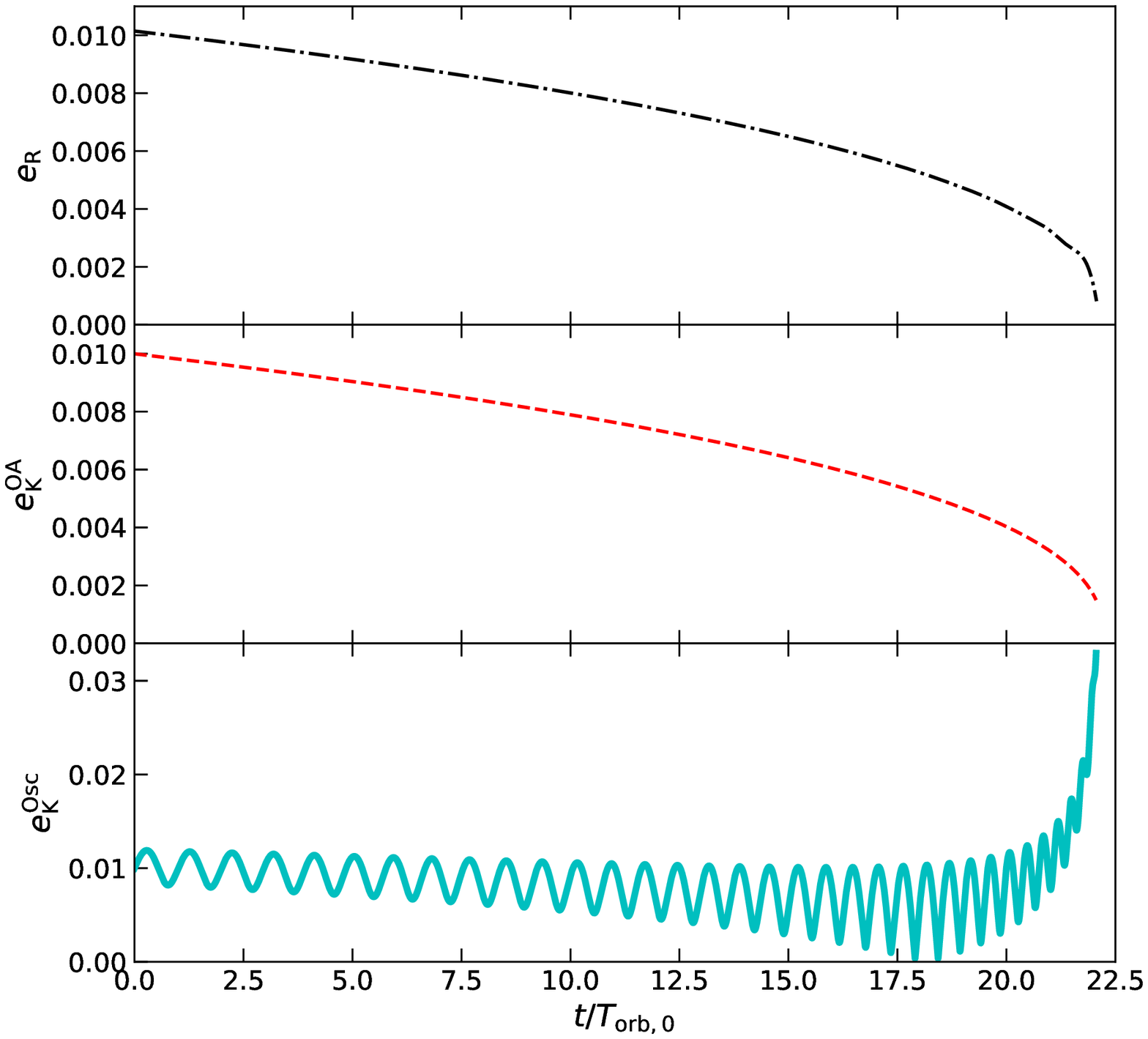}
\includegraphics[clip=true, scale=0.4]{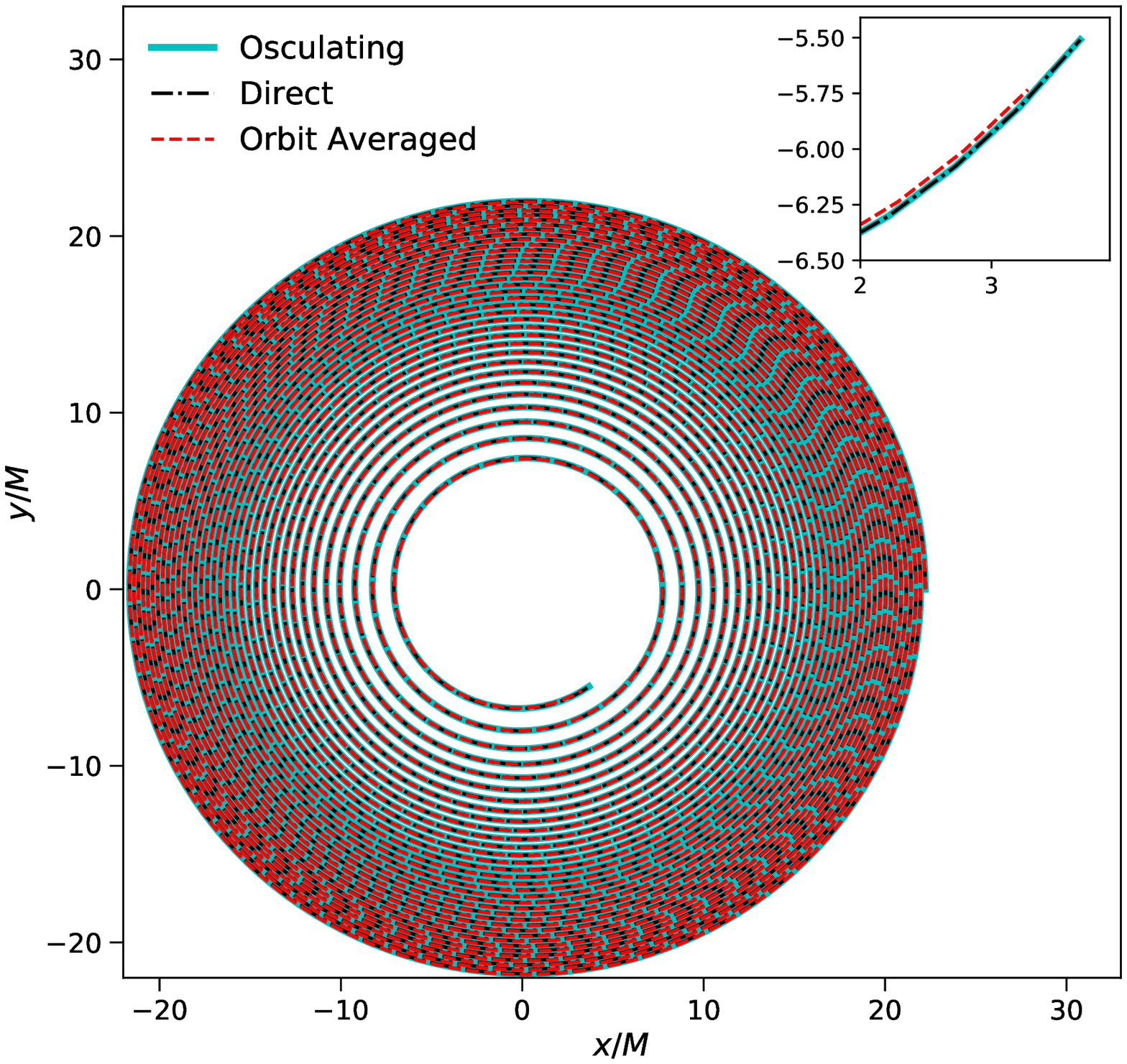}
\caption{\label{traj-fig} Left: Eccentricities of a BH binary system computed via direct-evolution (top), orbit-averaged (middle), and osculating (bottom) methods. For the direct evolution, we compute the radial velocity eccentricity using the method in Sec.~\ref{de}, while for the orbit-averaged and osculating methods, we compute the Keplerian eccentricity $e_{\K}$. Right: Trajectories of a binary system in an effective one-body frame using different methods of the equations of motion for the rN-RR problem: osculating (solid), direct (dot-dashed), and orbit-averaged (dashed). The inlay in the upper right of the plot shows a zoom in of the last stages of the evolution just before the systems reach ISCO.}
\end{figure*}

We provide two different comparisons in in Fig.~\ref{traj-fig}: a plot of the eccentricity parameters and another plot of the orbital trajectories in an effective one body frame. For the osculating and orbit-averaged methods, we need to reconstruct the relative coordinates of the binary to obtain the orbital trajectories. To do this, we use Eqs.~\eqref{eq:rkep} and~\eqref{eq:phidotkep} for the orbit-averaged method. For the osculating method, we also use these equations, but re-write them in terms of the components of the Runge-Lenz vector using
\begin{equation}
\label{eq:ew-to-ab}
e_{\K} = (\alpha^{2} + \beta^{2})^{1/2}\,, \qquad \omega = {\rm arccos}\left[\frac{\alpha}{(\alpha^{2} + \beta^{2})^{1/2}}\right]\,.
\end{equation}
In both plots, we study a BH binary system with masses $(m_{1}, m_{2}) = (10, 10) M_{\odot}$, and initial conditions $[p(0), e(0), \omega(0), \phi(0)] = (20M, 10^{-2}, \pi, 0)$ for the osculating and orbit-averaged method. For the direct evolution, we require the initial conditions $[r(0), \dot{r}(0), h(0), \phi(0)]$ match those for a Keplerian orbit with the same initial conditions given for the other two methods. 

The left plot of Fig.~\ref{traj-fig} displays the evolution of the eccentricities of the binary. The radial velocity eccentricity parameter $e_{\R}$ (top panel) and the orbit-averaged Keplerian eccentricity parameter $e_{\K}^{\OA}$ (middle panel) both exhibit the classic monotonic decrease through the inspiral. On the other hand, the osculating Keplerian eccentricity parameter $e_{\K}^{\Osc}$ displays two features not seen in the other two measures: it oscillates on the orbital time scale and it grows secularly late in the inspiral. The right plot of Fig.~\ref{traj-fig} shows the orbital trajectories of the binary in an effective one-body frame. The trajectories in the different methods are very similar, with the differences only visible if we greatly zoom in on the final part of the orbit (inlay). The direct evolution (dot-dashed line) and the osculating evolution (solid line) produce the same exact trajectory, but the orbit-averaged evolution dephases relative to these. 

We thus arrive at the main result of the comparison of the numerical evolutions: \textit{While the orbit-averaged eccentricity parameter of the binary decreases monotonically like the coordinate eccentricity parameter does, the trajectory in the orbit-averaged approximation exhibits a dephasing relative to the direct evolution, which will cause a dephasing between a waveform model used with the orbit-averaged approximation and an observed GW signal. To properly account for the phase of the orbits and the GWs emitted by the system, one must consider the full osculating behavior of the variational eccentricity, even though the evolution of the variational and coordinate eccentricities do not agree.}

While the osculating method produces the same trajectory as the direct evolution, this does not mean that the orbit is becoming more elliptical, or that the ellipticity of the orbit is oscillating. One has to disentangle the notions of ellipticity of an orbit with the specific eccentricity parameter chosen to characterize the orbit. While the osculating eccentricity is displaying a growth, the oscillations in the coordinate separation of the binary are actually decreasing, consistent with the system moving steadily toward a quasi-circular state. In reality, the ellipticity of the orbit is controlled by the extrinsic curvature of the spatial trajectory, which is slicing dependent, and not by the eccentricity parameter that one choses.

\begin{figure*}[ht]
\includegraphics[clip=true, scale=0.4]{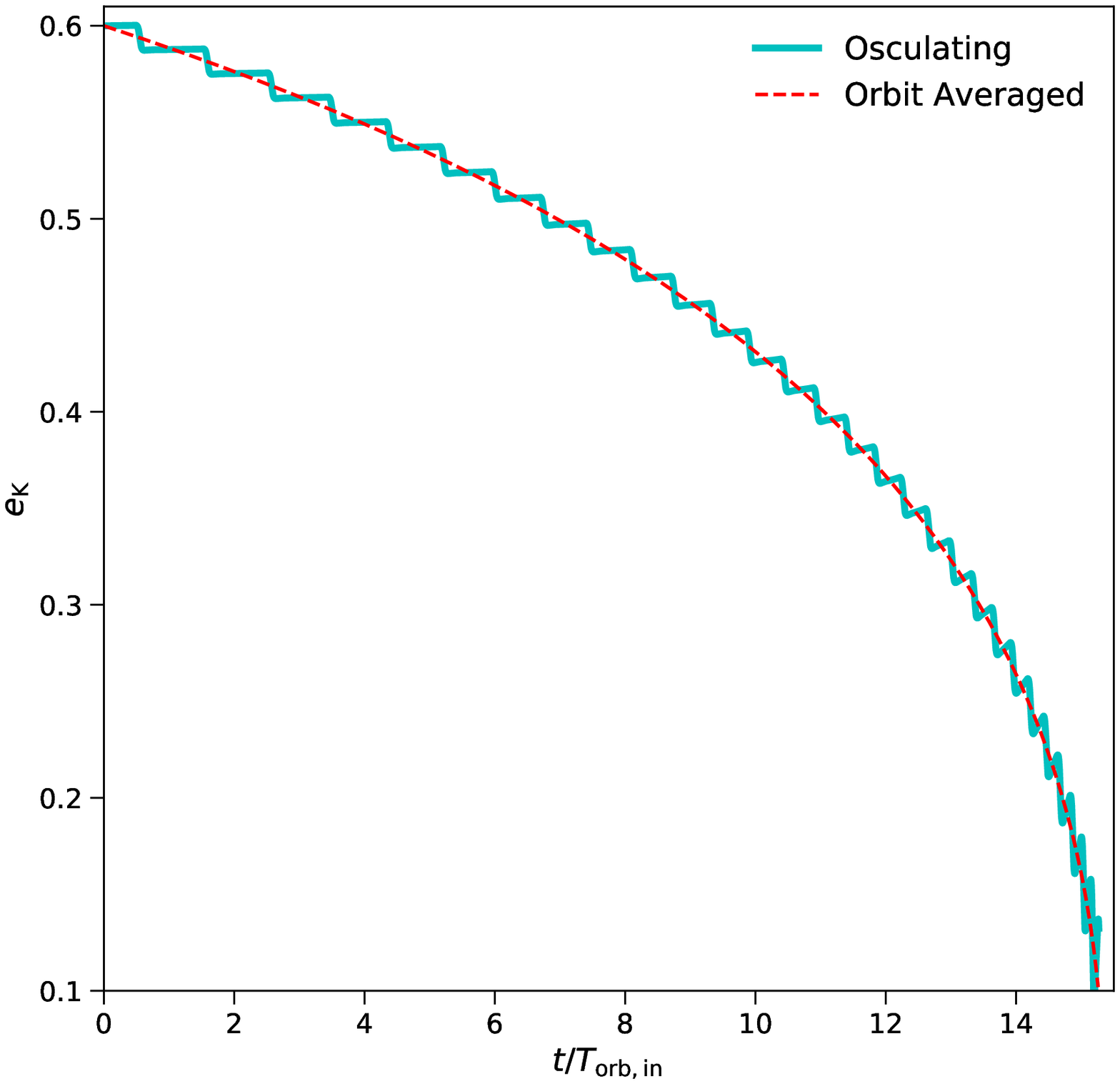}
\includegraphics[clip=true, scale=0.4]{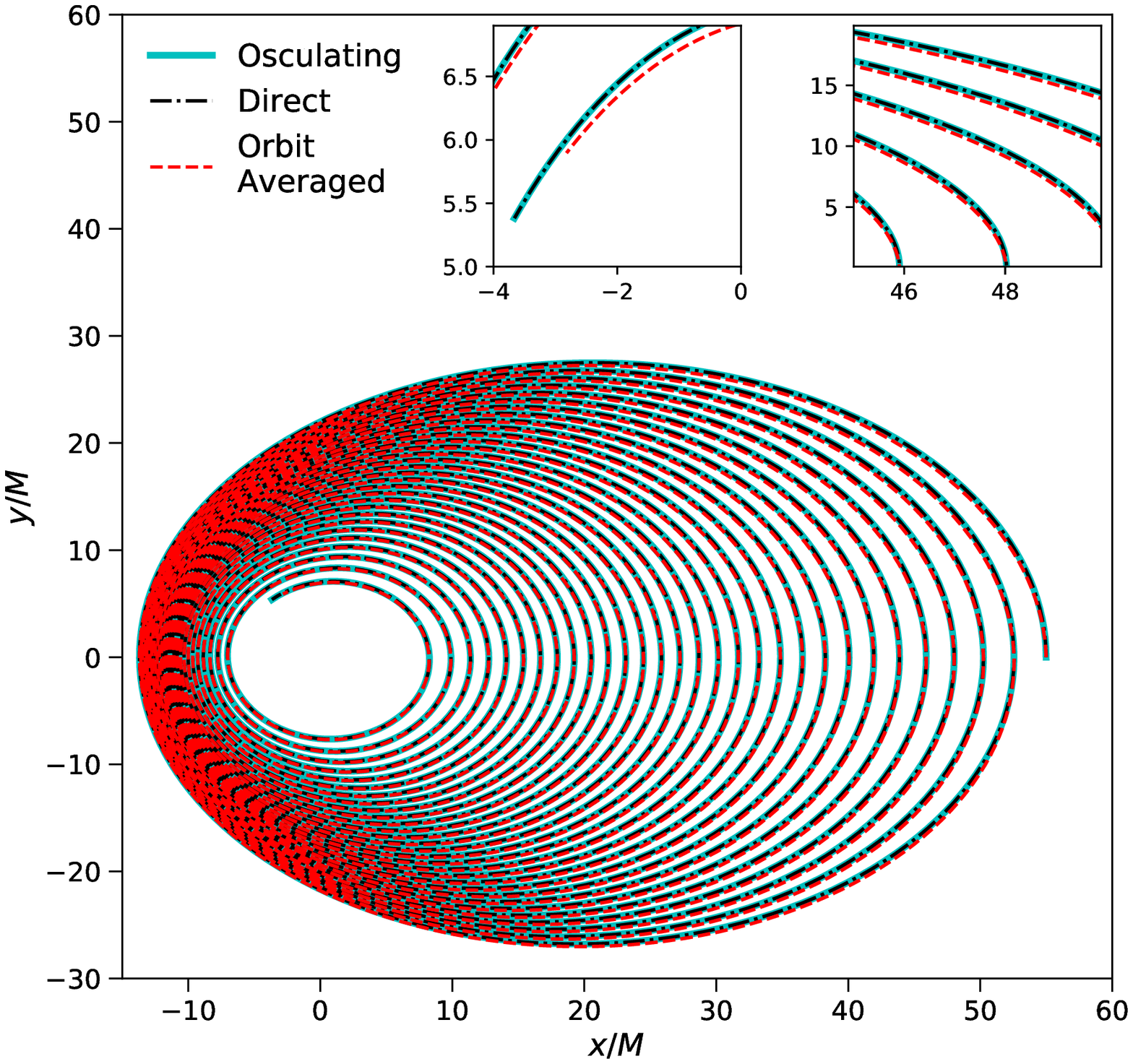}
\caption{\label{traj-fig2} Left: Evolution of the Keplerian eccentricity $e_{\K}$ of a BH binary system with initial eccentricity of 0.6, and computed via orbit-averaged (dashed line) and osculating (solid line) methods. Right: Trajectories of a binary system with initial Keplerian eccentricity $e_{\K}(0) = 0.6$ in an effective one-body frame using different methods of the equations of motion for the rN-RR problem: osculating (solid), direct (dot-dashed), and orbit-averaged (dashed). The inlay in the upper right of the plot shows a zoom of the region $\{x \in (40, 50)M, y \in (0, 20)M\}$ show the deviation of the orbit-averaged evolution from the osculating and direct methods. The inlay in the upper center of the plot shows last stages of the evolution just before the systems reach ISCO.}
\end{figure*}

Thus far, we have focused on binary systems with small initial Keplerian eccentricity, but how do the different methods compare for systems with moderate Keplerian eccentricity? In Fig.~\ref{traj-fig2}, we investigate this case for the same binary system as above, but with $e_{\K}(0) = 0.6$. In the left panel, we compare the evolution of $e_{\K}$ for this system using the osculating and orbit-averaged methods only. As opposed to the small eccentricity system, the evolution of $e_{\K}$ in the orbit-averaged method now agrees with the average of the osculating method, which is what we typically expect from these two approximations. The oscillations in the osculating $e_{\K}$ now resemble steps in the early evolution, since most of the GW power is being emitted at pericenter. While the eccentricity evolutions agree \textit{on average}, the trajectories do exhibit slightly different behavior as can be seen in the right panel of Fig.~\ref{traj-fig2}. Just like in the small eccentricity system of Fig.~\ref{traj-fig}, the trajectories as computed in the direct-evolution and osculating methods agree to the level of numerical error. But the orbit-averaged trajectory asymptotes to the other trajectories only at apocenter and pericenter, exhibiting a de-phasing in between that can be seen in the inlays of the right panel. This is not unexpected since the evolution of the Keplerian eccentricity $e_{\K}$ only intersects the evolution of the orbit-averaged eccentricity when the system is at pericenter or apocenter.

The comparison of methods at moderate eccentricities highlights the expected behavior of variational orbital elements within the orbit-averaged approximation, but it also indicates that the problems that plague the trajectories of binary systems when using orbit-averaging are not just restricted to systems with small Keplerian eccentricity. While the issue of de-phasing is potentially a major concern when it comes to detecting, and performing parameter estimation on, GW signals, the dephasing shown in the inlays of Figs.~\ref{traj-fig} and~\ref{traj-fig2} is very small, specifically $\delta \phi(e_{\K}=0.01) = \phi_{\DE}(t_{f}) - \phi_{\OA}(t_{f}) = -0.22$ radians and  $\delta \phi(e_{\K}=0.6) = \phi_{\DE}(t_{f}) - \phi_{\OA}(t_{f}) = 0.15$ radians, where $t_{f}$ is the time at which the numerical evolutions end. This estimate, of course, is only valid for the two systems we considered in this section, and it will change with the masses of the binary components and the initial conditions we use for the numerical evolution. We provide more direct comparisons of the waveform computed with these methods in Sec.~\ref{waves} in the context of observations with ground-based detectors.

\subsection{Eccentricity Growth in a Broader Context}
\label{comp2}

The secular growth we have described above was found when working with the leading PN order radiation-reaction force for comparable mass binaries, but this is certainly not the first time secular growth has been found. Secular growth has appeared before in EMRIs modeled through the self-force formalism~\cite{Cutler:1994pb}, and in the high PN order osculating calculations of~\cite{Lincoln:1990ji}. While we spare a detailed discussion of the analytic understanding of the secular growth we have found above until the next section, we wish to here clarify first the relationship between it and the growth found previously in the literature.

\subsubsection{EMRIs with Self-Force.} EMRIs occur when a small compact object (such as a solar mass BH or a neutron star) falls into a supermassive BH. Such events are expected to occur in the dense stellar environment of galactic nuclei, where scatterings force the densest objects toward the gravitational center of the environment. The small inspiraling mass is not strictly speaking a test mass, but rather it generates its own spacetime curvature, creating a so-called ``self-force" on its motion~\cite{Barack:2018yvs}. One must, thus, solve the Einstein field equations in a small mass ratio expansion to properly account for this self-force in the EMRI evolution. More specifically, one seeks a solution for the spacetime metric $g_{\mu \nu} = g_{\mu \nu}^{\BG} + q \; h_{\mu \nu} + {\cal{O}}(q^{2})$, where $g_{\mu \nu}^{\BG}$ is the background spacetime generated by the supermassive BH, $q = \mu/M$ with $(\mu,M)$ the mass of the small and supermassive BHs, respectively, and $h_{\mu \nu}$ is the metric perturbation induced by the small mass. For the sake of this discussion, we will assume that both BHs are non-spinning, so that the background spacetime in given by the Schwarzschild metric.

To leading order, the conservative dynamics of the small mass are governed by geodesics, while the dissipative self-force, induced by GW emission, enters at first order in the mass ratio. The GW fluxes can be computed in this relative leading order approximation with the aid of the Teukolsky formalism\footnote{Technically, for a non-spinning background, the metric perturbations can be solved in the Regge-Wheeler formalism~\cite{Regge:1957td, Vishveshwara:1970cc} for the even parity sector and the Zerilli-Moncrief formalism~\cite{Zerilli:1971wd, Moncrief:1974am} for the odd parity sector. The Teukolsky formalism is applicable for spinning backgrounds.}~\cite{Teukolsky:1973ha, Press:1973zz} , whereby the gravitational perturbation is governed by a linear wave equations for the scalar function $\Psi_{4}$, a particular projection of the Weyl tensor in a Newman-Penrose decomposition, and which is sourced by the geodesic motion of the small mass. The wave equation for $\Psi_{4}$ is separable on the Schwarzschild background and can be integrated using Green's functions. At spatial infinity, the scalar $\Psi_{4}$ is related to the two GW polarizations through $\Psi_{4} = (1/2) (\ddot{h}_{+} - i \ddot{h}_{\times})$, and the GW fluxes can be directly computed from this scalar.

If one works within an adiabatic approximation, where the gravitational perturbation evolves slowly compared to the orbital timescale of the geodesic motion, then the rate at which the orbital energy and angular momentum are lost is balanced by the averaged GW fluxes. This is equivalent to the balance-law statement of the PN formalism. Geodesics of the Schwarzschild background can alternatively be parameterized in terms of a semi-latus rectum $p$ and eccentricity $e$, with the mapping to orbital energy and angular momentum given by~\cite{Cutler:1994pb}
\begin{align}
E^{2} = \mu^{2} \frac{\left(\bar{p} - 2 - 2 e\right)\left(\bar{p} - 2 + 2 e\right)}{\bar{p} \left(\bar{p} - 3 - e^{2}\right)}\,, \qquad L^{2} = \frac{\mu^{2} M^{2} \bar{p}^{2}}{\bar{p} - 3 - e^{2}}\,,
\end{align}
where $\bar{p} = p/M$. Note that while the definitions for $(p,e)$ might be analogous to the Keplerian case, they are not the same orbital elements. The above mappings do, however, agree with the classical Kepler problem if one performs a week field expansion, specifically $\bar{p} \gg 1$, with the orbital energy shifted by the rest mass $\mu$. From these mappings, GW fluxes, and averaged balance laws, one can derive expressions for the rate of change of $(\bar{p}, e)$, which are given in Eqs.~(3.31) and~(3.32) in~\cite{Cutler:1994pb}. These expression help us elucidate the cause of secular growth of eccentricity within the self-force formalism, which was first presented and explained in~\cite{Cutler:1994pb}.

The motion of particles around a Schwarzschild BH have a separatrix in parameter space, dividing stable from unstable orbits. For eccentric orbits, this is given by the relationship $p_{\LSO} = 2 M(3 + e)$, which defines the last stable orbit of the inspiral of the mass $\mu$~\cite{Hansen:1971}. In a weak field expansion, and thus far from the separatrix, the expression for $\dot{e}$ is always negative, as can be seen from Eq.~(4.5) in~\cite{Cutler:1994pb}. It is also worth noting that this expression agrees with the leading PN order expression for $\dot{e}$, which was first derived by Peters and Mathews. In a strong field expansion, close to the separatrix, $\dot{e}$ is always positive, as can be seen from Eq.~(4.11) in~\cite{Cutler:1994pb}, while $\dot{p}$ is still negative. Thus, there is a point in the evolution where $de/dp$ changes sign, which occurs very close to the separatrix and induces a growth in the eccentricity parameter $e$. The secular growth in EMRIs observed within the self-force formalism is thus a strong field effect arising from the presence of the separatrix.

This is a different mechanism than what causes the secular growth observed in Fig.~\ref{traj-fig}. The secular growth observed in that figure occurs within the relative leading PN order radiation-reaction problem, where one does not use an orbit-averaged approximation. As we detailed in~\cite{Loutrel:2018ssg}, the growth we have seen can be understood as a second order effect in a MSA, which scales with the mass ratio squared. On the other hand, the growth seen in self-force calculations is (i) a strong field effect, where the PN expansion is not valid, (ii) it arises within the orbit-averaged approximation, within which the growth in Fig.~\ref{traj-fig} disappears, and (iii) it enters at first order in the mass ratio because it is caused by the first order dissipative self-force, at least to leading order.

An overlapping region where the PN and self-force formalisms agree of course exists. To find it, one simply takes the self force results for $\dot{p}$ and $\dot{e}$, and performs a weak field expansion. On the other hand, one can take the PN results for the orbit-averaged expression for $\dot{p}$ and $\dot{e}_{\K}$, and perform a small mass ratio expansion. The two expansions will agree, at least to the limit of the PN order they are known within the PN formalism. One could then take the expansion of the PN results, and perform a re-summation to leading order in the mass ratio, to thus recover the orbit-averaged self-force results. Since the effect found in Fig.~\ref{traj-fig} is second order in the mass ratio, we might expect that a similar expansion/re-summation could be used to match the PN growth observed here to a similar effect in the self-force formalism, provided the second order self-force were known. 

While our discussion here has focused on the adiabatic limit of the dissipative self-force, the state of the art calculations within this formalism have moved passed this approximation. Problems with the adiabatic limit of the self-force were first considered in~\cite{Pound:2005fs}. Post-adiabatic effects are generated by the leading order conservative self-force, an oscillatory component to the leading order dissipative self-force, and the secular part of second order in mass ration dissipative self-force~\cite{Osburn:2015duj}. Leverages hybrid techniques to compute the post-adiabatic self-force have resulted in exceptionally fast and accurate computations of EMRIs across large regions of the binary's parameter space~\cite{Osburn:2015duj, vandeMeent:2018rms}.

A consequence of working in the post-adiabatic limit of the self-force is that the orbital elements of the geodesic motion become oscillatory on the orbital timescale, in the exact same fashion shown here. These oscillatory effects encode a rich amount of information in the azimuthal and radial frequencies of the binary's motion, as can be seen from Fig.~11 in~\cite{Osburn:2015duj}, and which will ultimately be imprinted in the GWs observed from these systems. Just as in the case of the adiabatic limit of the dissipative self-force, we may expect a certain overlapping region where these oscillatory effects agree between the PN and self-force formalism, provided the self-force is computed to sufficiently high order in the mass ration and the PN force is computed to sufficiently high order in the velocity of the binary. We will explore a post-adiabatic formalism for the rN-RR problem in Section~\ref{msa}.

\subsubsection{Osculating Method at Higher PN Orders.} The osculating method is a generic perturbation theory method used to solve the motion of celestial objects, and thus, it is not surprising that it has found a home within the PN formalism, where PN corrections can be treated as perturbations of the Newtonian gravitational force. To understand the osculating method at higher PN orders, and the secular growth of Keplerian eccentricity often seen within this method, it is useful to consider the case of the 2.5PN accurate equations of motions, where $\vec{a} = \vec{f}_{\rm N} + \vec{f}_{\rm 1PN} + \vec{f}_{\rm 2PN} + \vec{f}_{\rm 2.5PN}$, which was first detailed in Lincoln and Will (LW)~\cite{Lincoln:1990ji}.

In this problem, the evolution equation of the Keplerian eccentricity, for example, may be written as
\begin{equation}
\frac{de_{\K}}{dt} = \xi G_{\rm 1PN}[\mu^{a}(t); \phi] + \xi^{2} G_{\rm 2PN}[\mu^{a}(t); \phi] + \xi^{5/2} G_{\rm 2.5PN}[\mu^{a}(t); \phi]
\end{equation}
where $\mu^{a}$ is the set of orbital elements, which become functions of time under the perturbing force, and $\xi$ is an order keeping parameter. LW sought to solve the osculating equations accurately, and consistently, to 2.5PN order in a MSA. As we detail in the next section, the source terms $G_{k}(\mu^{a}; \phi)$ do not just generate solutions for $e_{\K}(t)$ at the PN order they appear, but they also generate higher-order secular and oscillatory corrections, which enter at higher PN orders. Specifically, $G_{\rm 1PN}$ will source a leading order secular evolution of $e_{\K}$ at 1PN order, and first order oscillatory and secular MSA corrections at 2PN order (and higher). This also happens with $G_{\rm 2PN}$ and $G_{\rm 2.5PN}$, but these higher order MSA corrections enter beyond 2.5PN order. If we seek to solve the 2.5PN accurate equations of motion consistently, then the first order oscillatory and secular MSA corrections due to $G_{\rm 2PN}$ and $G_{\rm 2.5PN}$ can be neglected, just as LW have done. Thus, if one desires to solve the equations of motion consistently to 2.5PN order, one needs: (i) leading order secular, or orbit-averaged, evolutions from $(G_{\rm 1PN}, G_{\rm 2PN}, G_{\rm 2.5PN})$, and (ii) first order in MSA oscillatory and secular corrections from $G_{\rm 1PN}$.

The full osculating equations for $\mu^{a} = (e_{\K}, p, \omega)$ at 2.5 PN order are given by Eqs.~(2.11a)-(2.11c) in~\cite{Lincoln:1990ji}, but we can understand the effect of PN corrections already if we work to 1PN order. The 1PN order form of the evolution equations is 
\begin{align}
\frac{de_{\K}}{d\phi} &= \frac{M}{p} \left\{\left[3 - \eta + \frac{e_{\K}^{2}}{8} \left(56 - 47 \eta\right)\right] \sin V + (5 - 4 \eta) e_{\K} \sin(2V) - \frac{3}{8} e_{\K}^{2} \eta \sin(3V)\right\}\,,
\\
\frac{dp}{d\phi} &= 4 M (2 - \eta) e_{\K} \sin V\,,
\\
e_{\K} \frac{d \omega}{d\phi} &= \frac{M}{p} \left\{3 e_{\K} + \left[-(3 - \eta) + \frac{e_{\K}^{2}}{8} (8 + 21 \eta)\right] \cos V - (5 - 4 \eta) e_{\K} \cos(2V) 
\right.
\nn \\
&\left.
+ \frac{3}{8} e_{\K}^{2} \eta \cos(3V)\right\}\,,
\end{align}
and a simple application of the chain rule with Eq.~\eqref{eq:rkep} reveals that $\dot{r} \ne 0$ when $e_{\K} = 0$. However, the orbit is circular if $V=\pi$ regardless of the value of $e_{\K}$. This is a rather peculiar feature of circular orbits in the osculating formalism that leads to the following interpretation: when the growth of the Keplerian eccentricity occurs, the system transitions from inspiraling ellipses to a quasi-circular state, with the latter defined by Keplerian ellipses stuck in a perpetual state of apastron that precesses at the same rate as the orbital phase. 

The 1PN accurate solution of the above osculating equations can be obtained by analytic integration, taking the Keplerian orbital elements as constants for the time being, since they evolve on the longer radiation-reaction time scale. For the eccentricity, one obtains
\begin{align}
e_{\K}(\phi) &= e_{\K, 0} + \frac{M}{p_{0}} \left\{\left[3 - \eta + \frac{e_{\K,0}^{2}}{8} \left(56 - 47 \eta\right)\right] \cos V + \frac{1}{2}(5 - 4 \eta) e_{\K,0} \cos(2V) 
\right.
\nn \\
&\left.
- \frac{1}{8} e_{\K,0}^{2} \eta \cos(3V)\right\}
\end{align}
where $e_{\K,0} = e_{\K}(\phi = 0)$ and $p_{0} = p(\phi = 0)$. In the case of a system with initial zero Keplerian eccentricity, and restricting to the case of circular orbits with $V = \pi$, we obtain $e_{\K}(\phi) = (3-\eta) (M/p_{0})$. Thus, a circular orbit within the osculating approximation does not have zero Keplerian eccentricity.

The peculiarities of circular orbits within the osculating formalism have a profound impact on the inferred evolution of the Keplerian eccentricity when radiation reaction is included. If we work within the orbit-averaged approximation, radiation reaction can be included in the previous analysis by simply promoting the constants $(e_{\K,0}, p_{0})$ to be functions of time. Since we are considering a secular approximation, these functions will be non-oscillatory. It is convenient for this discussion to consider the Keplerian eccentricity defined through the components of the Runge-Lenz vector, specifically $e_{\K}^{2} = \alpha^{2} + \beta^{2}$, where recall that  $\alpha = e_{\K} \cos \omega$ and $\beta = e_{\K} \sin \omega$. To consider the secular evolution of this eccentricity parameter, LW constructed a mean-square eccentricity from this definition, or more specifically
\begin{equation}
\label{eq:e-LW}
\langle e_{\K}^{2} \rangle = e_{\K, \I}(t)^{2} + \frac{1}{2} (3 - \eta)^{2} \left[\frac{M}{p_{\I}(t)}\right]^{2} + {\cal{O}}\left[e_{\K,\I}^{2} \left(\frac{M}{p_{\I}}\right)^{2}\right]\,,
\end{equation}
where $e_{\K, \I}(t)$ and $p_{\I}(t)$ are obtained by solving the leading order PN radiation-reaction equations in Eqs.~\eqref{eq:dpdt-oa} and~\eqref{eq:dedt-oa}. This is an equivalent expression to Eq.~(3.14) in~\cite{Lincoln:1990ji}, and allows us to understand where the growth seen by LW comes from. The first term in Eq.~\eqref{eq:e-LW} obeys the usual orbit-averaged equations, and thus decreases monotonically as the binary inspirals. The second term scales as $p_{\I}^{-2}$, but the semi-latus rectum also monotonically decreases in the inspiral, and thus, this term is monotonically increasing. In fact, this term can become larger than the leading PN order term in Eq.~\eqref{eq:e-LW}, which induces a growth in the Keplerian eccentricity, producing the results of LW.

There are a few important distinctions between the growth seen by LW, and the growth we have observed in Fig.~\ref{traj-fig}. First, LW only considered the orbit-averaged radiation-reaction effects, whereas the growth in Fig.~\ref{traj-fig} is due to the full radiation-reaction force to leading PN order. Second, LW include the 1PN and 2PN conservative forces in the equations of motion, while in Fig.~\ref{traj-fig} we only considered the rN-RR equations of motion. It is these 1PN and 2PN forces combined with the orbit-averaged radiation-reaction force that produces the growth observed in LW. In our case, the secular growth of the Keplerian eccentricity is purely a dissipative effect, arising from the non-secular radiation-reaction force. In fact, as we have argued in~\cite{Loutrel:2018ssg} and will detail in the next section, the growth arises from non-linearities in a MSA, specifically through the square of first-order oscillatory terms. Thus, the growths seen by LW and that seen in Fig.~\ref{traj-fig} are different phenomena and arise from different aspects of the radiation-reaction problem, even though they are both formally a qualitative growth in an eccentricity parameter.

While the growths are different, there are some important similarities. First, when considering circular orbits, one can show that in the rN-RR problem considered in Fig.~\ref{traj-fig}, $\dot{r} = 0$ when $V=\pi$. This is the same behavior seen by LW, and our numerical calculations show that when the eccentricity begins to grow, $V$ becomes constant. The interpretation of the growth is thus the same as LW, specifically the system transitions from inspiraling ellipses to a quasi-circular state when the growth of the Keplerian eccentricity begins to occur. Further, one can also perform a direct integration of $de_{\K}/dt$, and construct a mean-squared eccentricity in the same way as LW, with the end result being
\begin{equation}
\langle e_{\K}^{2} \rangle = e_{\K, \I}(t)^{2} + \frac{2048}{25} \eta^{2} \left[\frac{M}{p_{\I}(t)}\right]^{5} + {\cal{O}}\left[e_{\K,\I} \left(\frac{M}{p_{\I}}\right)^{5}\right]\,.
\end{equation}
From this, however, we see an important difference: the growth, which is caused by the second term, scales with $\eta^{2}$ in the analysis that led to Fig.~\ref{traj-fig}, and it will occur later in the evolution than the LW growth due to its scaling with $M/p$. One might want to claim that the growth seen in the rN-RR problem is the same as the LW growth from this relationship, but this would not be correct: the second term above is the square of a 2.5PN order correction, and is thus purely dissipative, while the LW correction is purely due to conservative effects. Further, the above expression does not actually account for all of the growth; to do so, one must consider a full MSA of the osculating equations in the rN-RR problem, which we provide the details of in the next section.

\section{Multiple Scale Analysis and the Post-Adiabatic Approximation}
\label{msa}

With our comparison of numerical techniques for solving the radiation-reaction problem, we found that the secular growth in $e_{\K}^{\Osc}$ needs to be accounted for if we want an accurate calculation of the orbital phase of the binary. If we desire to create analytic Fourier domain waveforms that include this effect, how do we go about analytically describing this effect? This section provides the framework to do so through multiple scale analysis (MSA)~\cite{Bender}, which relies on there being a separation of timescales in the problem. For our purposes, these two timescales are the orbital period $T_{\rm orb} \sim M/v^{3}$ and the radiation-reaction timescales is $T_{\R\R} = p /|dp/dt| \sim M/v^{8}$. The ratio of these timescale $T_{\rm orb}/T_{\R\R} \sim v^{5}$, which is small when the orbital velocity is small; even close to the end of the inspiral, when the orbital velocity is about $1/3$ the speed of light, this ratio is still small, $3^{-5} \sim 0.004$. Thus, the MSA is well-justified in the inspiral phase of the coalescence and in the PN formalism. In~\cite{Loutrel:2018ssg}, we provided a brief introduction to the MSA to obtain an analytic explanation of the growth, although MSA is also detailed in~\cite{Bender, PW, Lincoln:1990ji, Pound:2007ti}. Below, we provide the full details of this analysis and how it goes beyond the orbit-averaged (or adiabatic) approximation. We refer to the application of the MSA as the post-adiabatic (PA) approximation. In analogy to how orders are counted in the PN formalism, terms that are first order in the MSA will be referred to 1PA corrections to the orbit-averaged approximation.

Let us then define $\textsf{p} = p/p_{\star}$ and $\textsf{t} = t/(p_{\star}^{3}/M)^{1/2}$, where $p_{\star}$ is a representative length scale of the system. We further define the parameter $\zeta = (8/5) \eta (M/p_{\star})^{5/2}$, which we take to be small. The parameter $p_{\star}$ is arbitrary, but should be chosen such that $\zeta \ll 1$. Previously, in~\cite{Loutrel:2018ssg} we chose $p_{\star} = M$, which allowed us to write the osculating equations in ``code" units, where effectively $M=1$. One could make a different choice, specifically $p_{\star} = p(t=0) = p_{\I}$. However, when computing something observable, $p_{\star}$ will drop out of the expression for the observable, so the particular choice is largely irrelevant. 

For our purposes, it is easier to choose the dependent variable as $\phi$ instead of $t$, since the osculating equations are written as harmonic functions of $\phi$ and the latter would require one to invert the complicated function $t(\phi)$. Thus, we choose $\mu^{a} = (\textsf{p}, \alpha, \beta)$, where recall that $\alpha = e_{\K} \cos\omega$ and $\beta = e_{\K} \sin\omega$, with the osculating equations becoming
\begin{align}
\label{eq:dpdphi}
\frac{d\textsf{p}}{d\phi} &= - \frac{\zeta}{\textsf{p}^{3/2}} \sum_{j=0}^{3} \left[\bar{C}_{p}^{j} \; \cos(j \phi) + \bar{S}_{p}^{j} \; \sin(j \phi)\right]\,,
\\
\label{eq:dadphi}
\frac{d\alpha}{d\phi} &= - \frac{\zeta}{\textsf{p}^{5/2}} \sum_{j=0}^{5} \left[\bar{C}_{\alpha}^{j} \; \cos(j \phi) + \bar{S}_{\alpha}^{j} \; \sin(j \phi)\right]\,,
\\
\label{eq:dbdphi}
\frac{d\beta}{d\phi} &= - \frac{\zeta}{\textsf{p}^{5/2}} \sum_{j=0}^{5} \left[\bar{C}_{\beta}^{j} \; \cos(j \phi) + \bar{S}_{\beta}^{j} \; \sin(j \phi)\right]\,,
\\
\label{eq:dtdphi}
\frac{d\textsf{t}}{d\phi} &= \frac{\textsf{p}^{3/2}}{\left[1 + \alpha \; \cos(\phi) + \beta \; \sin(\phi)\right]^{2}}\,.
\end{align}
We define the slow timescale $\tilde{\phi} = \zeta \phi$, and seek solutions of the form $\mu^{a} = \mu^{a}_{0}(\phi, \tilde{\phi}) + \zeta \mu^{a}_{1}(\phi, \tilde{\phi}) + \zeta^{2} \mu^{a}_{2}(\phi, \tilde{\phi}) + {\cal{O}}(\zeta^{3})$ and $\textsf{t} = \zeta^{-1} \textsf{t}_{-1}(\phi, \tilde{\phi}) + \textsf{t}_{0}(\phi, \tilde{\phi}) + \zeta \textsf{t}_{1}(\phi, \tilde{\phi}) + {\cal{O}}(\zeta^{2})$. Generally, the solution at each order can be written as an oscillatory contribution that depends on both $\phi$ and $\tilde{\phi}$, and a secular contribution which only depends on $\tilde{\phi}$, specifically $\mu^{a}_{j}(\phi, \tilde{\phi}) = \mu^{a}_{j, \osc}(\phi, \tilde{\phi}) + \mu^{a}_{j, \Sec}(\phi)$. Since we are interested in the secular growth observed at small Keplerian eccentricity, we will seek solutions to these contributions through 1PA order, and in a small eccentricity expansion to ${\cal{O}}(e_{\I}^{2})$, where $e_{\I}$ is the initial Keplerian eccentricity.

\subsection{Zeroth post-Adiabatic Order and the Orbit-Averaged Approximation}
\label{0pa}

At leading order (0PA), the osculating equations become
\begin{align}
\frac{\partial \mu^{a}_{0}}{\partial \phi} &= 0\,, \qquad \qquad \qquad \;\;\; \frac{\partial \textsf{t}_{-1}}{\partial \phi} = 0\,,
\\
\label{eq:msa-0}
\frac{\partial \mu_{0}^{a} }{\partial \tilde{\phi}} + \frac{\partial \mu_{1}^{a}}{\partial \phi} &= \textsf{F}^{a}(\mu^{a}_{0})\,, \qquad \frac{\partial \textsf{t}_{-1}}{\partial \tilde{\phi}} + \frac{\partial \textsf{t}_{0}}{\partial \phi} = \textsf{T}(\mu_{0}^{a})
\end{align}
where $\textsf{F}^{a}(\mu_{0}^{a})$ and $\textsf{T}(\mu_{0}^{a})$ are the right-hand sides of Eqs.~\eqref{eq:dpdphi}-\eqref{eq:dtdphi}. The first set of these equations implies the leading order contributions $(\mu_{0}^{a}, \textsf{t}_{-1})$ have no oscillatory terms and only depend on the long timescale $\tilde{\phi}$. This statement is consistent with the fact that, neglecting radiation reaction, the orbital elements $\mu^{a}$ are constant on Keplerian ellipses.

We are now left with solving Eqs.~\eqref{eq:msa-0}, which reduce to
\begin{equation}
\label{eq:msa-0-new}
\frac{d\mu_{0, \Sec}^{a}}{d\tilde{\phi}} + \frac{\partial \mu_{1,\osc}^{a}}{\partial \phi} = \textsf{F}^{a}(\mu^{a}_{0})\,, \qquad \frac{d \textsf{t}_{-1, \Sec}}{d \tilde{\phi}} + \frac{\partial \textsf{t}_{0, \osc}}{\partial \phi} = \textsf{T}(\mu_{0}^{a})\,.
\end{equation}
These equations may be solved by realizing that the second terms on the left-hand side of both equations are oscillatory and will vanish upon orbit averaging, specifically
\begin{equation}
\Big\langle \frac{\partial \mu_{1,\osc}^{a}}{\partial \phi} \Big\rangle = \frac{1}{2 \pi} \int_{0}^{2\pi} d\phi \frac{\partial \mu_{1,\osc}^{a}}{\partial \phi} = \mu_{1,\osc}^{a}(2\pi, \tilde{\phi}) - \mu_{1\osc}^{a}(0, \tilde{\phi}) = 0\,.
\end{equation}
After applying the orbit average to Eqs.~\eqref{eq:msa-0-new}, we finally arrive at the differential equations governing the secular terms,
\begin{align}
\label{eq:p0-msa}
\frac{d\textsf{p}_{0,\Sec}}{d\tilde{\phi}} &= -\frac{1}{\textsf{p}_{0,\Sec}^{3/2}} \left[8 + 7 \left(\alpha_{0,\Sec}^{2} + \beta_{0,\Sec}^{2}\right)\right]\,,
\\
\label{eq:a0-msa}
\frac{d\alpha_{0,\Sec}}{d\tilde{\phi}} &= - \frac{\alpha_{0,\Sec}}{24 \textsf{p}_{0,\Sec}^{5/2}} \left[304 + 121\left(\alpha_{0,\Sec}^{2} + \beta_{0,\Sec}^{2}\right)\right]\,,
\\
\label{eq:b0-msa}
\frac{d\beta_{0,\Sec}}{d\tilde{\phi}} &= - \frac{\beta_{0,\Sec}}{24 \textsf{p}_{0,\Sec}^{5/2}} \left[304 + 121\left(\alpha_{0,\Sec}^{2} + \beta_{0,\Sec}^{2}\right)\right]\,,
\\
\label{eq:t-1-msa}
\frac{d\textsf{t}_{-1,\Sec}}{d\tilde{\phi}} &= \left[\frac{\textsf{p}_{0,\Sec}}{1 - \left(\alpha_{0,\Sec}^{2} + \beta_{0,\Sec}^{2}\right)}\right]^{3/2}\,.
\end{align}
These equations can be combined to recover the results of Peters \& Mathews~\cite{PetersMathews, Peters:1964zz}, given in Eqs.~\eqref{eq:dpdt-oa}-\eqref{eq:dedt-oa}.

One could now solve the above equations numerically, but let us instead seek solutions in the small eccentricity limit only to gain some analytical insight. In this limit, $\alpha_{0,\Sec} \ll 1 \gg \beta_{0,\Sec}$ and we will work to second order in this expansion to recover the growth shown in Fig.~\ref{traj-fig}. When solving these equations, it is useful to choose one of the components of the Runge-Lenz vector as a proxy for the time variable, and solve for the remaining orbital elements in terms of this; we choose $\alpha_{0,\Sec}$ below. To obtain the differential equations for $d\mu_{0,\Sec}^{a}/d\alpha_{0,\Sec}$ one simply has to divide Eqs.~\eqref{eq:p0-msa}-\eqref{eq:t-1-msa} by Eq.~\eqref{eq:a0-msa}. First, consider $d\beta_{0,\Sec}/d\alpha_{0,\Sec}$,
\begin{equation}
\label{eq:dbda-0}
\frac{d\beta_{0,\Sec}}{d\alpha_{0,\Sec}} = \frac{\beta_{0,\Sec}}{\alpha_{0,\Sec}}\,,
\end{equation}
which can be immediately integrated to obtain
\begin{equation}
\label{eq:b0-sol}
\beta_{0,\Sec}(\tilde{\phi}) = \alpha_{0,\Sec}(\tilde{\phi}) \frac{\beta_{0,\Sec}(0)}{\alpha_{0,\Sec}(0)} = \alpha_{0,\Sec}(\tilde{\phi}) \; {\rm tan}(\omega_{\I})\,,
\end{equation}
where we have used $\alpha_{0,\Sec}(0) = e_{\I} \cos(\omega_{\I})$ and $\beta_{0,\Sec} = e_{\I} \sin(\omega_{\I})$, with $(e_{\I}, \omega_{\I})$ the initial values of the Keplerian eccentricity and longitude of pericenter, respectively. Equations~\eqref{eq:dbda-0} and~\eqref{eq:b0-sol} are valid to all orders in eccentricity and don't require any expansions in small eccentricity.

Next, consider the evolution of $\textsf{p}_{0,\Sec}$, which is governed by
\begin{equation}
\frac{d\textsf{p}_{0,\Sec}}{d\alpha_{0,\Sec}} = \frac{24 \textsf{p}_{0,\Sec}}{\alpha_{0,\Sec}} \left[\frac{8 + 7 \left(\alpha_{0,\Sec}^{2} + \beta_{0,\Sec}^{2}\right)}{304 + 121 \left(\alpha_{0,\Sec}^{2} + \beta_{0,\Sec}^{2}\right)}\right]\,.
\end{equation}
To solve this, we first insert Eq.~\eqref{eq:b0-sol} into the above expression and series expand about $\alpha_{0,\Sec} \ll 1$ to obtain
\begin{equation}
\frac{d\textsf{p}_{0,\Sec}}{d\alpha_{0,\Sec}} = \frac{12 \textsf{p}_{0,\Sec}}{19 \alpha_{0,\Sec}} + \frac{435 \textsf{p}_{0,\Sec}}{1444} \frac{\alpha_{0,\Sec}}{\cos^{2}(\omega_{\I})} + {\cal{O}}\left(\alpha_{0,\Sec}^{3}\right)\,.
\end{equation}
Once again, this equation can be directly integrated, and after applying the initial conditions $\textsf{p}_{0,\Sec}(0) = \textsf{p}_{\I}$ and $\alpha_{0,\Sec}(0) = e_{\I} \; \cos(\omega_{\I})$, we obtain
\begin{equation}
\label{eq:p0-sec}
\textsf{p}_{0,\Sec}(\tilde{\phi}) = \textsf{p}_{\I} \sigma(\tilde{\phi})^{12/19} \left\{1 - \frac{435}{2888} e_{\I}^{2} \left[1 - \sigma(\tilde{\phi})^{2}\right]\right\}\,.
\end{equation}
where $\sigma(\tilde{\phi}) = \alpha_{0,\Sec}(\tilde{\phi})/[e_{\I} \cos(\omega_{\I})]$.

The above procedure can be applied to the solve the remaining equations for $\textsf{t}_{-1,\Sec}(\alpha_{0,\Sec})$ and $\tilde{\phi}(\alpha_{0,\Sec})$, specifically
\begin{align}
\label{eq:phi-1-sec}
\tilde{\phi} &= \tilde{\phi}_{\cj} + \frac{\textsf{p}_{\I}^{5/2}}{20} \Bigg\{1 - \sigma(\tilde{\phi})^{30/19} + e_{\I}^{2} \left[-\frac{105}{272} + \frac{2175}{5776} \sigma(\tilde{\phi})^{30/19} + \frac{465}{49096} \sigma(\tilde{\phi})^{68/19} \right]\Bigg\}\,,
\\
\label{eq:t-1-sec}
\textsf{t}_{-1,\Sec}(\tilde{\phi}) &= \textsf{t}_{\cj} + \frac{\textsf{p}_{\I}^{4}}{32} \Bigg\{ 1 - \sigma(\tilde{\phi})^{48/19} + e_{\I}^{2} \left[\frac{14}{53} + \frac{435}{722} \sigma(\tilde{\phi})^{48/19} - \frac{29535}{31046} \sigma(\tilde{\phi})^{86/19}\right]\Bigg\}\,,
\end{align}
where $(\tilde{\phi}_{\cj}, \textsf{t}_{\cj})$ are overall integration constants. These equations can be inverted to write the evolution of the orbital elements in terms of the secular variables $\tilde{\phi}$ or $\textsf{t}_{-1}$, if one desires. The solutions given above are equivalent to the post-circular framework of~\cite{PhysRevD.80.084001}, and the results of Peters and Mathews in the small eccentricity limit.

\subsection{First Post-Adiabatic Order}
\label{1pa}

As discussed in~\cite{Loutrel:2018ssg}, the secular growth in eccentricity can be recovered if one goes to higher order in a MSA of the osculating equations. We here provide an analytic analysis of the osculating equations to first order beyond the adiabatic approximation, or 1PA order. This requires us to solve for both the oscillatory and secular contributions to $\mu_{1}^{a}$ and $\textsf{t}_{0}$. We will begin with the oscillatory terms $(\mu_{1,\osc}^{a}, \textsf{t}_{0,\osc})$, which require us to return to Eq.~\eqref{eq:msa-0-new}. The general procedure for obtaining the oscillatory terms is to move the first term on the left-hand side of these equations to the right-hand side, apply the equations governing the adiabatic approximation $d\mu_{0,\Sec}/d\tilde{\phi}$, and then integrate with respect to $\phi$, specifically
\begin{align}
\label{eq:mu1-osc}
\mu_{1,\osc}^{a}(\phi, \tilde{\phi}) &= \int d\phi \left\{\textsf{F}^{a}[\mu_{0,\Sec}^{a}(\tilde{\phi}), \phi] - \langle \textsf{F}^{a} \rangle [\mu_{0,\Sec}^{a}(\tilde{\phi})]\right\} \,,
\\
\label{eq:t0-osc}
\textsf{t}_{0,\osc}(\phi, \tilde{\phi}) &= \int d\phi \left\{\textsf{T}^{a}[\mu_{0,\Sec}^{a}(\tilde{\phi}), \phi] - \langle \textsf{T}^{a} \rangle [\mu_{0,\Sec}^{a}(\tilde{\phi})]\right\}\,.
\end{align}

First, consider the oscillatory corrections to the orbital elements $\mu_{1,\osc}^{a}$. The forcing functions $\textsf{F}^{a}$ are generally given by the right hand side of Eqs.~\eqref{eq:dpdphi}-\eqref{eq:dbdphi}. Inserting these into Eq.~\eqref{eq:mu1-osc} and integrating, we arrive at
\begin{align}
\textsf{p}_{1,\osc}(\phi, \tilde{\phi}) &= - \frac{1}{\textsf{p}_{0,\sec}(\tilde{\phi})^{3/2}} \sum_{j=1}^{3} \frac{1}{j} \left\{\bar{S}_{p}^{j}\left[\alpha_{0,\sec}(\tilde{\phi}), \beta_{0,\sec}(\tilde{\phi})\right] \cos(j\phi) 
\right.
\nn \\
&\left.
\qquad \qquad \qquad \qquad- \bar{C}_{p}^{j}\left[\alpha_{0,\sec}(\tilde{\phi}), \beta_{0,\sec}(\tilde{\phi})\right] \sin(j\phi)\right\}\,,
\\
\alpha_{1,\osc}(\phi, \tilde{\phi}) &= - \frac{1}{\textsf{p}_{0,\sec}(\tilde{\phi})^{5/2}} \sum_{j=1}^{5} \frac{1}{j} \left\{\bar{S}_{\alpha}^{j}\left[\alpha_{0,\sec}(\tilde{\phi}), \beta_{0,\sec}(\tilde{\phi})\right] \cos(j\phi) 
\right.
\nn \\
&\left.
\qquad \qquad \qquad \qquad- \bar{C}_{\alpha}^{j}\left[\alpha_{0,\sec}(\tilde{\phi}), \beta_{0,\sec}(\tilde{\phi})\right] \sin(j\phi)\right\}\,,
\\
\beta_{1,\osc}(\phi, \tilde{\phi}) &= - \frac{1}{\textsf{p}_{0,\sec}(\tilde{\phi})^{5/2}} \sum_{j=1}^{5} \frac{1}{j} \left\{\bar{S}_{\beta}^{j}\left[\alpha_{0,\sec}(\tilde{\phi}), \beta_{0,\sec}(\tilde{\phi})\right] \cos(j\phi) 
\right.
\nn \\
&\left.
\qquad \qquad \qquad \qquad- \bar{C}_{\beta}^{j}\left[\alpha_{0,\sec}(\tilde{\phi}), \beta_{0,\sec}(\tilde{\phi})\right] \sin(j\phi)\right\}\,.
\end{align}
These resemble the integrated (with respect to $\phi$) version of Eqs.~\eqref{eq:dpdphi}-\eqref{eq:dbdphi}, except that the sums begin at $j=1$ instead of $j=0$. These expressions are also general since they apply for arbitrary eccentricity. 

In order to gain some analytical insight, let us again focus on the small eccentricity limit. To obtain this limit, one can insert the solutions for $[\textsf{p}_{0,\Sec}(\tilde{\phi}), \beta_{0,\Sec}(\tilde{\phi})]$ into the above equations, and expand about $\alpha_{0,\Sec} \ll 1 \gg e_{\I}$ to obtain
\begin{align}
\label{eq:p1-osc}
\textsf{p}_{1,\osc}(\phi, \tilde{\phi}) &= - \frac{18}{\textsf{p}_{\I}^{3/2}} \left\{e_{\I} \sigma(\tilde{\phi})^{1/19} \; \sin(\phi - \omega_{\I}) + \frac{5}{36} e_{\I}^{2} \sigma(\tilde{\phi})^{20/19} \;\sin[2(\phi - \omega_{\I})]\right\}\,,
\\
\label{eq:a1-osc}
\alpha_{1,\osc}(\phi, \tilde{\phi}) &= - \frac{8}{\textsf{p}_{\I}^{5/2}} \left\{\frac{\sin(\phi)}{\sigma(\tilde{\phi})^{30/19}} \left(1 + \frac{2175}{5776} e_{\I}^{2}\right) + \frac{5}{6} \frac{e_{\I}}{\sigma(\tilde{\phi})^{11/19}} \; \sin(2 \phi - \omega_{\I}) 
\right.
\nn \\
&\left.
\qquad + e_{\I}^{2} \sigma(\tilde{\phi})^{8/19}\left[\frac{9377}{5776} \sin(\phi) + \frac{77}{96} \sin(\phi - 2 \omega_{\I}) + \frac{91}{288} \sin(3 \phi - 2 \omega_{\I})\right] \right\}\,,
\\
\label{eq:b1-osc}
\beta_{1,\osc}(\phi, \tilde{\phi}) &= \frac{8}{\textsf{p}_{\I}^{5/2}} \left\{\frac{\cos(\phi)}{\sigma(\tilde{\phi})^{30/19}} \left(1 + \frac{2175}{5776} e_{\I}^{2}\right) + \frac{5}{6} \frac{e_{\I}}{\sigma(\tilde{\phi})^{11/19}} \cos(2 \phi - \omega_{\I}) 
\right.
\nn \\
&\left.
\qquad + e_{\I}^{2} \sigma(\tilde{\phi})^{8/19}\left[\frac{9377}{5776} \cos(\phi) - \frac{77}{96} \cos(\phi - 2 \omega_{\I}) + \frac{91}{288} \cos(3 \phi - 2 \omega_{\I})\right] \right\}\,.
\end{align}
For the solutions $\textsf{t}_{0,\osc}(\phi, \tilde{\phi})$, it is actually easier to simply start with such an expansion from Eq.~\eqref{eq:t0-osc}, rather than solving for the general expression and performing the expansion afterward. Doing so, we obtain
\begin{align}
\label{eq:t0-osc}
\textsf{t}_{0,\osc}(\phi, \tilde{\phi}) &= \textsf{p}_{0,\Sec}(\tilde{\phi})^{3/2} \left\{2 \left[ \beta_{0,\Sec}(\tilde{\phi}) \; \cos(\phi) - \alpha_{0, \Sec}(\tilde{\phi}) \; \sin(\phi)\right] 
\right.
\nn \\
&\left.
+ \frac{3}{4} \left[\left(\alpha_{0,\Sec}(\tilde{\phi})^{2} - \beta_{0,\Sec}(\tilde{\phi})^{2}\right) \; \sin(2\phi) - 2 \alpha_{0,\Sec}(\tilde{\phi}) \beta_{0,\Sec}(\tilde{\phi}) \; \cos(2 \phi)\right] \,.\right\}
\end{align}
After inserting the solutions from the adiabatic approximation, we finally obtain
\begin{align}
\label{eq:t0-osc}
\textsf{t}_{0,\osc}(\phi, \tilde{\phi}) &= \textsf{p}_{\I}^{3/2} \left\{ - 2 e_{\I} \sigma(\tilde{\phi})^{37/19} \; \sin(\phi -\omega_{\I}) + \frac{3}{4} e_{\I}^{2} \sigma(\tilde{\phi})^{56/19} \; \sin[2(\phi -\omega_{\I})]\right\}\,,
\end{align}
which completes the calculation of the 1PA oscillatory terms in the small eccentricity limit.

The first order computation still isn't complete, however. While we have now exhausted Eqs.~\eqref{eq:msa-0}, we still do not have the first order secular contributions $(\mu_{1,\Sec}^{a}, \textsf{t}_{0,\Sec})$. To obtain these, one must go to next order in the MSA, which gives
\begin{equation}
\frac{\partial \mu_{1}^{a}}{\partial \tilde{\phi}} + \frac{\partial \mu_{2}^{a}}{\partial \phi} = \mu^{b}_{1} [\partial_{b} \textsf{F}^{a}](\mu_{0}^{a}, \phi)\,, \qquad \frac{\partial \textsf{t}_{0}}{\partial \tilde{\phi}} + \frac{\partial \textsf{t}_{1}}{\partial \phi} = \mu_{1}^{b} [\partial_{b} \textsf{T}^{a}](\mu_{0}^{a}, \phi)
\end{equation}
where $\partial_{b} = \partial/\partial \mu^{b}$. The general procedure to solve these equations follows the exact same steps as used at 0PA order. Thus, the equations governing the first order secular corrections are
\begin{align}
\frac{d \mu_{1,\Sec}^{a}}{d\tilde{\phi}} &= \mu_{1,\Sec}^{b} \langle \partial_{b} \textsf{F}^{a} \rangle(\mu_{0}^{a}) + \langle \mu_{1,\osc}^{b} \partial_{b} \textsf{F}^{a} \rangle(\mu_{0}^{a})
\\
\frac{d \textsf{t}_{0,\Sec}}{d\tilde{\phi}} &= \mu_{1,\Sec}^{b} \langle \partial_{b} \textsf{T}^{a} \rangle(\mu_{0}^{a}) + \langle \mu_{1,\osc}^{b} \partial_{b} \textsf{T}^{a} \rangle(\mu_{0}^{a})
\end{align}
where we have used the fact that $\mu_{1}^{a} = \mu_{1,\Sec}^{a}(\tilde{\phi}) + \mu_{1,\osc}^{a}(\phi, \tilde{\phi})$ to expand the right-hand side. Naively, it might seem like the second term will vanish upon averaging, but it actually does not since oscillatory terms in the forcing functions will combine with the oscillatory $\mu_{1,\osc}^{a}$, producing terms that are non-oscillatory. This is a general feature of higher order computations.

The expressions resulting from the above orbit average procedure are rather lengthy and not necessary for our purposes. If we expand in the low eccentricity limit, the resulting differential equations become
\allowdisplaybreaks[4]
\begin{align}
\frac{d \textsf{p}_{1,\Sec}}{d\alpha_{0,\Sec}} &= - \frac{18}{19 e_{\I} \cos(\omega_{\I})} \frac{\textsf{p}_{1,\Sec}(\tilde{\phi})}{\sigma(\tilde{\phi})} \left[1 + \frac{145}{304} e_{\I}^{2} \sigma(\tilde{\phi})^{2}\right] + \frac{21}{19} \textsf{p}_{\I} \left[ \alpha_{1, \Sec}(\tilde{\phi}) + \beta_{1,\Sec}(\tilde{\phi}) \tan(\omega_{\I})\right]\,,
\\
\frac{d \alpha_{1,\Sec}}{d\alpha_{0,\Sec}} &=\frac{1}{e_{\I} \cos(\omega_{\I})} \frac{\alpha_{1,\Sec}(\tilde{\phi})}{\sigma(\tilde{\phi})} \left\{1 + \frac{121}{152} e_{\I}^{2} \cos^{2}(\omega_{\I}) \sigma(\tilde{\phi})^{2} \left[1 + \frac{\beta_{1, \Sec}(\tilde{\phi})}{\alpha_{1,\Sec}(\tilde{\phi})} \tan(\omega_{\I})\right] \right\} 
\nn \\
& \qquad - \frac{5 \textsf{p}_{1,\Sec}}{2 \textsf{p}_{\I} \sigma(\tilde{\phi})^{12/19}} - \frac{296 \tan(\omega_{\I})}{57 \textsf{p}_{\I}^{5/2} \sigma(\tilde{\phi})^{30/19}}\,,
\\
\frac{d \beta_{1,\Sec}}{d \alpha_{0,\Sec}} &= \frac{1}{e_{\I} \cos(\omega_{\I})} \frac{\beta_{1, \Sec}(\tilde{\phi})}{\sigma(\tilde{\phi})} \left\{1 + \frac{121}{152} e_{\I}^{2} \cos(\omega_{\I}) \sin(\omega_{\I}) \sigma(\tilde{\phi})^{2} \left[\frac{\alpha_{1,\Sec}(\tilde{\phi})}{\beta_{1,\Sec}(\tilde{\phi})} + \tan(\omega_{\I})\right]\right\}
\nn \\
& - \frac{5 \textsf{p}_{1,\Sec} \tan(\omega_{\I})}{2 \textsf{p}_{\I} \sigma(\tilde{\phi})^{12/19}} + \frac{296}{57 \textsf{p}_{\I}^{5/2} \sigma(\tilde{\phi})^{30/19}}\,.
\end{align}
The above differential equations may seem singular in the limit $\omega_{\I} = n \pi/2$ where $n \in \mathbb{Z}$. However, this is an artifact of our use of $\alpha_{0,\Sec}$ as a proxy for time, since in this limit $\alpha_{0,\Sec}(0) = 0$. For this case, it would be more appropriate to use $\beta_{0,\Sec}$ as the evolution variable instead of $\alpha_{0,\Sec}$.

To solve for the 1PA secular contributions, we require an ansatz of the form $\mu_{1,\Sec}^{a} = \mu_{1, \sec}^{a, (0)} + e_{\I} \mu_{1,\Sec}^{a, (1)} + e_{\I}^{2} \mu_{1, \Sec}^{a, (2)}$, and then we need to solve the above equations order by order in $e_{\I}$. For initial conditions, we require that $\mu_{1}^{a}(0,0) = 0$, which implies $\mu_{1,\Sec}^{a}(0) = - \mu_{1,\osc}^{a}(0,0)$. These conditions are then applied at each order in $e_{\I}$ using Eqs.~\eqref{eq:p1-osc}-\eqref{eq:b1-osc}. The end result for the 1PA secular contributions are
\begin{align}
\label{eq:p1-sec}
\textsf{p}_{1,\Sec}(\tilde{\phi}) &= - \frac{6 e_{\I} \sin(\omega_{\I})}{17 \textsf{p}_{\I}^{3/2}} \left[\frac{44 + 7 \sigma(\tilde{\phi})^{68/19}}{\sigma(\tilde{\phi})^{18/19}}\right] - \frac{5 e_{\I}^{2} \sin(2 \omega_{\I})}{408 \textsf{p}_{\I}^{3/2}} \left[\frac{36 + 168 \sigma(\tilde{\phi})^{68/19}}{\sigma(\tilde{\phi})^{18/19}}\right] 
\\
\alpha_{1,\Sec}(\tilde{\phi}) &= \frac{4 e_{\I} \sin(\omega_{\I})}{45 \textsf{p}_{\I}^{5/2}} \left[\frac{37 - 112 \sigma(\tilde{\phi})^{30/19}}{\sigma(\tilde{\phi})^{11/19}}\right] 
\nn \\
& \qquad - \frac{e_{\I}^{2} \sin(2 \omega_{\I})}{5814 \textsf{p}_{\I}^{5/2}} \left[\frac{71478 - 19754 \sigma(\tilde{\phi})^{30/19} + 279 \sigma(\tilde{\phi})^{68/19}}{\sigma(\tilde{\phi})^{11/19}}\right]
\\
\beta_{1,\Sec}(\tilde{\phi}) &= - \frac{8}{\textsf{p}_{\I}^{5/2}} \sigma(\tilde{\phi}) - \frac{4 e_{\I} \cos(\omega_{\I})}{45 \textsf{p}_{\I}^{5/2}} \left[\frac{37 + 38 \sigma(\tilde{\phi})^{30/19}}{\sigma(\tilde{\phi})^{11/19}}\right]
\nn \\
& \qquad + \frac{e_{\I}^{2}}{5814 \textsf{p}_{\I}^{5/2}} \frac{1}{\sigma(\tilde{\phi})^{11/19}} \left\{ 7942 + 2363 \sigma(\tilde{\phi})^{30/19} + 31 \sigma(\tilde{\phi})^{68/19} 
\right.
\nn \\
&\left.
\qquad + \left[71478 - 49147 \sigma(\tilde{\phi})^{30/19} + 279 \sigma(\tilde{\phi})^{68/19}\right] \cos(2\omega_{\I})\right\}
\end{align}
Following the same procedure for $d\textsf{t}_{0,\Sec}/d\alpha_{0,\Sec}$, and using the above solutions, we find
\begin{align}
\label{eq:t0-sec}
\textsf{t}_{0,\Sec}(\tilde{\phi}) &= \textsf{p}_{\I}^{3/2} \left\{ e_{\I} \sin(\omega_{\I}) \left[-\frac{761}{172} + \frac{33}{17} \sigma(\tilde{\phi})^{18/19} + \frac{1413}{294} \sigma(\tilde{\phi})^{86/19}\right] 
\right.
\nn \\
&\left.
\qquad + e_{\I}^{2} \sin(2\omega_{\I}) \left[\frac{201}{688} + \frac{15}{272} \sigma(\tilde{\phi})^{18/19} + \frac{2355}{5848} \sigma(\tilde{\phi})^{86/19}\right]\right\}\,.
\end{align}
%

\subsection{Reconstructed Eccentricity}
\label{eofF}

\begin{figure*}[ht]
\includegraphics[clip=true, scale=0.4]{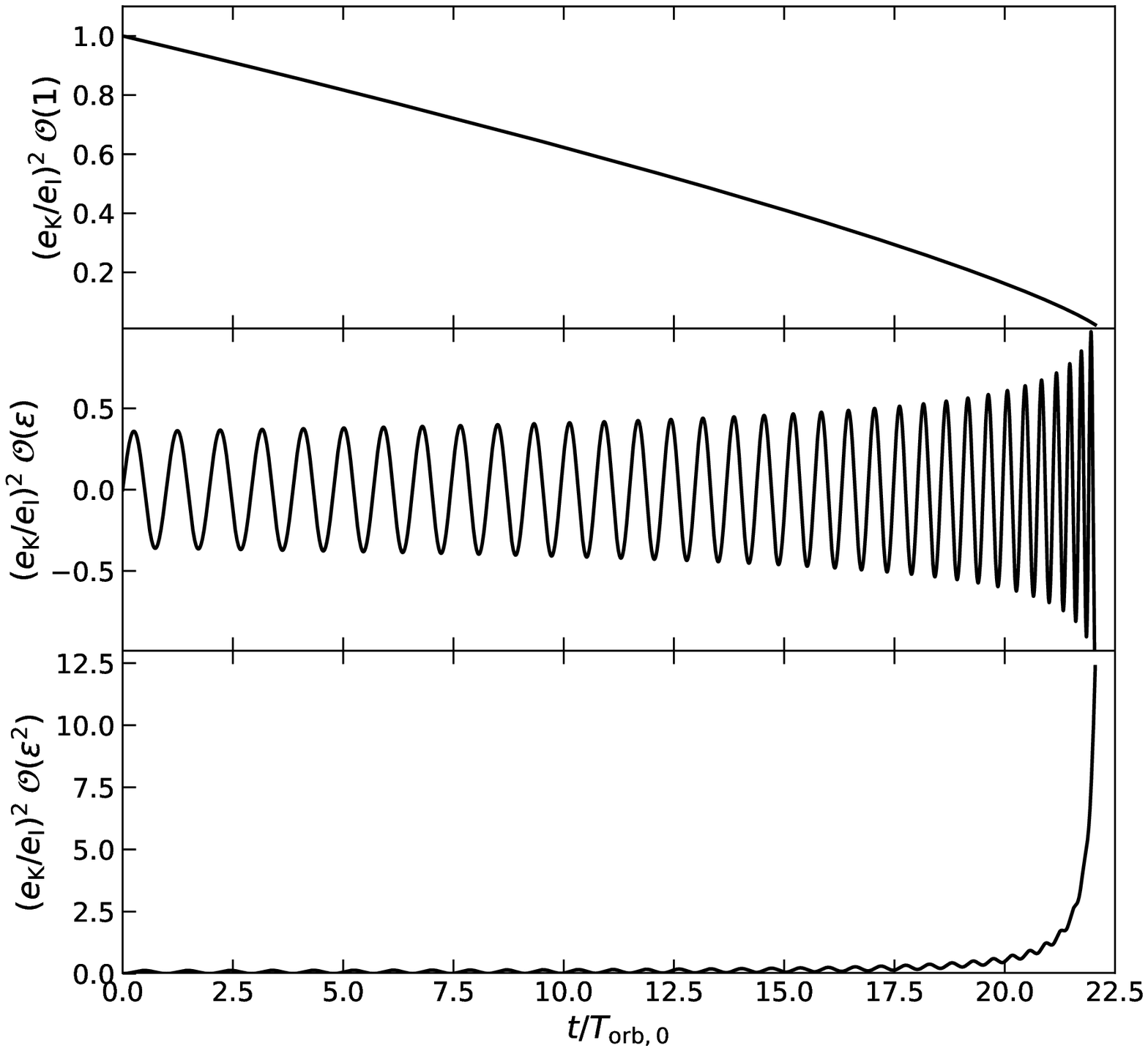}
\includegraphics[clip=true, scale=0.4]{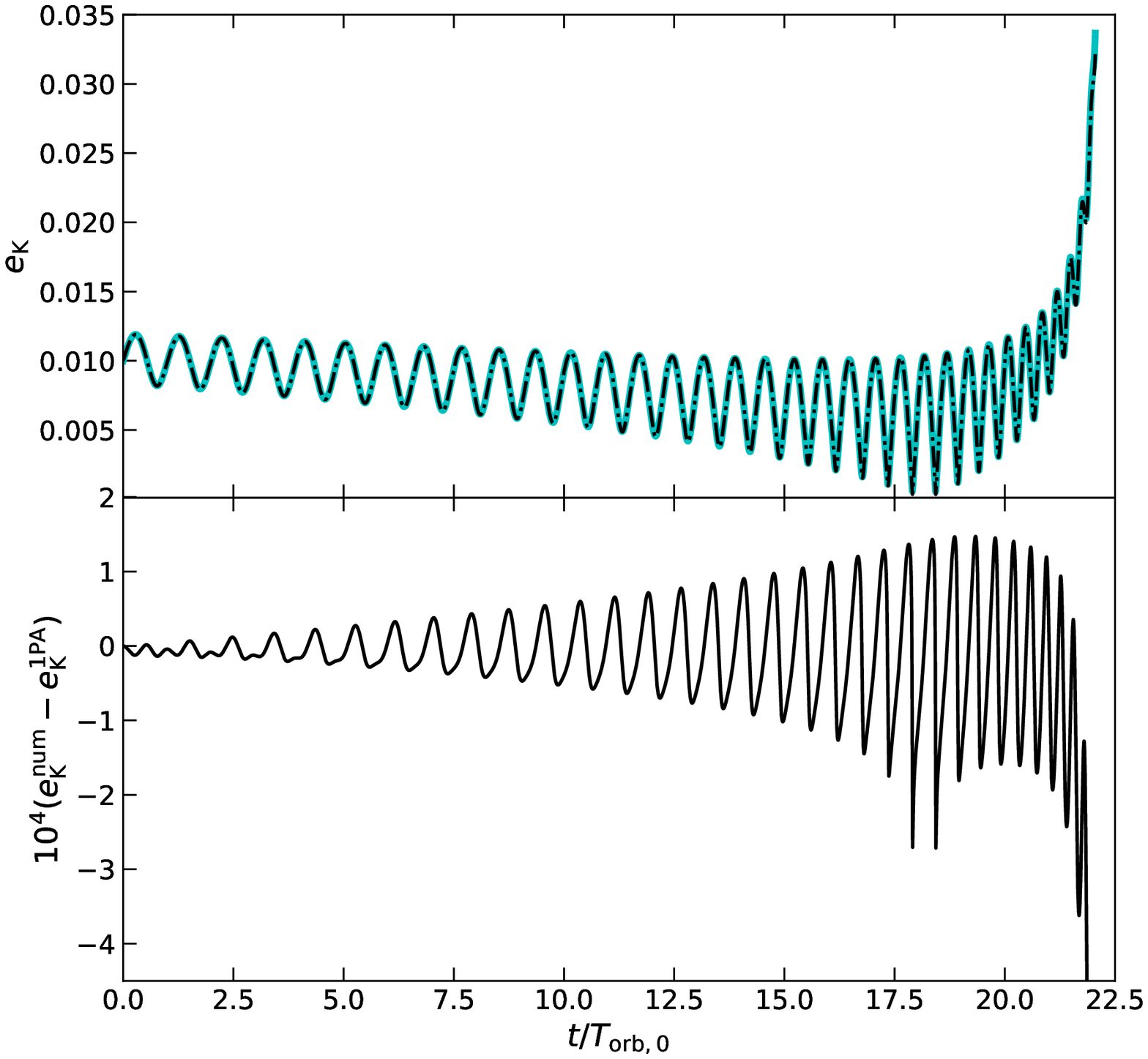}
\caption{\label{pa-fig} Left: Plot of the terms at order ${\cal{O}}(\epsilon^{0})$ (top), ${\cal{O}}(\epsilon)$ (middle), ${\cal{O}}(\epsilon^{2})$ (bottom) in the 1PA reconstruction of $e_{\K}^{2}$. Right: Plot of $e_{\K}$ from the numerical evolution of the osculating equations (solid line) and at 1PA order (dot-dashed). The bottom panel displays the difference between the numerical evolution and the 1PA calculation, multiplied by $10^{4}$.}
\end{figure*}

Now that the PA expansion has been carried out to zeroth- and first-order, let us use the solutions to reconstruct physical observables. The Keplerian eccentricity is given as a function of the Runge-Lenz vector in Eq.~\eqref{eq:ew-to-ab}, which upon expanding is
\begin{align}
\label{eq:rec-e}
e_{\K}(\phi, \tilde{\phi})^{2} &= \alpha_{0,\Sec}(\tilde{\phi})^{2} + \beta_{0,\Sec}(\tilde{\phi})^{2} + 2 \zeta \left[\alpha_{0,\Sec}(\tilde{\phi}) \alpha_{1}(\phi, \tilde{\phi}) + \beta_{0,\Sec}(\tilde{\phi}) \beta_{1}(\phi, \tilde{\phi})\right] 
\nn \\
&\qquad + \zeta^{2} \left[\alpha_{1}(\phi, \tilde{\phi})^{2} + \beta_{1}(\phi,\tilde{\phi})^{2}\right]\,,
\end{align}
where we have not truncated at ${\cal{O}}(\zeta)$. We plot the contributions at each order in $\zeta$ in the left panel of Fig.~\ref{pa-fig}. The leading order contribution reproduces the well known orbit-averaged results, while the first order ${\cal{O}}(\zeta)$ term, is dominated by oscillatory behavior, even though it is a sum of both oscillatory and secular terms. This should not be unexpected, however. The purely secular function $\sigma(\tilde{\phi})$ decreases as the binary inspirals, since it scales like $e_{\K}^{\OA}(t) / e_{\I}$. The 1PA oscillatory terms scale as $\alpha_{1,\osc} \sim \sigma^{-30/19} \sim \beta_{1,\osc}$ while the 1PA secular terms scale as $\alpha_{1,\Sec} \sim \sigma^{-11/19} \sim \beta_{1,\Sec}$ to leading order in $\sigma$. Thus, the 1PA oscillatory terms dominate the linear-in-$\zeta$ corrections to $e_{\K}^{2}$. Finally, we plot the ${\cal{O}}(\zeta^{2})$ corrections in the bottom panel of Fig.~\ref{pa-fig}. These contributions contain both the expected secular corrections from $(\alpha_{1,\Sec}, \beta_{1,\Sec})$, but also secular contributions that come from the square of the 1PA oscillatory terms. These contributions show strong growth in the late inspiral and account for much of the secular growth. We compare the 1PA reconstructed eccentricity from Eq.~\eqref{eq:rec-e} to the eccentricity as computed by numerically solving the osculating equations in the left plot of Fig.~\ref{pa-fig}. The difference between the numerical eccentricity and the 1PA eccentricity does exhibit a secular trend in the late inspiral, indicating that one could create an improved measure of eccentricity by going to higher PA order. However, the difference seen here is less than one hundred times smaller than $e_{\K}$ at the end of the inspiral, so including higher PA order terms will only result in marginal improvement compared to the numerical computation.

Since we have left out the 2PA contributions to $(\alpha, \beta)$ in Eq.~\eqref{eq:rec-e}, one may wonder whether these terms have a large impact on the secular growth described here. While it is difficult to compute these contribution analytically at 2PA due to the increased complexity of the 2PA differential equations, we have numerically investigated the impact of these effects on Eq.~\eqref{eq:rec-e} by numerically solving the MSA equations through 2PA order. We found that the 2PA terms do not significantly affect the secular growth. This can be understood through the analytic calculations carried out here. The contribution to Eq.~\eqref{eq:rec-e} from the 2PA contributions of the Runge-Lenz vector enter as $\alpha_{2}(\phi, \tilde{\phi}) \alpha_{0,\Sec}(\tilde{\phi}) + \beta_{2}(\phi, \tilde{\phi}) \beta_{0,\Sec}(\tilde{\phi})$. As we pointed out previously, $\alpha_{0,\Sec} \sim \sigma(\tilde{\phi}) \sim \beta_{0,\Sec}$, and thus the 2PA contributions to the Runge-Lenz vector are suppressed relative to the 1PA-squared terms. This is further compounded by the fact that $\sigma(\tilde{\phi})$ decreases monotonically as the binary inspirals.

As a final step in our computation of the 1PA approximation, we compute the transformation $\sigma(\chi)$, where $\chi = F_{\I}/F_{0,\Sec}(\tilde{\phi})$ with $F=1/T_{\orb}$ the orbital frequency. We are interested in determining how the inclusion of secular growth affects our ability to measure orbital eccentricity from GW observations, and central to this goal will be the creation of a post-adiabatic Fourier domain waveform. To do this, one needs to know the mapping $e_{\K}(F)$ in order to apply the stationary-phase approximation~\cite{Bender, PhysRevD.80.084001, Moore:2018kvz}. For our calculation, this amounts to finding $\sigma(\chi)$. Writing out the orbital frequency and evaluating it with the leading-order secular contributions $\mu^{a}_{0,\Sec}$ from Eqs.~\eqref{eq:b0-sol} and~\eqref{eq:p0-sec}, we find
\begin{equation}
\label{eq:F0-of-sigma}
\textsf{F}_{0, \Sec} = \textsf{F}_{\I} \left\{\sigma(\tilde{\phi})^{-18/19} - \frac{9969}{5776} e_{\I}^{2} \sigma(\tilde{\phi})^{-18/19} \left[-1 + \sigma(\tilde{\phi})^{2}\right]\right\}
\end{equation}
where $\textsf{F} = (p_{\star}^{3}/M)^{1/2} F$ and $\textsf{F}_{\I} = (1/2\pi) [(1-e_{\K,\I}^{2})/\textsf{p}_{\I}]^{3/2}$. We seek a perturbative inversion of this in the limit $e_{\I} \ll 1$, specifically $\sigma(\chi) = \sigma_{0}(\chi) + e_{\I}^{2} \sigma_{1}(\chi)$. Inserting this ansatz into Eq.~\eqref{eq:F0-of-sigma}, expanding about $e_{\I}$, and solving gives
\begin{equation}
\label{eq:sigma-of-F}
\sigma(\chi) = \chi^{19/18} - \frac{3323}{1824} e_{\I}^{2} \chi^{19/18} \left(-1 + \chi^{19/9}\right)\,,
\end{equation}
where $\chi = \textsf{F}_{\I}/\textsf{F}_{0, \Sec}$. This expression is equivalent to Eq.~(3.11) in~\cite{PhysRevD.80.084001}. We can now re-express the 1PA approximation in terms of $\chi$ by using this expression and re-expanding about $e_{\I} \ll 1$. 

\section{Toward Hybridization of Eccentric Inspiral-Merger-Ringdown Waveforms}
\label{waves}

We have shown throughout this work that the secular growth of the Keplerian eccentricity in the late inspiral is unavoidable. Although the eccentricity itself is a coordinate dependent quantity and is not an observable, it does induce observable effects into the GW emission of the binary. We here explore the impact of the secular growth of the eccentricity parameter that enters PN waveform models on the GW evolution.

\subsection{Waveform Harmonics}
\label{harm}

How does the eccentricity impact the GW emission of a binary system? For circular binaries, and at leading PN order, GWs are emitted at twice the orbital frequency. Once the binary becomes slightly eccentric, the system also emits GWs at the first and third harmonics of the orbital frequency, but the power is still dominated by the second harmonic. As the eccentricity increases, the GW power can be spread over all possible harmonics of the orbital frequency. If one does not properly account for this extra harmonic structure, one could bias the parameters of the recovered signal, or lose detection efficiency.

\begin{figure}[ht]
\includegraphics[clip=true, width=\columnwidth]{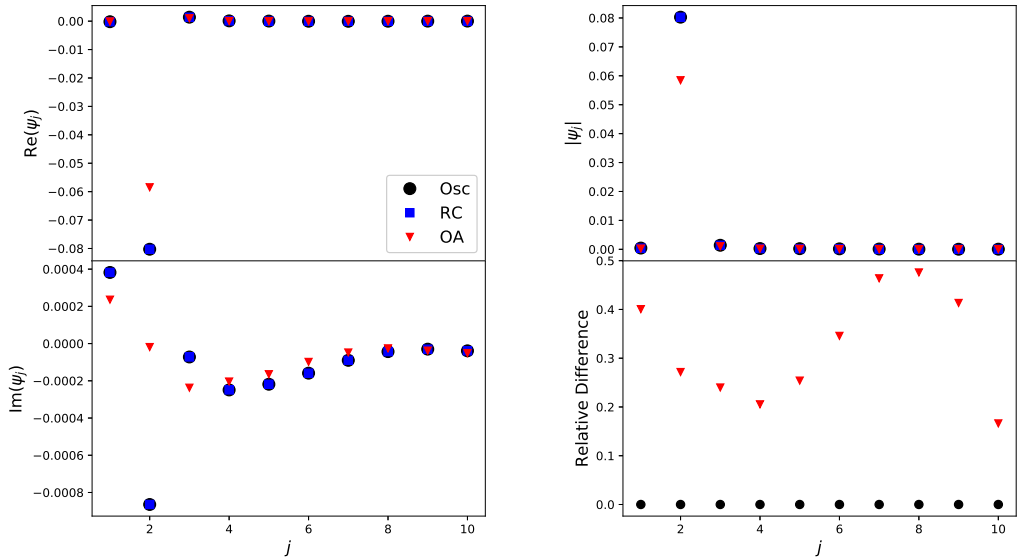}
\caption{\label{psi-fig-1} Waveform harmonic coefficients $\psi_j$ for the three evolution methods considered: osculating (circles), direct evolution of the relative coordinates (squares), and orbit-averaged (triangles). The binary system has initial conditions $(p_{\I}, e_{\I}, \omega_{\I}, \phi_{\I}) = (20 M, 10^{-2}, \pi, 0)$ and masses $(m_{1}, m_{2}) = (10,10) M_{\odot}$. The bottom right plot shows the difference in the norm of the waveform coefficients relative to the direct evolution method. The coefficients of the osculating and direct-evolution waveforms are identical, while the coefficients of the orbit-averaged waveforms display differences of tens of percent when compared to the direct evolution method.}
\end{figure}

To quantify the effect of the eccentricity evolution on the observed GWs, we compute the harmonic coefficients
\begin{equation}
\psi_{j} = \frac{D_{\lum}}{2 \eta M} \int_{t_{0}}^{t_{f}} dt \left[h_{+}(t) - i h_{\times}(t)\right] e^{i j \ell(t)}
\end{equation}
where $i$ is the imaginary unit, and $\ell$ is the mean anomaly, which in the presence of radiation reaction becomes
\begin{equation}
\ell(t) = 2\pi \int dt \; T_{\orb}(t)^{-1}\,.
\end{equation}
Formally, we take the limits of integration $(t_{0}, t_{f})$ to be the start and end times of our numerical evolutions, specifically $t_{0} = 0$ and $t_{f} = t(p=6M)$. We plot the coefficients $\psi_{j}$ in Fig.~\ref{psi-fig-1} for the same system as Figs.~\ref{traj-fig}, and for the three methods of evolving the binary. From the norm of the coefficients $|\psi_{j}|$ (top right panel), we see the typical behavior one would expect from a slightly eccentric system: the $j=2$ harmonic dominates the power. However, one can see clear differences between the coefficients in the different methods by computing the difference between the coefficients relative to the direct evolution method. The coefficients computed in both the osculating and direct-evolution methods agree, which is to be expected since these methods produce the same trajectories for the binary. However, the coefficients in the orbit-averaged method show small differences when compared to the direct evolution method, with the coefficients differing by roughly 15-45$\%$ depending on the harmonic.

\begin{figure}[ht]
\includegraphics[clip=true, width=\columnwidth]{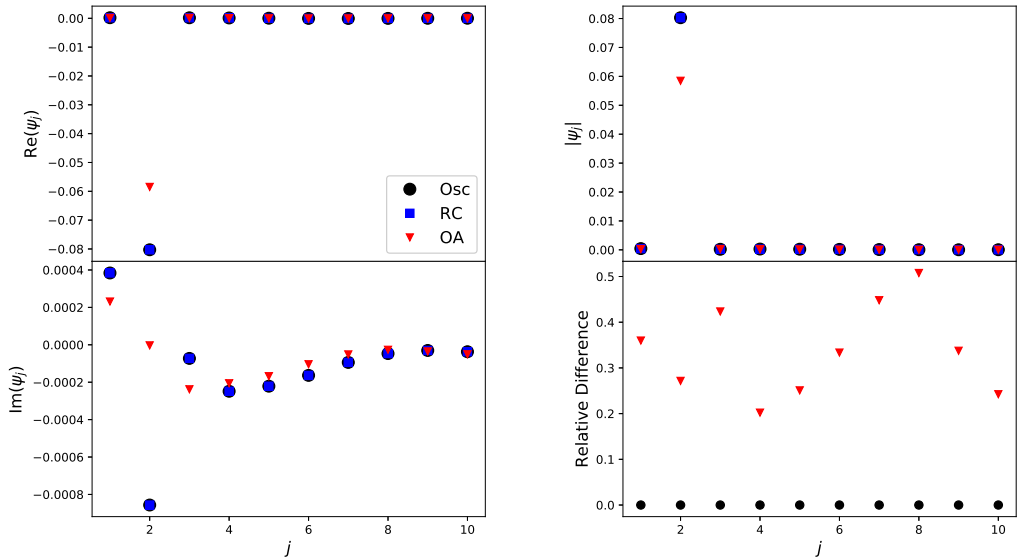}
\caption{\label{psi-fig-2} The same as Fig.~\ref{psi-fig-1}, but with $e_{\I} = 0$. In this case, the Keplerian eccentricity $e_{\K}$ grows secularly throughout the inspiral, while in the orbit-averaged approximation, it remains zero. As with the previous case, the coefficients in the orbit averaged approximation show differences of tens of percent relative to the direct evolution method, while they are identical in the osculating approximation.}
\end{figure}

We consider another scenario of interest in Fig.~\ref{psi-fig-2}, a binary with $e_{\I} = 0$. Note that even though we have set the initial Keplerian eccentricity parameter to zero, the orbit is not per se circular, since oscillations will be present in both the radial separation and velocity at the beginning of the evolution. In the orbit-averaged approximation, the Keplerian eccentricity $e_{\K}$ remains zero throughout the inspiral. On the other hand, in the osculating method, the eccentricity parameter grows, which is consistent with quasi-circular notion that the binary has a non-zero radial velocity. Once again, the coefficients from the orbit-averaged evolution do not match those of the direct evolution method, but those from the osculating approximation do. In principle, one could thus ``search" for the presence of eccentricity growth in a detected signal, or an NR waveform, by computing the harmonic coefficients in this way. 

There is one problem that complicates such a computation. Consider the imaginary part of $\psi_j$, specifically within the orbit-averaged approximation and for the case of $e_{\I} = 0$. In the bottom left panel of Fig.~\ref{psi-fig-2}, we see that the first and third harmonics have a small deviation from zero. However, the conventional wisdom about the orbit-averaged approximation for such a system is that only the second harmonic should be non-zero, since the amplitude of all other harmonics scales with $e_{\K}$. We have verified that the small imaginary component of $\psi_1$ and $\psi_3$ are not due to numerical error. Instead, they result from something more fundamental, specifically the breaking of orthogonality in the radiation-reaction problem. In the modeling of eccentric systems, one often performs a Fourier decomposition, such that the motion, and as a result the waveform, can be written as harmonics of $\ell$. However, this decomposition only works provided the frequency of these harmonics is fixed, and thus, orthogonality between different harmonics is preserved. When radiation reaction is included, the frequency of each harmonic evolves in time, and this orthogonality is broken. In fact this has already been studied in the context of GW modeling in~\cite{Moore:2018kvz}.

For systems with small eccentricity, this can complicate any comparisons to actual detections or to NR waveforms. Detections contain uncertainties due to detector noise, while NR simulations contain numerical error. The harmonic coefficients $\psi_{j\ne2}$ are very small for the systems considered here, and thus may be completely contaminated by numerical error or systematic uncertainties if one attempted to apply these measures to detections or NR waveforms. Once detections are achieved with sufficiently high SNR, or NR simulations reduce numerical error sufficiently, the computation of $\psi_j$ will be possible, and direct comparisons could be made to the PN waveforms considered here.

\subsection{Faithfulness}
\label{sec:match}

An alternative method of comparing different waveform families is through the use of the match. One can define the noise weighted inner product~\cite{cutlerflanagan} between waveforms to be
\begin{equation}
\label{eq:in-prod}
\big( \tilde{h}_{1} | \tilde{h}_{2} \big) = 4  \Re \int df \; \frac{\tilde{h}_{1} \tilde{h}_{2}^{*}}{S_{n}(f)}
\end{equation}
where $\tilde{h}(f)$ is the Fourier transform of the time domain waveform $h(t)$, * corresponds to the complex conjugation, $\Re$ is the real part operator, and $S_{n}(f)$ is the power spectral density of the detector considered. The match between waveforms is then defined as the normalized inner product, maximized over time and phase offsets, specifically
\begin{equation}
{\cal{M}}_{1,2} = \underset{\Delta t, \Delta\phi}{\max} \frac{\big(\tilde{h}_{1} | \tilde{h}_{2} e^{2\pi i f \Delta t + i \Delta \phi}\big)}{\big(\tilde{h}_{1} | \tilde{h}_{1}\big)^{1/2} \big(\tilde{h}_{2} | \tilde{h}_{2}\big)^{1/2}}\,.
\end{equation}
When comparing PN, analytic, Fourier-domain waveform computed in the stationary-phase approximation to numerical waveforms, $(\Delta t, \Delta \phi)$ are $(t_{c}, \phi_{c})$, the time and phase of coalescence, which enter the analytic template as integration constants. In our case, we are comparing numerical waveforms, so we simply apply a time and phase offset to one of the waveform through the overall factor $e^{2\pi i f \Delta t + i \Delta \phi}$, and maximize over these.

The match ${\cal{M}}_{1,2}$ is always in the range $[0,1]$, and is related to the parameter bias induced by using one waveform $\tilde{h}_{1}$ to detect another $\tilde{h}_{2}$. In order for the recovered parameters to be within one-sigma of the injected value, the match must be
\begin{equation}
\label{eq:match-pe}
{\cal{M}}_{1,2} > 1 - \frac{D}{2 \rho^{2}}\,,
\end{equation}
where $D$ is the dimensionality of the intrinsic parameters of the system, and $\rho = (\tilde{h} | \tilde{h})^{1/2}$ is the signal-to-noise ratio~\cite{Baird:2012cu, Lindblom:2008cm, Chatziioannou:2017tdw}. For the eccentric systems considered, $D = 5$, with the intrinsic parameters being $(p_{\I}, e_{\K,\I}, \omega_{\I}, m_{1}, m_{2}).$

We desire to determine at what SNR using orbit-averaged waveforms will produce parameter biases relative to the waveforms computed via direct evolution. For comparison, we also desire to quantify whether the osculating approximation will produce any biases relative to the direct evolution. We thus compute the match between the waveforms in these two approximations versus the direct-evolution waveforms, taking the initial conditions for the numerical evolution of the radiation-reaction equations to be the same across all methods. Specifically, we require the initial values of the relative coordinates to be the same between the various methods. We also require that the initial orbital frequency $F=5$ Hz, and evolve the binaries up to ISCO. We then take the limits of integration in the inner product to be $f_{\min} = 10$Hz and $f_{\max} = 2 F_{\rm ISCO}$.

The results of this computation are show in Table~\ref{match}, where we vary the total mass, mass ratio, and initial eccentricity. For all cases, the match between osculating and direct-evolution waveforms is always unity to the level of numerical error, i.e. ${\cal{M}}_{\Osc, \DE}=1$. This is expected since the osculating and direct-evolution methods produce the same trajectories for the binary up to our numerical accuracy. The match between the orbit-averaged and direct-evolution waveforms, however, is less than unity but still ${\cal{M}}_{\OA, \DE} > 0.9997$ in all cases. Using Eq.~\eqref{eq:match-pe}, the SNR at which a match less than one starts to introduce parameter bias is $\rho_{\rm crit}^{2} = (D/2) (1 - {\cal{M}}_{\OA, \DE})^{-1}$. For the systems that we are considering, and taking ${\cal{M}}_{\OA, \DE} = 0.9997$, this corresponds to $\rho_{\rm crit} \approx 90$. Systems with SNR above this value will have their parameters biased if one performs parameter estimation on eccentric systems using orbit-averaged templates. While ground-based detectors have not detected systems with such a high SNR, these detectors are still undergoing improvement, and it is not unreasonable to expect $\rho \sim 100$ events in the near future. Further, third generation detectors are expected to detect events with $\rho \sim 1000$, thus increasing possible biases in recovered parameters with orbit-averaged templates. Thus, the inclusion of radiation-reaction effects beyond the adiabatic approximation will be necessary in the not too distant future to ensure parameter estimation on eccentricity signals is not biased. We consider the construction of analytic Fourier domain templates in the post-adiabatic approximation in the next section.

%
\begin{table}
\centering
\begin{centering}
\begin{tabular}{c|ccccc}
\hline
\hline
	 $ e_{\K,\I} = 10^{-2}, q = 1 $  & $ M=10 M_{\odot} $ & $ M = 20 M_{\odot} $ & $ M = 40 M_{\odot} $ & $ M = 60 M_{\odot}$ \\ 
\hline
	$ {\cal{M}}_{\OA, \DE} $ & $ 0.9997 $ & $ 0.9998 $  & $ 0.9998 $ & $ 0.9998 $\\
	$ {\cal{M}}_{\Osc, \DE} $ & $ 1.0000 $ & $ 1.0000 $ & $ 1.0000 $ & $ 1.0000 $\\
\hline
\hline
$ e_{\K,\I} = 10^{-2}, m_{1} = 5 M_{\odot} $  & $ m_{2} = 5 M_{\odot} $ & $ m_{2} = 10 M_{\odot} $ & $ m_{2} = 20 M_{\odot} $ & $ m_{2} = 30 M_{\odot}$ \\ 
\hline
	$ {\cal{M}}_{\OA, \DE} $ & $ 0.9997 $ & $ 0.9998 $  & $ 0.9999 $ & $ 0.9999 $\\
	$ {\cal{M}}_{\Osc, \DE} $ & $ 1.0000 $ & $ 1.0000 $ & $ 1.0000 $ & $ 1.0000 $\\
\hline
\hline
	$ m_{1} = 5 M_{\odot}, m_{2} = 5 M_{\odot} $  & $ e_{\K,\I} = 0 $ & $ e_{\K,\I} = 10^{-3} $ & $ e_{\K,\I} = 10^{-2} $ & $ e_{\K,\I} = 10^{-1} $ \\ 
\hline
	$ {\cal{M}}_{\OA, \DE} $ & $ 0.9997 $ & $ 0.9997 $  & $ 0.9997 $ & $ 0.9997 $\\
	$ {\cal{M}}_{\Osc, \DE} $ & $ 1.0000 $ & $ 1.0000 $ & $ 1.0000 $ & $ 1.0000 $\\
\hline
\hline
\end{tabular}
\end{centering}
\caption{\label{match} Matches between waveforms computed in the rN-RR problem using the direct-evolution (DE), osculating (Osc), and orbit-averaged (OA) methods. We vary the total mass (top), mass ratio (middle), and initial Keplerian eccentricity (bottom) of the binary, while the remaining parameters are held fixed, i.e. $(\omega_{\I}, \iota, \beta) = (\pi, 0, 0)$.}
\end{table}
%

\subsection{Post Adiabatic, eCcentric, Multiscale Analysis, Next-to-leading-order (PACMAN) Waveforms}
\label{pacman}

The preceding sections give the following picture: the secular growth in the Keplerian eccentricity has an observable impact on the GWs emitted by binary systems, and post-adiabatic effects that cause this growth will be needed in waveform models in the future to limit biasing recovered parameters. In this section, we detail the construction of such a model, relying on the stationary-phase approximation (SPA)~\cite{Bender}, which has found wide application in the construction of analytic Fourier domain waveforms for GWs from binary systems (see, for example,~\cite{Moore:2018kvz, Chatziioannou:2017tdw, PhysRevD.80.084001, Klein:2018ybm}). The SPA is a method of approximating integrals of the form
\begin{equation}
\label{eq:I-int}
I(f) = \int dt A(t) e^{i f \Psi(t)}\,,
\end{equation}
and relies on the phase $\Psi(t)$ varying more rapidly than the amplitude $A(t)$. More precisely, this requires $A^{-1} dA/dt \ll d\Psi(t)/dt$. Within the orbit-averaged approximation, this relationship is well known to hold for both circular and eccentric systems. However, this is not necessarily true in the post-adiabatic approximation, since the orbital elements themselves become oscillatory.

\begin{figure*}[ht]
\includegraphics[clip=true, scale=0.4]{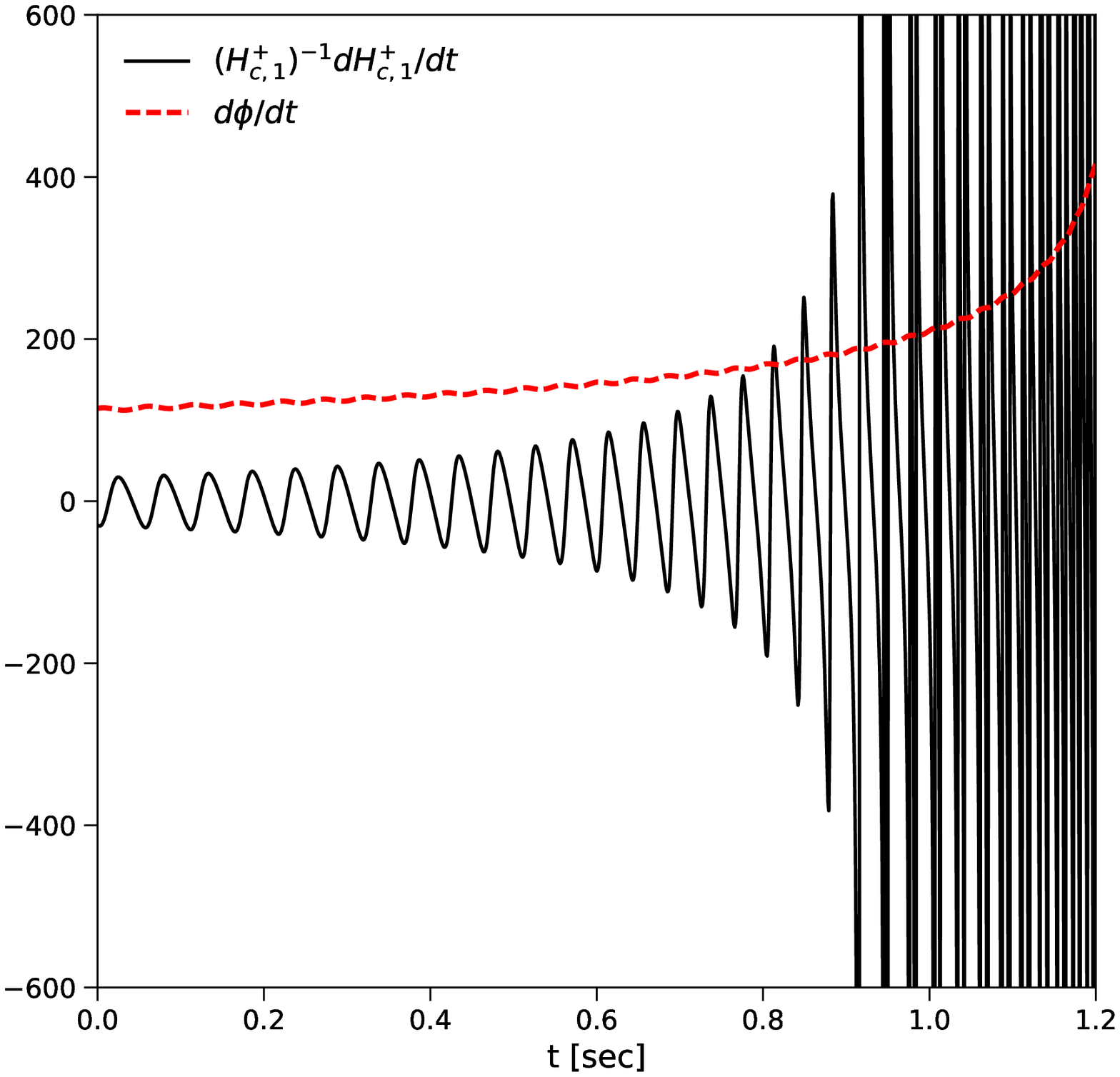}
\includegraphics[clip=true, scale=0.4]{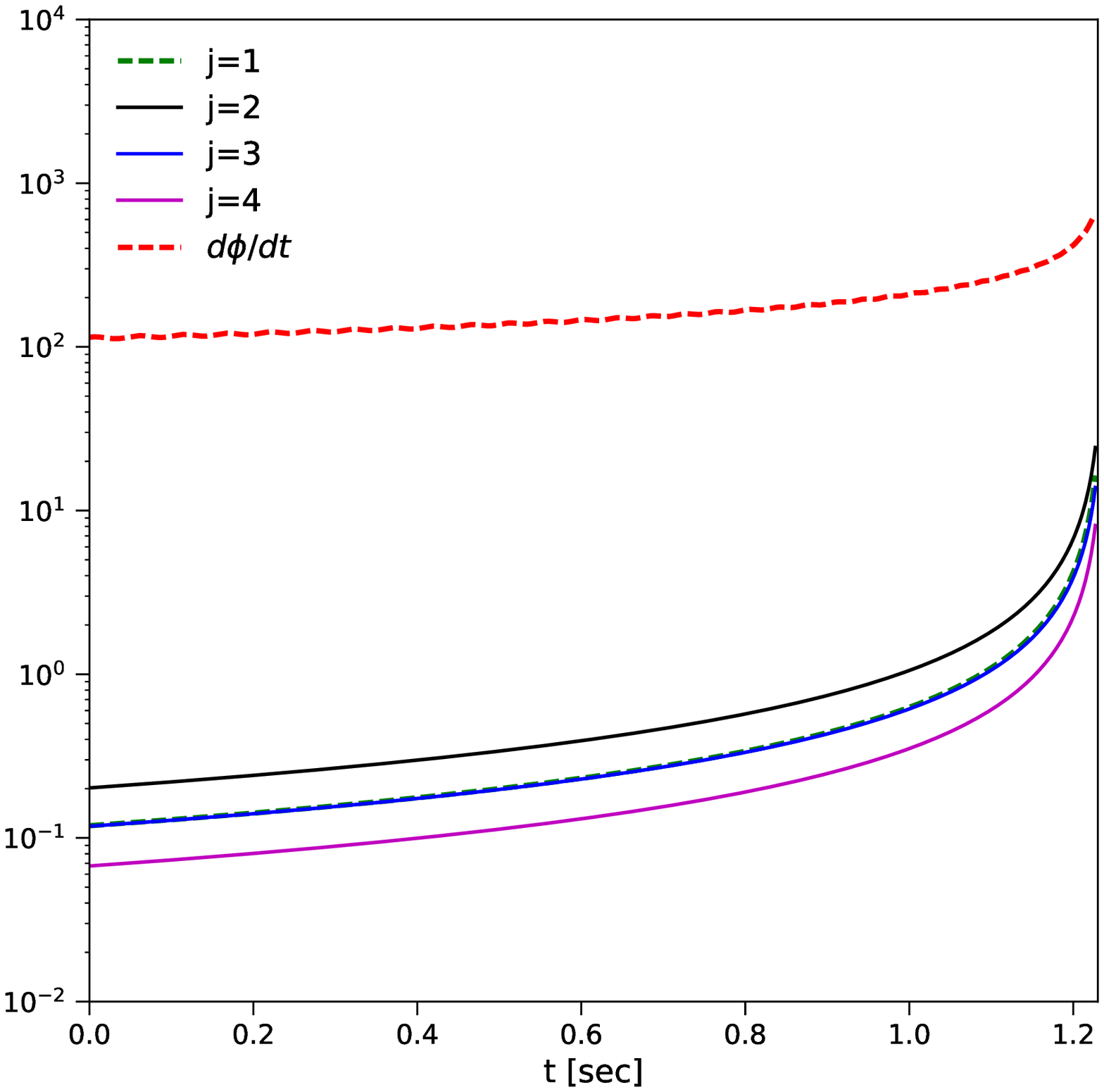}
\caption{\label{spa-fig} Check of the SPA condition $A^{-1} dA/dt \ll d\phi/dt$, for the BH binary in Fig.~\ref{traj-fig}, with (right) and without (left) the 1PA oscillatory contributions removed from the amplitude.}
\end{figure*}

To test whether the SPA condition still holds in the PA approximation, we numerically solve the osculating equations for $[p(t), \alpha(t), \beta(t), \phi(t)]$, and insert them into the time domain waveform coefficients of each harmonic given in Eqs.~\eqref{eq:h-osc}-\eqref{eq:h-osc-coeffs2}. We plot the amplitude of the $\cos(\phi)$ harmonic in $h_{+}$ as a function of time, against the time derivative of the orbital phase in the left plot of Fig.~\ref{spa-fig}. We see from this that the oscillations in the orbital elements cause the SPA condition to be violated. Normally, this would present a problem if we sought to analytically evaluate the Fourier transform of the time domain waveform. However, an oscillatory part of an amplitude can always be recast as part of the phase. For example, in Eq.~\eqref{eq:I-int}, if $A(t) = A'(t) e^{i \Psi'(t)}$ where $A'(t)$ is a non-oscillatory function, then the integral can be rewritten as
\begin{equation}
I(f) = \int dt A'(t) e^{i [f \Psi(t) + \Psi'(t)]}\,,
\end{equation}
and one can apply the SPA to the above integral. Note that, at this stage, the application of the SPA requires that the two phases add constructively. If the phase add destructively, then there can exists points where the first and second derivatives of the phase both vanish, creating catastrophes and breaking the stationary phase condition. The SPA method can be modified in these cases to obtain the Fourier transform~\cite{Klein:2014bua}. We have verified that catastrophes do not exist when we separate the oscillatory contributions from the amplitude.

The PA approximation provides us a means to separate out the oscillatory behavior of the orbital elements in the osculating method from the amplitudes of the waveform. To understand how to do this, it is useful to consider just one of the harmonics, since the procedure is sufficiently general that it can be applied to all of them. Consider, for example, the third harmonics of the orbital phase in $h_{+}$, or more specifically,
\begin{equation}
h_{+}^{(j=3)} = - \frac{\eta M^{2}}{p D_{\lum}} \left[\left(H^{+}_{3,c} + i H^{+}_{3,s}\right) e^{-3i\phi} + \text{c.c.}\right]
\end{equation}
where we have re-written the trigonometric functions in terms of complex exponentials, and c.c. is shorthand for the complex conjugate of the preceding term. The above harmonic contains three pieces: an overall amplitude $h_{0} = - \eta M^{2}/p D_{\lum}$, a harmonic coefficient $\bar{H}^{+}_{3} = H^{+}_{3,c} + i H^{+}_{3,s}$, which depends on the components of the Runge-Lenz vector $(\alpha, \beta)$ and the polarization angles $(\iota, \upsilon)$, and the exponential $e^{-3i\phi}$. The exponential will automatically become part of the phase in the Fourier transform of the waveform, so it does not require any special treatment. The overall amplitude $h_{0}$ and the harmonic coefficient, on the other hand, require us to perform a PA expansion. First, consider $h_{0}$, which only depends on the semi-latus rectum $p$. Using the PA approximation, and the fact that $p = p_{\star} \textsf{p}$, we can write this as
\begin{equation}
h_{0} = - \frac{\eta M^{2}}{p_{\star} D_{\lum}} \frac{1}{\textsf{p}(\phi, \tilde{\phi})} = - \frac{\eta M^{2}}{p_{\star} D_{\lum}} \frac{1}{\textsf{p}_{0,\Sec}(\tilde{\phi})} \left[1 + \zeta \frac{\textsf{p}_{1, \osc}(\phi, \tilde{\phi}) + \textsf{p}_{1,\Sec}(\tilde{\phi})}{\textsf{p}_{0,\Sec}(\tilde{\phi})} + {\cal{O}}(\zeta^{2})\right]\,,
\end{equation}
where we have expanded in $\zeta \ll 1$ to obtain the second equality. Now, we may apply the solution for the 1PA oscillatory contribution $\textsf{p}_{1,\osc}(\phi, \tilde{\phi})$ given in Eq.~\eqref{eq:p1-osc} to re-write the above expression as
\begin{equation}
\label{eq:h0-PA}
h_{0} = - \frac{\eta M^{2}}{p_{\star} D_{\lum}} \frac{1}{\textsf{p}_{0,\Sec}(\tilde{\phi})} \left\{1 + \zeta \frac{\textsf{p}_{1,\Sec}(\tilde{\phi})}{\textsf{p}_{0,\Sec}(\tilde{\phi})} + \zeta \sum_{k=1}^{2} \left[\textsf{h}_{k}(\tilde{\phi}) e^{i k \phi} - \text{c.c.}\right]\right\}\,,
\end{equation}
where $\textsf{h}_{k}$ are functions that only depend on time through $\tilde{\phi}$. In this manner, we have thus separated $h_{0}$ into parts that only depend on $\tilde{\phi}$, and parts that contain oscillatory contributions, which are to be re-combined into the phase of the Fourier transform.

Now consider the harmonic coefficients $\bar{H}^{+}_{3}$, which, using Eqs.~\eqref{eq:hpc3} and~\eqref{eq:hps3}, can be written as
\begin{equation}
\bar{H}^{+}_{3} = \frac{1}{8} \left[3 + \cos(2\iota)\right] e^{2i\upsilon} \left[\alpha(\phi, \tilde{\phi}) + i \beta(\phi, \tilde{\phi})\right]\,.
\end{equation}
Once again, we expand this expression out using the 1PA expressions for $\alpha(\phi, \tilde{\phi})$ and $\beta(\phi, \tilde{\phi})$ given in Eqs.~\eqref{eq:a1-osc} and~\eqref{eq:b1-osc}, obtaining
\begin{equation}
\label{eq:harm-PA}
\bar{H}^{+}_{3} = \frac{1}{8} \left[3 + \cos(2\iota)\right] e^{2i\upsilon} \left\{\alpha_{0,\Sec}(\tilde{\phi}) + i \beta_{0,\Sec}(\tilde{\phi}) + \zeta \left[\alpha_{1,\Sec}(\tilde{\phi}) + i \beta_{1,\Sec}(\tilde{\phi})\right] + \zeta \sum_{l=-1}^{3} \textsf{a}_{l}^{+}(\tilde{\phi}) e^{il\phi}\right\}\,.
\end{equation}
We may now recombine this with the PA expansion of the overall factor $h_{0}$ to re-express $h_{+}^{(j=3)}$ as
\begin{equation}
h_{+}^{(j=3)} = - \frac{\eta M^{2}}{p_{\star} D_{\lum}} \frac{1}{\textsf{p}_{0,\Sec}(\tilde{\phi})} \sum_{j=0}^{4} \left[\textsf{H}^{+}_{3}(\tilde{\phi}) e^{-ij\phi} + \text{c.c.}\right]
\end{equation}
where the coefficients $\textsf{H}_{3}^{+}$ are easily constructed from Eqs.~\eqref{eq:h0-PA} and~\eqref{eq:harm-PA}. This can be generalized to any of the harmonics of $h_{+}$. Thus, the general structure of $h_{+}$ and $h_{\times}$ takes the form 
\begin{equation}
\label{eq:hpc-PA}
h_{+,\times} = - \frac{\eta M^{2}}{p_{\star} D_{\lum}} \frac{1}{\textsf{p}_{0,\Sec}(\tilde{\phi})} \sum_{j=0}^{6} \left[\textsf{H}^{+,\times}_{j}(\tilde{\phi}) e^{-ij\phi} + \text{c.c.}\right]\,.
\end{equation}
The question is now whether the new amplitudes, which only depend on $\tilde{\phi}$, satisfy the SPA condition. We investigate this in the left plot of Fig.~\ref{spa-fig} for the first four harmonics of the above expression. We see from this that the large oscillations which cause the violation of the SPA before the PA expansion are now gone, and the new amplitudes are several orders of magnitude smaller than the time derivative of the orbital phase. This implies that the PA expansion is naturally well suited to the application of the SPA.

Before we consider the application of the SPA to the Fourier transform of Eq.~\eqref{eq:hpc-PA}, there are two subtle details in the PA expansion that need to be pointed out. First is the presence of $p_{\star}$ is the waveform amplitude. This also enters the waveform through any factor of $\zeta$. However, $p_{\star}$ is an arbitrary scale that was introduced in order to perform the MSA, and thus, should not be a true parameter in the waveform. We have verified that once the waveform is written in terms of the physical semi-latus rectum, or more specifically $p_{0,\Sec}(\tilde{\phi})$, that all factors of $p_{\star}$ cancel and the waveform is completely independent of this arbitrary scale. We will show this explicitly when we write down the Fourier domain waveform later in this section. Second, in Sec.~\ref{msa}, we solved for the evolution of the orbital elements to 1PA order. However, the secular growth results form 1PA-squared terms in the eccentricity given by Eq.~\eqref{eq:rec-e}. Since we desire to capture the effect of the secular growth in the Fourier domain waveform, we do not truncate the amplitudes $\textsf{H}_{j}^{+,\times}$ at 1PA order, but also allow them to contain 2PA terms that result from the square of 1PA terms. As a result, these amplitudes \textit{will not} be complete to 2PA order, but will contain the effects of secular growth. Recall that, as we showed before, 2PA corrections to the Keplerian eccentricity through Eq.~\eqref{eq:rec-e} are suppressed relative to the 1PA-squared terms, and thus should not have significant impact on the waveform. 

Now that we have a PA time domain waveform, we may consider the practical problem of applying the SPA to the Fourier transform of this waveform. We desire to construct a frequency domain template that captures PA effects like the secular growth in the waveforms observed by GW detectors. For this reason, we consider the Fourier transform of the detector response, $h(t) = F_{+} h_{+}(t) + F_{\times} h_{\times}(t)$, where $(F_{+}, F_{\times})$ are the beam pattern functions, specifically
\begin{align}
F_{+} &= \frac{1}{2} \left(1 + \cos^{2}\Theta\right) \cos(2\Phi)\,,
\\
F_{\times} &= \cos\Theta \sin(2\Phi)\,,
\end{align}
where $(\Theta, \Phi)$ are the sky location of the source. The Fourier transform of $h(t)$, specifically $\tilde{h}(f) = {\cal{F}}[h(t)]$, takes the form
\begin{equation}
\label{eq:hoff-int}
\tilde{h}(f) = - \frac{\eta M^{2}}{p_{\star} D_{\lum}} \sum_{j=0}^{4} \int_{-\infty}^{\infty} dt \left[\frac{\textsf{H}_{j}(\tilde{\phi})}{\textsf{p}_{0,\Sec}(\tilde{\phi})} e^{i \Psi_{j}^{-}(f,t)} + \frac{\textsf{H}_{j}^{*}(\tilde{\phi})}{\textsf{p}_{0,\Sec}(\tilde{\phi})} e^{i \Psi_{j}^{+}(f,t)}\right]\,,
\end{equation}
where $\textsf{H}_{j}(\tilde{\phi}) = F_{+} \textsf{H}_{j}^{+}(\tilde{\phi}) + F_{\times} \textsf{H}_{j}^{\times}(\tilde{\phi})$, $\Psi_{j}^{\pm}(f,t) = 2 \pi f t \pm j \phi(t)$, and $^{*}$ corresponds to complex conjugation. The application of the SPA requires the existence of a stationary point in the phases $\Psi_{j}^{\pm}(f,t)$. However, the PA expansion in Sec.~\ref{msa} is given in terms of the variable $\tilde{\phi}$, and not $t$. One could derive the relationship $\phi(t)$, or $\tilde{\phi}(t)$, and re-express the 1PA solutions in terms of $t$ in this manner. Although, from a practical standpoint, it is much cleaner to continue to work in terms of $\tilde{\phi}$ instead of $t$. This does require a few extra steps to modify Eq.~\eqref{eq:hoff-int} so that it becomes an integral of $\tilde{\phi}$.

Let us begin by considering the phase $\Psi_{j}^{\pm}(f,t)$. In a PA expansion, this can be written as
\begin{equation}
\Psi_{\pm}(f,\phi, \tilde{\phi}) = \zeta^{-1} \left[2 \pi \textsf{f} \; \textsf{t}_{-1,\Sec}(\tilde{\phi}) \pm j \tilde{\phi}\right] + 2 \pi \textsf{f} \left[\textsf{t}_{0,\osc}(\phi, \tilde{\phi}) + \textsf{t}_{0,\Sec}(\tilde{\phi})\right] + {\cal{O}}(\zeta)\,,
\end{equation}
where $\textsf{f} = (p_{\star}^{3}/M)^{1/2} f$, and we have used the fact that $\tilde{\phi} = \zeta \phi$. This phase contains oscillatory terms through $\textsf{t}_{0,\osc}(\phi, \tilde{\phi})$, which can complicate finding the stationary points. In the preceding discussion, we re-expressed amplitudes that contained oscillatory components into new amplitudes, that were only secularly evolving, and a correction to the phase of $h_{+,\times}(t)$. To handle the oscillatory terms in $\Psi_{j}^{\pm}$, we can apply a similar procedure. We consider the function $e^{i \Psi_{j}^{\pm}}$, separate out the oscillatory terms, and perform a small eccentricity expansion, specifically
\begin{equation}
e^{i \Psi_{j}^{\pm}(f,\phi,\tilde{\phi})} = e^{i \Psi_{j}^{\pm}(f,\tilde{\phi})} e^{2\pi i \textsf{f} \; \textsf{t}_{0,\osc}(\phi, \tilde{\phi})} = e^{i \Psi_{j}^{\pm}(f,\tilde{\phi})} \left\{1 + \sum_{k=1}^{2} \left[\textsf{T}_{k}(f,\tilde{\phi}) e^{-i k \phi} + \text{c.c}\right]\right\}
\end{equation}
where
\begin{equation}
\Psi_{j}^{\pm}(f,\tilde{\phi}) = \zeta^{-1} \left[2 \pi \textsf{f} \; \textsf{t}_{-1,\Sec}(\tilde{\phi}) \pm j \tilde{\phi}\right] + 2\pi \textsf{f} \; \textsf{t}_{0,\Sec}(\tilde{\phi})\,,
\end{equation}
and $\textsf{T}_{k}(f,\tilde{\phi})$ are complex functions that arise from Eq.~\eqref{eq:t0-osc} and are accurate to ${\cal{O}}(e_{\I}^{2})$. The terms $e^{ik\phi}$ can now be recombined in the Fourier phase $\Psi_{j}^{\pm}(f,\tilde{\phi})$ by using $\tilde{\phi} = \zeta \phi$. 

This expansion and the treatment detailed above is similar to what one normally does when considering the SPA for eccentricity binaries in the post-circular limit. The Fourier phase $\Psi_{j}^{\pm}$, when written in terms of the orbital phase $\phi(t)$, can have many stationary points for each value of $j$. On the other hand, one can re-write the Fourier phase in terms of the secularly evolving mean anomaly, $\ell(t)$, which ensures that each harmonic has only one stationary point. The oscillatory terms in the relationship between $\phi$ and $\ell$ are re-expressed as part of the amplitude using a Bessel decomposition. 

Now, consider the differential $dt$. This can be immediately converted using $dt = (dt/d\phi) d\phi$. However, we once again have to perform a PA expansion, ensuring that oscillatory terms are properly separated out and combined into the phase. More specifically,
\begin{equation}
\label{eq:dt-PA}
dt = \zeta^{-1} \left(\frac{p_{\star}^{3}}{M}\right)^{1/2} \left[\frac{\partial \textsf{t}(\phi, \tilde{\phi})}{\partial\phi} + \zeta \frac{\partial \textsf{t}(\phi, \tilde{\phi})}{\partial \tilde{\phi}}\right] d\tilde{\phi}\,,
\end{equation}
where $\textsf{t}(\phi,\tilde{\phi})$ is given to 1PA order through Eqs.~\eqref{eq:t-1-sec},~\eqref{eq:t0-osc}, and~\eqref{eq:t0-sec}. The process of separating out the oscillatory terms coming from $\textsf{t}_{0,\osc}(\phi, \tilde{\phi})$ follows the exact same procedures detailed above, so we will not repeat them here. After re-inserting all of this back into Eq.~\eqref{eq:hoff-int}, we finally obtain the Fourier integral
\begin{equation}
\label{eq:hoff-int2}
\tilde{h}(f) = - \zeta^{-1} \left(\frac{p_{\star}}{M}\right)^{1/2} \frac{\eta M^{2}}{D_{\lum}} \sum_{j=0}^{6} \int_{-\infty}^{\infty} d\tilde{\phi} \left[\textsf{A}_{j}(\tilde{\phi}) e^{i \Psi_{j}^{-}(f,\tilde{\phi})} + \textsf{A}_{j}^{*}(\tilde{\phi}) e^{i \Psi_{j}^{+}(f,\tilde{\phi})}\right]\,.
\end{equation}
The process of re-arranging the amplitudes and phase, as well as writing the integral over $\tilde{\phi}$ instead of $t$, ensures that each harmonic in the integrand will only have one stationary point, and we thus may now apply the SPA to the above integral.

To solve for the stationary points $\tilde{\phi}_{s}$, we seek the points where $d\Psi_{j}^{\pm}(f,\tilde{\phi})/d\tilde{\phi} = 0$. Taking the necessary derivative, we have
\begin{equation}
\frac{d\Psi_{j}^{\pm}(f,\tilde{\phi})}{d\tilde{\phi}} = \zeta^{-1} \left[2 \pi \textsf{f} \; \frac{d\textsf{t}_{-1,\Sec}}{d\tilde{\phi}} \pm j\right] + 2 \pi \textsf{f} \; \frac{d \textsf{t}_{0,\Sec}}{d\tilde{\phi}}=0\,.
\end{equation}
This can be directly solved to obtain
\begin{align}
\textsf{f} &= \mp \frac{j}{2\pi} \left(\frac{d\textsf{t}_{-1,\Sec}}{d\tilde{\phi}}\right)_{\tilde{\phi} = \tilde{\phi}_{s}}^{-1} \left[1 - \zeta \left(\frac{d\textsf{t}_{-1,\Sec}}{d\tilde{\phi}}\right)_{\tilde{\phi} = \tilde{\phi}_{s}}^{-1} \left(\frac{d\textsf{t}_{0,\Sec}}{d\tilde{\phi}}\right)_{\tilde{\phi} = \tilde{\phi}_{s}} + {\cal{O}}(\zeta^{2})\right]
\nn \\
& = \mp \frac{j}{\textsf{T}_{\rm orb}(\tilde{\phi}_{s})} \left[1 - \zeta \frac{2\pi}{\textsf{T}_{\rm orb}(\tilde{\phi}_{s})} \left(\frac{d\textsf{t}_{0,\Sec}}{d\tilde{\phi}}\right)_{\tilde{\phi} = \tilde{\phi}_{s}} + {\cal{O}}(\zeta^{2})\right]
\end{align}
where in the second line we have applied Eq.~\eqref{eq:t-1-msa}, and realized that this is simply the orbital period normalized by $(p_{\star}^{3}/M)^{1/2}$. The right-hand side of this expression can be straightforwardly evaluated, using the results of Sec.~\ref{msa}, to obtain $\tilde{\phi}_{s}(\textsf{f})$. The first term, which is ${\cal{O}}(\zeta^{0})$, is the condition for the stationary point in the orbit-averaged approximation, which has been found in previous work. We find that this is modified by a 1PA order term, which acts to blueshift (increase) the frequency in the case of $\Psi_{j}^{-}$, and redshift the frequency in the case of $\Psi_{j}^{+}$, since $d\textsf{t}_{0,\Sec}/d\tilde{\phi} < 0$. Further, this additional term scales like a 2.5PN order correction relative to the leading-order term, due to how the PA expansion works in the problem we are considering. Thus, for inspiraling binaries, this term is always guaranteed to be much smaller than the leading order term. This ensures that the frequency for $\Psi_{j}^{+}$ is always negative in the inspiral, and can be neglected for the sources under consideration.

To evaluate the Fourier integral in Eq.~\eqref{eq:hoff-int2}, we expand both the amplitude and phase about $\tilde{\phi}_{s}$, specifically
\begin{equation}
\tilde{h}^{\spa}(f) = - \zeta^{-1} \left(\frac{p_{\star}}{M}\right)^{1/2} \frac{\eta M^{2}}{D_{\lum}} \sum_{j=1}^{6} \textsf{A}_{j}(\tilde{\phi}_{s}) e^{i \Psi_{j}^{-}(\tilde{\phi}_{s})} \int_{-\infty}^{\infty} d\tilde{\phi} \; e^{(i/2)(\Psi_{j}^{-})''(\tilde{\phi}_{s}) (\tilde{\phi} - \tilde{\phi}_{s})^{2}}\,.
\end{equation}
The above integral can be easily evaluated, and after applying $\tilde{\phi}_{s}(\textsf{f})$, we obtain
\begin{equation}
\label{eq:h-pacman}
\tilde{h}^{\spa}(f) = - \left(\frac{5}{384}\right)^{1/2} \frac{{\cal{M}}^{5/6}}{\pi^{2/3} f^{7/6} D_{\lum}} \sum_{j=1}^{6} {\cal{A}}_{j}(f) e^{i [\tilde{\Psi}_{j}^{0}(f) + \delta \tilde{\Psi}^{\PA}_{j}(f)]}\,,
\end{equation}
where the Fourier phase is
\begin{align}
\label{eq:psi-adiab}
\tilde{\Psi}_{j}^{0}(f) &= 2\pi f t_{c} - j \phi_{c} - \frac{\pi}{4} + \frac{3 j^{8/3}}{2^{2/3} 512} \left(\pi {\cal{M}} f\right)^{-5/3} \left[1 - \frac{2355}{1462} e_{\I}^{2} \left(\frac{j F_{\I}}{f}\right)^{19/9}\right]\,,
\\
\label{eq:psi-pa}
\delta \tilde{\Psi}^{\PA}_{j}(f) &= j \left(\frac{33}{17} e_{\I} \sin\omega_{\I} + \frac{15}{136} e_{\I}^{2} \cos\omega_{\I} \sin\omega_{\I}\right) + j^{3} \left[\frac{1413}{2924} e_{\I} \sin\omega_{\I} \left(\frac{j F_{\I}}{f}\right)^{34/9} 
\right.
\nn \\
&\left.
\qquad + \frac{2355}{2924} e_{\I}^{2} \cos\omega_{\I} \sin\omega_{\I} \left(\frac{j F_{\I}}{f}\right)^{34/9}\right]\,,
\end{align}
with ${\cal{M}} = M \eta^{3/5}$ the chirp mass of the binary, $F_{\I}$ the initial orbital frequency, $t_{c} = \zeta^{-1} \textsf{t}_{c}$, $\phi_{c} = \zeta^{-1} \tilde{\phi}_{c}$, and the Fourier amplitudes ${\cal{A}}_{j}(f)$ are given explicitly in~\ref{fas}. This completes the derivation of the PACMAN waveform.

Before continuing, it is useful to note a few properties of the Fourier phase of the PACMAN waveform. First, the term proportional to $j$ in $\delta \tilde{\Psi}_{j}^{\PA}$ is a constant, and thus, completely degenerate with the arbitrary phase of coalescence $\phi_{c}$. In fact, this term can be removed from the total phase by re-defining $\phi_{c}$. Second, the term proportional to $j^{3}$, is actually a 2.5PN correction to the adiabatic part of the phase $\tilde{\Psi}_{j}^{0}$. This can be seen be realizing that the orbital velocity is $v = (\pi {\cal{M}} f)^{1/3}$. The frequency dependent part of the adiabatic phase scales as $v^{-5}$, a well known result of the orbit-averaged approximation. Meanwhile, the frequency dependent part of $\delta \tilde{\Psi}_{j}^{\PA}$ scales like $v^{0}$, and is thus $v^{5}$, or 2.5PN order, higher than the leading PN order term in the adiabatic phase. The amplitudes display a similar structure, which can be seen from Eq.~\eqref{eq:calA-pa}. This is exactly the behavior we expect from the PA approximation considered here. 

Before we consider parameter estimation with the PACMAN waveform, we would like to note a few key aspects about the model. In developing this waveform, we worked to second order in $e_{\I}$, and as a result, the amplitudes and phase of the waveform are only accurate to this order. We only considered the small eccentricity to this order, since it is the minimum order that is required to understand the secular growth reported here. We do, however, have analytic control over these expressions, and one could easily extend them to any order in $e_{\I}$ if one desired to consider binaries with moderate values of the Keplerian eccentricity parameter. For an example of this higher order computation in the adiabatic limit, see~\cite{PhysRevD.80.084001}.

Further, we have thus far only considered the rN-RR problem, and as a result, the adiabatic part of the PACMAN waveform's phase, $\tilde{\Psi}_{j}^{0}$, is only accurate to leading PN order. This would normally limit the waveform's usefulness for source of ground-based detectors, since higher PN order effects can have a significant impact on the phase of the GW. However, the benefit of the PA expansion considered here is that the adiabatic part of the phase is identical to the phase one would find by working in the orbit averaged approximation, with corrections only appearing at ${\cal{O}}(v^{5n})$, where $n$ is the PA order one is working to. As a result, one can very easily write down the adiabatic part of the phase to 3PN order, which is given to second order in $e_{\I}$ in Eq.~(6.26) in~\cite{Moore:2016qxz}. The next order PA effects from the 2.5PN radiation reaction force would enter the phase at relative 5PN order, or ${\cal{O}}(v^{10})$ relative to the leading PN order term in the adiabatic part of the phase. In reality, the ``true" next order PA effects will actually enter at relative 3.5PN order, or ${\cal{O}}(v^{7})$, since the radiation reaction force also contains corrections at 3.5PN order, or 1PN order relative to $\vec{f}_{\mbox{\tiny 2.5PN}}$, specifically $\vec{f}_{\mbox{\tiny 3.5PN}}$. In order to include these effects, one would have to repeat the analyses carried out in this section, as well as Sec.~\ref{msa}, starting with the osculating method for the equations of motion $\vec{a} = \vec{f}_{\N} + \vec{f}_{\mbox{\tiny 1PN}} + \vec{f}_{\mbox{\tiny 2.5PN}} + \vec{f}_{\mbox{\tiny 3.5PN}}$, which is the relative 1PN order radiation reaction problem. We considered the presence of eccentricity growth in this problem numerically in Fig.~3 of~\cite{Loutrel:2018ssg}, with the PA framework for tackling this problem developed in~\cite{Damour:2004bz} and 1PA oscillatory corrections considered in~\cite{Damour:2004bz, Konigsdorffer:2006zt, Moore:2016qxz}.

\subsection{Parameter Estimation}
\label{pe}

Now that we have obtained the Fourier transform of the detector response $\tilde{h}(f)$ in the SPA, we can begin to consider how PA effects will impact parameter estimation in a more rigorous sense. For the purposes of this study, we seek to determine how the PACMAN waveform may improve our ability to measure the parameters of the binary system emitting the GWs. Specifically, we focus on how accurately a ground based detector, aLIGO in this instance, can recover the initial eccentricity of the binary. To do this, we perform a Fisher analysis on the PACMAN waveforms, and compare to the same analysis in the adiabatic limit. 

We will here use a Fisher analysis~\cite{Finn:1992xs, cutlerflanagan, Cutler:2007mi, Gondan:2017hbp, Vallisneri:2007ev} to estimate the accuracy to which parameters can be extracted. This analysis relies on the Fisher information matrix, whose elements are given by
\begin{equation}
\Gamma^{ab} = \Bigg(\frac{\partial \tilde{h}}{\partial \lambda^{a}} \Bigg| \frac{\partial \tilde{h}}{\partial \lambda^{b}} \Bigg)\,,
\end{equation}
where $\lambda^{a}$ is the set of parameters characterizing the detector response, and recall the noise-weighted inner product is given in Eq.~\eqref{eq:in-prod}. The accuracy to which one can measure the parameters $\lambda^{a}$ can then be estimated by the diagonal components of the inverse Fisher matrix, specifically
\begin{equation}
\Delta \lambda^{a} = \left[\left(\Gamma^{-1}\right)^{aa}\right]^{1/2}\,.
\end{equation}
This approximation works well provided the SNR is high enough and the noise is stationary and Gaussian~\cite{Vallisneri:2007ev}. Given that secular effects only begin to matter with SNRs of ${\cal{O}}(100)$, the Fisher approximation should be reliable. 

There are some practical considerations that can make computations of the Fisher matrix somewhat tricky. First, the integrands inside of the inner product are, typically, highly oscillatory integrals, due to the fact that the inner product produces ``cross" terms between different harmonics, whose phases will not cancel. One has to be carefully when numerically integrating these highly oscillatory integrands to avoid introducing uncontrolled numerical error in the elements of the Fisher matrix. We use \texttt{Mathematica}'s \texttt{NIntegrate} command to perform the numerical integration, and have verified that the results are robust to numerical error, with each element of the Fisher matrix computed to a relative accuracy of $10^{-12}$. The second consideration is the inversion of the Fisher matrix. Matrices, especially those whose elements have been computed using numerical methods like we do here, are often badly conditioned for numerical inversion routines, introducing large errors in the elements of the inverse matrix. For the inversion, we use \texttt{Mathematica}'s \texttt{Inverse} routine, and compute the product between the Fisher matrix and its inversion using this method. If the inversion were exact, this would produce the identity matrix. We have verified that the deviation in the elements of the matrix product from those of the identity matrix are of the same level as the numerical error in our integration methods, and thus the inversion does not introduce any erroneous errors in our results. 

For the comparison, we consider binaries with different component masses and initial eccentricities. We take the initial orbital frequency to be $F_{\I} = 5$Hz, while the initial longitude of pericenter is $\omega_{\I} = \pi/4$. We take the orientation of the binary to be ``face-on" so that $\iota = 0$, and we set the polarization angle $\upsilon = \pi/2$. We make this choice since most of the GW power is emitted along the axis defined by the orbital angular momentum, creating an observational bias where systems that are face-on are more likely to be observed than systems that are not. Further, we take the luminosity distance of the source to be $D_{\lum} = 100$ Mpc, and we set the sky location of the source to $\Theta = 0.7923$ and $\Phi = 1.4293$. For $S_{n}(f)$, we use the expected aLIGO noise spectrum at design sensitivity~\cite{Matt-priv, Aasi:2013wya}. Given this, we will estimate the accuracy to which the parameters $\lambda^{a} = (\ln {\cal{M}}, e_{\I}, t_{c}, \phi_{c}, \ln D_{\lum})$ can be measured. The remaining parameters, specifically $(F_{\I}, \omega_{\I}, \Theta, \Phi)$ cause the Fisher matrices to be badly conditioned and the inversion to contain significant numerical error. We take the limits of integration in the inner product to be $f_{1} = 20$ Hz and $f_{2} = f_{\ISCO} = 2 F_{\ISCO}$.

The results for the measurability of the initial eccentricity are given in Table~\ref{fisher-tab}. We study several BH binaries, with varying total masses and mass ratios as can be seen from the first two columns. For convenience, we provide the SNR of the signals in the third column. We also vary the initial eccentricity for each set of masses over the three values listed in the fourth column. We list the accuracy to which the eccentricity can be measured using the PACMAN waveform is the fifth column. Generally, the eccentricity can be measured using aLIGO to $10^{-3} - 10^{-2}$. The accuracy increases with increasing SNR and increasing initial eccentricity, which matches previous studies investigating the measurability of the initial eccentricity~\cite{PhysRevD.92.044034}. We compare the accuracy between the PA and orbit-averaged waveforms in the sixth column by computing the difference between these values. For the systems studied, this difference is at most ${\cal{O}}(10^{-6})$, but is typically $\lesssim {\cal{O}}(10^{-7})$. This indicates that the PA effects that capture the secular growth of eccentricity investigated here do not significantly improve the precision to which we can measure eccentricity with aLIGO.

%
\begin{table}
\centering
\begin{centering}
\begin{tabular}{ccc|cccc}
\hline
\hline
	 $ m_{1}/M_{\odot} $  & $ m_{2}/M_{\odot} $ & SNR & $ e_{\I} $ & $ \Delta e_{\I}^{\mbox{\tiny PACMAN}}$ & $\Delta e_{\I}^{\mbox{\tiny PACMAN}} - \Delta e_{\I}^{\OA}$ \\ 
\hline
	$ $ & $ $ & $ $ & $0$ & $9.1\times10^{-3}$ & $-9.8\times10^{-9}$\\
	$ 5 $ & $ 10 $ & $ 137 $ & $10^{-3}$ & $9.1\times10^{-3}$ & $8.5\times10^{-7}$\\
	$ $ & $ $ & $ $ & $10^{-2}$ & $8.6\times10^{-3}$ & $7.9\times10^{-6}$\\
\hline
	$ $ & $ $ & $ $ & $0$ & $6.8\times10^{-3}$ & $-1.7\times10^{-7}$\\
	$ 10 $ & $ 10 $ & $ 178 $ & $10^{-3}$ & $6.8\times10^{-3}$ & $1.5\times10^{-7}$\\
	$ $ & $ $ & $ $ & $10^{-2}$ & $6.7\times10^{-3}$ & $3.2\times10^{-6}$\\
\hline
	$ $ & $ $ & $ $ & $0$ & $6.1\times10^{-3}$ & $-7.9\times10^{-8}$\\
	$ 5 $ & $ 30 $ & $ 176 $ & $10^{-3}$ & $6.1\times10^{-3}$ & $8.0\times10^{-8}$\\
	$ $ & $ $ & $ $ & $10^{-2}$ & $6.1\times10^{-3}$ & $1.7\times10^{-6}$\\
\hline
	$ $ & $ $ & $ $ & $0$ & $2.9\times10^{-3}$ & $-7.7\times10^{-7}$\\
	$ 30 $ & $ 30 $ & $ 318 $ & $10^{-3}$ & $2.9\times10^{-3}$ & $-7.5\times10^{-7}$\\
	$ $ & $ $ & $ $ & $10^{-2}$ & $2.9\times10^{-3}$ & $-5.7\times10^{-7}$\\
\hline
\hline
\end{tabular}
\end{centering}
\caption{\label{fisher-tab} Accuracy to which eccentricity can be measured by aLIGO using a Fisher analysis for various BH binaries. The fourth column displays the accuracy to which the initial eccentricity can be recovered using the PA waveform. The same value for the orbit-averaged waveform does not differ significantly from this, as can be seen from the fifth column.}
\end{table}

This result is different from what we found when we studied the match between waveforms in Sec.~\ref{sec:match}. However, the Fisher analysis and match calculation help to quantify separate features of this problem. The match helps us to determine when parameters will become biased due to systematics in waveform modeling. As a result, the recovered parameters will not be the same as the true parameters of the system observed. Thus, the match helps us determine the \textit{accuracy} of recovered parameters relative to their true values. On the other hand, the Fisher analysis helps us quantify the \textit{precision} to which we can measure parameters. In the context of posterior probability distributions, the match gives us information about the peak of the posterior, while the Fisher analysis, specifically $\Delta \lambda^{a}$, provides us with the variance of the posterior.

\section{Discussion}

We have here expanded on the discussion started in~\cite{Loutrel:2018ssg} by investigating the effects of the secular growth on the observation of GWs from inspiraling binaries. We have shown that the orbit-averaged approximation does indeed break down in the low eccentricity limit when compared to the direct integration of the relative acceleration of the binary. Meanwhile, the osculating approximation, which exhibits the secular growth, reproduces the direct integration to double precision. Thus, the reason for the discrepancy between the orbit-averaged and osculating approximation is a result of oscillations that are assumed to average out as the binary inspirals. While this holds approximately for any one given orbit, the violation of this averaging will build over many orbits and become non-negligible in the late inspiral, producing the growth in eccentricity.

Having verified this, we investigated the effect that secular growth would have on observations of GWs from systems where this effect is seen in the osculating approximation. First, we considered the match between numerical time domain waveforms computed via the three methods considered here: direct integration, osculating approximation, and the orbit-averaged approximation. We have shown that the match between the waveforms computed using the orbit-averaged approximation and direct integration are not identical, with mismatches around $1-M \sim 10^{-4}$. The nominal SNR at which this will begin to bias parameter estimation is $\rho \sim 100$ for the systems considered here. However, this statement is for the recovered parameters to be within 1-$\sigma$ of there ``true" values. As the SNR increases, parameters will be systematically more biased, and thus less accurately recovered, relative to injected parameters. One avenue of future work is to perform parameter estimation using Bayesian inference to determine how biased recovered parameters of a binary will be if one uses orbit-averaged waveforms, where the injected waveform would be those found via our direct integration method. Thanks to the work carried out here, the same parameter estimation study could be achieved with the PACMAN waveform, and using Markov chain Monte-Carlo methods to explore the parameter space of the model.

Second, we considered the accuracy to which the initial eccentricity of the source can be measured using observations by aLIGO at design sensitivity. We showed that the difference between the accuracy as measured by PA and orbit-averaged waveforms is $\lesssim {\cal{O}}(10^{-6})$, indicating that PA effects will not significantly improve the precision to which we can measure eccentricity. We would expect that this holds even if one were to consider the more rigorous parameter estimation study detailed above, specifically, the ``width" of posteriors should not change when using orbit-averaged or PA waveforms. 

Eccentricity has a rich and non-trivial impact on the dynamics of binary systems, which, as has been shown here, must be investigated carefully. While our analyses indicate that the violation of the orbit-averaged approximation, and thus the secular growth of eccentricity, will not significantly impact observations, we have only focused our efforts on small eccentricity systems. For highly eccentric binaries, the orbital period diverges as $(1-e^{2})^{-3/2}$, which may become comparable to the radiation-reaction timescale if the eccentricity is sufficiently large. This would indicate a severe breakdown in the orbit-averaged approximation for such systems, which warrants further investigation.

\section*{Acknowledgements}

We would like to thank Katerina Chatziioannou, Peter Diener, Davide Gerosa, Achamveedu Gopakumar, Scott Hughes, Sylvain Marsat, Amos Ori, Frans Pretorius, Leo Stein, Niels Warburton, Clifford Will, and Aaron Zimmerman for insightful suggestions and other comments. N. L. acknowledges support from NSF grant PHY-1607449, the Simons Foundation, and the Canadian Institute For Advanced Research (CIFAR). N.Y. acknowledges support from NSF CAREER grant PHY-1250636 and NASA grants NNX16AB98G and 80NSSC17M0041. N. J. C. acknowledges support from the NSF award PHY-1306702 and NASA award NNX16AB98G.

\appendix
\section{$(C_{a}^{j},S_{a}^{j})$-Coefficients in the Osculating Equations of the Radiation-Reaction Problem}
\label{coeffs}

We here provide the harmonic coefficients $(C_{a}^{j}, S_{a}^{j})$ from the osculating equations in Eqs.~\eqref{eq:dpdt-osc}-\eqref{eq:dphidt-osc}. In the harmonic gauge, the non-zero coefficients are
\allowdisplaybreaks[4]
\begin{align}
C_{p}^{0} &= -192 - 696 \alpha^2 - 162 \alpha^4 - 696 \beta^2 - 324 \alpha^2 \beta^2 - 162 \beta^4\,,
\\
C_{p}^{1} &= -816 \alpha - 828 \alpha^3 - 36 \alpha^5 - 828 \alpha \beta^2 - 72 \alpha^3 \beta^2 - 36 \alpha \beta^4\,,
\\
C_{p}^{2} &= -648 \alpha^2 - 192 \alpha^4 + 648 \beta^2 + 192 \beta^4\,,
\\
C_{p}^{3} &= -228 \alpha^3 - 12 \alpha^5 + 684 \alpha \beta^2 + \beta24 \alpha^3 \beta^2 + 36 \alpha \beta^4\,,
\\
C_{p}^{4} &= -30 \alpha^4 + 180 \alpha^2 \beta^2 - 30 \beta^4\,,
\\
S_{p}^{1} &= -816 \beta - 828 \alpha^2 \beta - 36 \alpha^4 \beta - 828 \beta^3 - 72 \alpha^2 \beta^3 - 36 \beta^5\,,
\\
S_{p}^{2} &= -1296 \alpha \beta - 384 \alpha^3 \beta - 384 \alpha \beta^3\,,
\\
S_{p}^{3} &= -684 \alpha^2 \beta - 36 \alpha^4 \beta + 228 \beta^3 - 24 \alpha^2 \beta^3 + 12 \beta^5\,,
\\
S_{p}^{4} &= -120 \alpha^3 \beta + 120 \alpha \beta^3\,,
\\
C_{\alpha}^{0} &= -1984 \alpha - 3564 \alpha^3 - 542 \alpha^5 - 3564 \alpha \beta^2 - 1084 \alpha^3 \beta^2 - 542 \alpha \beta^4\,,
\\
C_{\alpha}^{1} &= -768 - 6440 \alpha^2 - 3604 \alpha^4 - 108 \alpha^6 - 2392 \beta^2 - 4608 \alpha^2 \beta^2 - 252 \alpha^4 \beta^2
\nn \\
&- 1004 \beta^4 - 180 \alpha^2 \beta^4 - 36 \beta^6\,,
\\
C_{\alpha}^{2} &= -2048 \alpha - 4848 \alpha^3 - 829 \alpha^5 - 768 \alpha \beta^2 - 878 \alpha^3 \beta^2 - 49 \alpha \beta^4\,,
\\
C_{\alpha}^{3} &= -2200 \alpha^2 - 1834 \alpha^4 - 66 \alpha^6 + 2200 \beta^2 + 1524 \alpha^2 \beta^2 - 30 \alpha^4 \beta^2 + 1326 \beta^4 
\nn \\
&+ 90 \alpha^2 \beta^4 + 54 \beta^6\,,
\\
C_{\alpha}^{4} &= -1188 \alpha^3 - 322 \alpha^5 + 3564 \alpha \beta^2 + 844 \alpha^3 \beta^2 + 766 \alpha \beta^4\,,
\\
C_{\alpha}^{5} &= -322 \alpha^4 - 18 \alpha^6 + 1932 \alpha^2 \beta^2 + 90 \alpha^4 \beta^2 - 322 \beta^4 + 90 \alpha^2 \beta^4 - 18 \beta^6\,,
\\
C_{\alpha}^{6} &= -35 \alpha^5 + 350 \alpha^3 \beta^2 - 175 \alpha \beta^4\,,
\\
S_{\alpha}^{1} &= -4048 \alpha \beta - 2600 \alpha^3 \beta - 72 \alpha^5 \beta - 2600 \alpha \beta^3 - 144 \alpha^3 \beta^3 - 72 \alpha \beta^5\,,
\\
S_{\alpha}^{2} &= -2048 \beta - 6888 \alpha^2 \beta - 1219 \alpha^4 \beta - 2808 \beta^3 - 1658 \alpha^2 \beta^3 - 439 \beta^5\,,
\\
S_{\alpha}^{3} &= -4400 \alpha \beta - 4176 \alpha^3 \beta - 144 \alpha^5 \beta - 2144 \alpha \beta^3 - 240 \alpha^3 \beta^3 - 96 \alpha \beta^5\,,
\\
S_{\alpha}^{4} &= -3564 \alpha^2 \beta - 1016 \alpha^4 \beta + 1188 \beta^3 - 344 \alpha^2 \beta^3 + 272 \beta^5\,,
\\
S_{\alpha}^{5} &= -1288 \alpha^3 \beta - 72 \alpha^5 \beta + 1288 \alpha \beta^3 + 72 \alpha \beta^5\,,
\\
S_{\alpha}^{6} &= -175 \alpha^4 \beta + 350 \alpha^2 \beta^3 - 35 \beta^5\,,
\\
C_{\beta}^{0} &= -1984 \beta - 3564 \alpha^2 \beta - 542 \alpha^4 \beta - 3564 \beta^3 - 1084 \alpha^2 \beta^3 - 542 \beta^5\,,
\\
C_{\beta}^{1} &= -4048 \alpha \beta - 2600 \alpha^3 \beta - 72 \alpha^5 \beta - 2600 \alpha \beta^3 - 144 \alpha^3 \beta^3 - 72 \alpha \beta^5\,,
\\
C_{\beta}^{2} &= 2048 \beta + 768 \alpha^2 \beta + 49 \alpha^4 \beta + 4848 \beta^3 + 878 \alpha^2 \beta^3 + 829 \beta^5\,,
\\
C_{\beta}^{3} &= 4400 \alpha \beta + 2144 \alpha^3 \beta + 96 \alpha^5 \beta + 4176 \alpha \beta^3 + 240 \alpha^3 \beta^3 + 144 \alpha \beta^5\,,
\\
C_{\beta}^{4} &= 3564 \alpha^2 \beta + 766 \alpha^4 \beta - 1188 \beta^3 + 844 \alpha^2 \beta^3 - 322 \beta^5\,,
\\
C_{\beta}^{5} &= 1288 \alpha^3 \beta + 72 \alpha^5 \beta - 1288 \alpha \beta^3 - 72 \alpha \beta^5\,,
\\
C_{\beta}^{6} &= 175 \alpha^4 \beta - 350 \alpha^2 \beta^3 + 35 \beta^5\,,
\\
S_{\beta}^{1} &= -768 - 2392 \alpha^2 - 1004 \alpha^4 - 36 \alpha^6 - 6440 \beta^2 - 4608 \alpha^2 \beta^2 - 180 \alpha^4 \beta^2 
\nn \\
&- 3604 \beta^4 - 252 \alpha^2 \beta^4 - 108 \beta^6\,,
\\
S_{\beta}^{2} &= -2048 \alpha - 2808 \alpha^3 - 439 \alpha^5 - 6888 \alpha \beta^2 - 1658 \alpha^3 \beta^2 - 1219 \alpha \beta^4\,,
\\
S_{\beta}^{3} &= -2200 \alpha^2 - 1326 \alpha^4 - 54 \alpha^6 + 2200 \beta^2 - 1524 \alpha^2 \beta^2 - 90 \alpha^4 \beta^2 + 1834 \beta^4 
\nn \\
&+ 30 \alpha^2 \beta^4 + 66 \beta^6\,,
\\
S_{\beta}^{4} &= -1188 \alpha^3 - 272 \alpha^5 + 3564 \alpha \beta^2 + 344 \alpha^3 \beta^2 + 1016 \alpha \beta^4\,,
\\
S_{\beta}^{5} &= -322 \alpha^4 - 18 \alpha^6 + 1932 \alpha^2 \beta^2 + 90 \alpha^4 \beta^2 - 322 \beta^4 + 90 \alpha^2 \beta^4 - 18 \beta^6\,,
\\
S_{\beta}^{6} &= -35 \alpha^5 + 350 \alpha^3 \beta^2 - 175 \alpha \beta^4\,,
\\
C_{\phi}^{0} &= 1 + \frac{1}{2} \left(\alpha^2 + \beta^2\right)\,,
\\
C_{\phi}^{1} &= 2 \alpha\,,
\\
C_{\phi}^{2} &= \frac{1}{2} \left(\alpha^2 - \beta^2\right)\,,
\\
S_{\phi}^{1} &= 2 \beta\,,
\\
S_{\phi}^{2} &= \alpha \beta\,.
\end{align}
%

\section{Waveforms for Eccentric Systems}
\label{pols}

Typically, when one considers waveforms for eccentric systems, one usually uses the well known polarization given in Eqs.~(3.1) and~(3.2) in~\cite{PhysRevD.80.084001}, for example. These waveforms are often constructed by working in an orbital parameterization where the longitude of pericenter is fixed, and thus, one can rotate the orbital plane such that $\omega=0$. However, in the osculating evolution, this is no longer the case since $\omega$ obtains a non-trivial oscillatory evolution. Thus, if we desire to make comparisons of the waveforms computed using various methods of solving the equations of motion, we require more general waveform polarizations.

The GWs we are considering will propagate along the direction $\vec{N} = (\sin\iota \cos\upsilon, \sin\iota \cos\upsilon, \cos{\iota})$, where $(\iota, \upsilon)$ are the inclination angle of the binary and an arbitrary polarization angle, respectively. We define two vectors which span the polarization sub-space, specifically $\vec{\cal{I}} = (\cos\iota \cos\upsilon, \cos\iota \sin\upsilon, - \sin\upsilon)$ and $\vec{\Upsilon} = (-\cos\upsilon, \sin\upsilon,0)$. The projectors into the plus and cross polarizations are then
\begin{equation}
e_{+}^{ij} = \frac{1}{2} \left[{\cal{I}}^{i} {\cal{I}}^{j} - \Upsilon^{i} \Upsilon^{j}\right]\,, \qquad e_{\times}^{ij} = \frac{1}{2} \left[{\cal{I}}^{i} \Upsilon^{j} + \Upsilon^{i} {\cal{I}}^{j}\right]\,,
\end{equation}
where $({\cal{I}}^{i}, \Upsilon^{i})$ are the components of $\vec{{\cal{I}}}$ and $\vec{\Upsilon}$, respectively. Our starting points will be the waveform polarizations in terms of the relative coordinates. The metric components describing the waveform are given by the second time derivative of the binary system's quadrupole moment, which becomes $h_{ij} = (4 \eta M/D_{\lum}) [v_{i} v_{j} - (M/r) n_{i} n_{j}]$, where $\vec{v}$ and $\vec{n}$ are the relative velocity and the unit radial vector, respectively. Applying the above polarization projectors to this waveform, we obtain
\begin{align}
h_{+} &= \frac{2 M \eta}{D_{\lum}}\Bigg\{2 \left(1 + \cos^{2}\iota\right) \cos\upsilon \sin\upsilon \left[r \dot{r} \dot{\phi} \cos(2\phi) + \frac{1}{2}\left(\dot{r}^{2} - \frac{M}{r} - r^{2} \dot{\phi}^{2}\right) \sin(2\phi)\right]
\nn \\
&+ \left(\cos^{2}\iota \cos^{2}\upsilon - \sin^{2}\upsilon\right) \left[\frac{1}{2} \left(\dot{r}^{2} + r^{2} \dot{\phi}^{2} - \frac{M}{r}\right) + \frac{1}{2} \left(\dot{r}^{2} - r^{2} \dot{\phi}^{2} - \frac{M}{r}\right) \cos(2\phi)
\right.
\nn\\
&\left.
- r \dot{r} \dot{\phi} \sin(2\phi)\right] + \left(\cos^{2}\iota \sin^{2}\upsilon - \cos^{2}\upsilon\right)\left[\frac{1}{2}\left(\dot{r}^{2} + r^{2} \dot{\phi}^{2} - \frac{M}{r}\right)
\right.
\nn \\
&\left.
- \frac{1}{2} \left(\dot{r}^{2} - r^{2} \dot{\phi}^{2} - \frac{M}{r}\right) \cos(2\phi) + r \dot{r} \dot{\phi} \sin(2\phi) \right]\Bigg\}\,,
\\
h_{\times} &= \frac{2 \eta M}{D_{\lum}} \cos\iota \left\{2 r \dot{r} \dot{\phi}  \cos[2 (\phi - \upsilon)] + \left(\dot{r}^{2} - r^{2} \dot{\phi}^{2} - \frac{M}{r}\right) \sin[2(\phi - \upsilon)]\right\}\,.
\end{align}

To obtain the waveforms needed for the orbit-averaged and osculating methods, one must insert the Keplerian parameterization in Sec.~\ref{newt} into the above waveforms. For the orbit-averaged approximation, we write the polarization in terms of the Keplerian eccentricity and longitude of pericenter, specifically
\begin{align}
h_{+} &= - \frac{2 \eta M^{2}}{p D_{\lum}} \left[ \left(1 + \cos^{2}\iota\right)\Bigg\{\cos[2(\phi-\upsilon)] + \frac{5 e_{\K}}{4} \cos(\phi - 2 \upsilon + \omega) + \frac{e_{\K}}{4} \cos(3 \phi -2 \upsilon - \omega) 
\right.
\nn \\
&\left.
\qquad + \frac{e_{\K}^{2}}{2} \cos[2(\upsilon - \omega)]\Bigg\} + \frac{1}{2} \sin^{2}\iota \left[e_{\K}^{2} + e_{\K} \; \cos(\phi - \omega)\right]\right]\,,
\\
h_{\times} &= - \frac{2 \eta M^{2}}{p D_{\lum}} \cos\iota \left\{2 \sin[2(\phi - \upsilon)] + \frac{5e_{\K}}{2} \sin(\phi - 2\upsilon + \omega) + \frac{e_{\K}}{2} \sin(3 \phi - 2 \upsilon - \omega) 
\right.
\nn \\
&\left.
\qquad - e_{\K}^{2} \; \sin[2 (\upsilon - \omega)]\right\}\,.
\end{align}
Note that these waveforms reduce to those of~\cite{PhysRevD.80.084001} when $\omega=0$. For the osculating method, we reparameterize these polarization in term of the components of the Runge-Lenz vector, specifically,
\begin{align}
\label{eq:h-osc}
h_{+,\times} &= - \frac{2 \eta M^{2}}{p D_{\lum}} \sum_{j=0}^{3} \left[H^{+,\times}_{j, \cj} \cos(j \phi) + H^{+,\times}_{j, \sj} \sin(j \phi)\right]\,,
\end{align}
with
\allowdisplaybreaks[4]
\begin{align}
\label{eq:h-osc-coeffs1}
H^{+}_{0,\cj} &= \frac{1}{8} \left\{2 \alpha^2 + 2 \beta^2 + (-2 \alpha^2 - 2 \beta^2) \cos(2 \iota) + (\alpha^2 - \beta^2) \cos(2 \iota - 2 \upsilon)
\right.
\nn \\
&\left.
+ (6 \alpha^2 - 6 \beta^2) \cos(2 \upsilon) + (\alpha^2 - \beta^2) \cos(2 \iota + 2 \upsilon) - 2 \alpha \beta \sin(2 \iota - 2 \upsilon) + 
\right.
\nn \\
&\left.
12 \alpha \beta \sin(2 \upsilon) + 2 \alpha \beta \sin(2 \iota + 2 \upsilon)\right\}
\\
H^{+}_{1,\cj} &= \frac{1}{16} \left\{4 \alpha + 5 \alpha \cos[2 (\iota - \upsilon)] + 30 \alpha \cos(2 \upsilon) + 5 \alpha \cos[2 (\iota + \upsilon)] + 30 \beta \sin(2 \upsilon) + 
\right.
\nn \\
&\left.
\cos(2 \iota) \left[-4 \alpha + 10 \beta \sin(2 \upsilon)\right]\right\}\,,
\\
H^{+}_{2,\cj} &= \frac{1}{2} \left[3 + \cos(2\iota)\right] \cos(2 \upsilon)\,,
\\
\label{eq:hpc3}
H^{+}_{3,\cj} &= \frac{1}{8} \left[3 + \cos(2\iota)\right] \left[\alpha \cos(2\upsilon) - \beta \sin(2\upsilon)\right]
\\
H^{+}_{1,\sj} &= \frac{1}{16} \left\{4 \beta - 5 \beta \cos[2 (\iota - \upsilon)] - 30 \beta \cos(2 \upsilon) - 5 \beta \cos[2 (\iota + \upsilon)] + 30 \alpha \sin(2 \upsilon) + 
\right.
\nn \\
&\left.
\cos(2 \iota) (-4 \beta + 10 \alpha \sin[2 \upsilon])\right\}\,,
\\
H^{+}_{2,\sj} &= \frac{1}{2} \left[3 + \cos(2\iota)\right] \sin(2 \upsilon)\,,
\\
\label{eq:hps3}
H^{+}_{3,\sj} &= \frac{1}{8} \left[3 + \cos(2\iota)\right] \left[\beta \cos(2\upsilon) + \alpha \sin(2\upsilon)\right]\,,
\\
H^{\times}_{1,\cj} &= \frac{5}{2} \cos\iota \left[ \beta \cos(2\upsilon) - \alpha \sin(2\upsilon)\right]
\\
H^{\times}_{2,\cj} &= -2 \cos\iota \sin(2\upsilon)
\\
H^{\times}_{3,\cj} &= - \frac{1}{2} \cos\iota \left[ \beta \cos(2\upsilon) + \alpha \sin(2\upsilon)\right]
\\
H^{\times}_{0,\cj} &= \cos\iota \left[2 \alpha \beta \cos(2\upsilon) + (- \alpha^{2} + \beta^{2}) \sin(2\upsilon)\right]
\\
H^{\times}_{1,\sj} &= \frac{5}{2} \cos\iota \left[ \alpha \cos(2\upsilon) + \beta \sin(2\upsilon)\right]
\\
H^{\times}_{2,\sj} &= 2 \cos\iota \cos(2\upsilon)
\\
\label{eq:h-osc-coeffs2}
H^{\times}_{3,\sj} &= \frac{1}{2} \cos\iota \left[\alpha \cos(2\upsilon) - \beta \sin(2\upsilon)\right]
\end{align}
%

\section{PACMAN Fourier Amplitudes}
\label{fas}

The PACMAN waveforms can be expressed in the form given in Eq.~\eqref{eq:h-pacman}, with the phase given by Eq.~\eqref{eq:psi-adiab}-\eqref{eq:psi-pa}. In this appendix, we list the amplitudes ${\cal{A}}_{j}(f)$. For simplicity, we assume we are observing the binary face on so that $\iota = 0$, and we set the polarization angle $\upsilon = \pi/2$, so as to match the commonly used waveforms in~\cite{PhysRevD.80.084001}. Generically, the amplitudes can be written schematically as ${\cal{A}}_{j}(f) = F_{+} {\cal{A}}_{j}^{+}(f) + F_{\times} {\cal{A}}_{j}^{\times}(f)$, where the plus and cross amplitudes can be written in a PA style expansion as 
\begin{equation}
\label{eq:calA-pa}
{\cal{A}}_{j}^{+,\times}(f) = {\cal{A}}_{j,0}^{+,\times}(f) + (\pi {\cal{M}} f)^{5/3} {\cal{A}}_{j,\PA}^{+,\times}(f) + (\pi {\cal{M}} f)^{10/3} {\cal{A}}_{j,{\mbox{\tiny 2}}\PA}^{+,\times}(f)\,,
\end{equation}
representing the adiabatic, 1PA, and 2PA terms, respectively. To non-zero amplitude functions for the plus polarization are as follows, with $\tilde{\chi} = F_{\I}/f$:
\begin{align}
\label{eq:calA-1-0}
{\cal{A}}_{1,0}^{+}(f) &= \frac{3}{2^{2/3}} e_{\I} \tilde{\chi}^{19/18} e^{-i \omega_{\I}}\,, 
\\
{\cal{A}}_{2,0}^{+}(f) &= -4 + \frac{2^{1/9} 277}{3} e_{\I}^{2}\tilde{\chi}^{19/9}\,, 
\\
{\cal{A}}_{3,0}^{+}(f) &= - \frac{3^{13/18} 27}{2^{2/3}} e_{\I} \tilde{\chi}^{19/18} e^{i \omega_{\I}}\,, 
\\
{\cal{A}}_{4,0}^{+}(f) &= - 2^{8/9} 256 e_{\I}^{2} \tilde{\chi}^{19/19} e^{2 i \omega_{\I}}\,,
\\
{\cal{A}}_{1,\PA}^{+}(f) &= - 128 i \tilde{\chi}^{49/18} + i e_{\I} \left[\frac{11456}{45} \tilde{\chi}^{19/18} e^{-i\omega_{\I}} -\frac{64}{45} \tilde{\chi}^{49/18} e^{-i\omega_{\I}}(-37 + 75 e^{2i\omega_{\I}})\right]
\nn \\
& + i e_{\I}^{2} \left[\frac{52096}{85} \tilde{\chi}^{19/18} (1 - e^{-2i\omega_{\I}}) - \frac{4}{171} \tilde{\chi}^{49/18} e^{-2i\omega_{\I}} \left(4053 + 26151 e^{2i\omega_{\I}} + 1729 e^{4i\omega_{\I}}\right) 
\right.
\nn \\
&\left.
- \frac{4}{4845} \tilde{\chi}^{29/6} e^{-2i\omega_{\I}} \left(-893151 + 22910 e^{2i\omega_{\I}} + 19380 e^{4i\omega_{\I}}\right)\right]\,,
\\
{\cal{A}}_{2,\PA}^{+}(f) &= \frac{256i}{5} + \frac{2^{7/9} 128i}{15} e_{\I} \tilde{\chi}^{34/9} e^{-i\omega_{\I}} (-145 + 49 e^{2i\omega_{\I}}) + i e_{\I}^{2} \left[ \frac{2^{1/9} 393152}{225} \tilde{\chi}^{19/9} 
\right.
\nn \\
&\left.
+ \frac{2^{7/9} 64}{225} \tilde{\chi}^{34/9} e^{-2 i \omega_{\I}} \left(-3625 + 1184 e^{2i\omega_{\I}} + 1225 e^{4i\omega_{\I}} \right)\right]\,,
\\
{\cal{A}}_{3,\PA}^{+}(f) &= \frac{3^{13/18} 384i}{5}\tilde{\chi}^{49/18} + i e_{\I} \left[- \frac{24512}{3^{17/18} 75} \tilde{\chi}^{19/18} e^{i\omega_{\I}} 
\right.
\nn \\
&\left.
- \frac{64}{3^{5/18} 25} \tilde{\chi}^{49/18} e^{-i \omega_{\I}} (-75 + 37 e^{2i\omega_{\I}})\right]\,,
\\
{\cal{A}}_{4,\PA}^{+}(f) &= \frac{2^{5/9}12288i}{5} e_{\I} \tilde{\chi}^{34/9} e^{i\omega_{\I}} + i e_{\I}^{2} \left[-\frac{2^{2/9}726944}{225} \tilde{\chi}^{19/9} e^{2i\omega_{\I}} 
\right.
\nn \\
&\left.
- \frac{2^{5/9} 2048}{75} \tilde{\chi}^{34/9} (-75 + 37 e^{2i\omega_{\I}})\right]\,,
\\
{\cal{A}}_{5,\PA}^{+}(f) &= 5^{5/6} 20400 i e_{\I}^{2} \tilde{\chi}^{29/6} e^{2i\omega_{\I}}\,,
\\
{\cal{A}}_{1,{\mbox{\tiny 2}}\PA}^{+}(f) &= - \frac{2^{2/3} 65536}{25} \tilde{\chi}^{49/18} + e_{\I} \left[ \frac{2^{2/3} 8249344}{1125} \tilde{\chi}^{19/18} e^{-i\omega_{\I}} 
\right.
\nn \\
&\left.
- \frac{2^{2/3}4096}{25} \tilde{\chi}^{49/18} e^{-i\omega_{\I}} (2393 + 2775 e^{2i\omega_{\I}}) + \frac{2^{2/3} 439296}{85} \tilde{\chi}^{13/2} e^{-i\omega_{\I}}(-1 + e^{2i\omega_{\I}})\right] 
\nn \\
&+ e_{\I}^{2} \left[- \frac{2^{2/3} 157696}{75} \tilde{\chi}^{19/18} e^{-2i\omega_{\I}} (-1 + e^{2i\omega_{\I}}) 
\right.
\nn \\
&\left.
+ \frac{2^{2/3} 256}{72675} \tilde{\chi}^{49/18} (-378441 - 4118328 e^{2i\omega_{\I}} + 153881 e^{4i\omega_{\I}}) 
\right.
\nn \\
&\left.
- \frac{2^{2/3} 512}{1425} \tilde{\chi}^{29/6} e^{-2i\omega_{\I}} (9036 + 33171 e^{2i\omega_{\I}} + 37525 e^{4i\omega_{\I}}) 
\right.
\nn \\
&\left.
+ \frac{2^{2/3} 146432}{1275} \tilde{\chi}^{13/2} e^{-2i\omega_{\I}} (-19 - 56 e^{2i\omega_{\I}} + 75 e^{4i\omega_{\I}})\right]
\\
{\cal{A}}_{2,{\mbox{\tiny 2}}\PA}^{+}(f) &= \frac{8192}{25} + e_{\I} \left[-\frac{270336i}{425} \sin\omega_{\I} + \frac{2^{7/9} 4096}{1275} \tilde{\chi}^{34/9} e^{-i\omega_{\I}} (4291 + 3019 e^{2i\omega_{\I}})\right] 
\nn \\
&+ e_{\I}^{2} \left[ - \frac{256}{195075} e^{-2i\omega_{\I}} (698787 - 1411344 e^{2i\omega_{\I}} + 712557 e^{4i\omega_{\I}}) + \frac{2^{1/9} 20621312}{675} \tilde{\chi}^{19/9} 
\right.
\nn \\
&\left.
+ \frac{2^{7/9} 2048}{195075} \tilde{\chi}^{34/9}e^{-2i\omega_{\I}} (142617 - 228182 e^{2i\omega_{\I}} + 1030017 e^{4i\omega_{\I}}) 
\right.
\nn \\
&\left.
+ \frac{2^{5/9} 49152}{7225} \tilde{\chi}^{68/9} e^{-2i\omega_{\I}} (-12545 - 35158 e^{2i\omega_{\I}} + 47703 e^{4i\omega_{\I}}) \right]
\\
{\cal{A}}_{3,{\mbox{\tiny 2}}\PA}^{+}(f) &= e_{\I} \left[ - \frac{2^{2/3} 155648}{3^{11/19} 75} \tilde{\chi}^{19/18} e^{i\omega_{\I}} - \frac{2^{2/3} 3^{1/18} 135168}{425} \tilde{\chi}^{49/18} e^{-i\omega_{\I}} (-1 + e^{2i\omega_{\I}}) 
\right.
\nn \\
&\left.
- \frac{2^{2/3} 3^{5/6} 11860992}{425} \tilde{\chi}^{13/2} e^{-i\omega_{\I}} (-1 + e^{2i\omega_{\I}})\right] + e_{\I}^{2} \left[-\frac{2^{2/3} 72472576}{3^{11/18} 19125} \tilde{\chi}^{19/18} (-1 + e^{2i\omega_{\I}}) 
\right.
\nn \\
&\left.
+ \frac{2^{2/3} 1792}{3^{17/18} 2125} \tilde{\chi}^{49/18} e^{-2i\omega_{\I}} (975 - 1408 e^{2i\omega_{\I}} + 433 e^{4i\omega_{\I}}) 
\right.
\nn \\
&\left.
+ \frac{2^{2/3} 3^{1/6} 512}{2125} \tilde{\chi}^{29/6} (1129173 + 397172 e^{2i\omega_{\I}}) 
\right.
\nn \\
&\left.
- \frac{2^{2/3} 3^{5/6} 1317888}{2125} \tilde{\chi}^{13/2} e^{-2i\omega_{\I}} (-75 + 56 e^{2i\omega_{\I}} + 19 e^{4i\omega_{\I}})\right]
\\
{\cal{A}}_{4,{\mbox{\tiny 2}}\PA}^{+}(f) &= - \frac{2^{8/9} 827392}{75} e_{\I} \tilde{\chi}^{34/9} e^{i\omega_{\I}} + e_{\I}^{2} \left[- \frac{2^{5/9} 47217664}{3375} \tilde{\chi}^{19/9} e^{2i\omega_{\I}} 
\right.
\nn \\
&\left.
- \frac{2^{8/9} 1024}{57375} \tilde{\chi}^{34/9} (87420 + 173564 e^{2i\omega_{\I}}) - \frac{2^{4/9} 1843396608}{425} \tilde{\chi}^{68/9} (-1 + e^{2i\omega_{\I}}) \right]\,,
\\
\label{eq:calA-5-2}
{\cal{A}}_{5,{\mbox{\tiny 2}}\PA}^{+}(f) &= - \frac{2^{2/3} 5^{1/6} 900608}{3} \tilde{\chi}^{29/6} e^{2i\omega_{\I}}\,.
\end{align}
The following relations hold between the amplitude functions of the plus and cross polarizations:${\cal{A}}_{j,0}^{\times} = i {\cal{A}}_{j,0}^{+}$, ${\cal{A}}_{j\ne1,(\PA,{\mbox{\tiny 2}}\PA)}^{\times} = i {\cal{A}}_{j\ne1,(\PA,{\mbox{\tiny 2}}\PA)}^{+}$, and
\begin{align}
\label{eq:calA-pa-rel}
{\cal{A}}_{1,\PA}^{\times} - i {\cal{A}}_{1,\PA}^{+} &= - 32 e_{\I}^{2} \tilde{\chi}^{29/6} e^{2i\omega_{\I}}\,,
\\
\label{eq:calA-2pa-rel}
{\cal{A}}_{1,{\mbox{2}}\PA}^{\times} - i {\cal{A}}_{1,{\mbox{2}}\PA}^{+} &= \frac{2^{2/3} 80896}{3} e_{\I}^{2} \tilde{\chi}^{29/6} e^{2i\omega_{\I}}\,,
\end{align}
which can be used to obtain the amplitude functions for the cross polarization.

\section*{References}
\bibliography{master}
\end{document}